\documentclass{article}

\usepackage[english]{babel}
\usepackage[utf8]{inputenc}
\usepackage[letterpaper,top=2cm,bottom=2cm,left=3cm,right=3cm,marginparwidth=1.75cm]{geometry}

\usepackage{amsmath,amssymb,amsthm}
\usepackage{latexsym}            
\usepackage{multirow}
\usepackage{epsfig}
\usepackage{epstopdf}
\usepackage{amsfonts,amscd, makecell}
\usepackage[ruled,vlined]{algorithm2e}
\usepackage{color,comment}
\usepackage{amsbsy,bm}
\usepackage{subfigure}
\usepackage{epstopdf}
\usepackage[utf8]{inputenc}          
\usepackage[T1]{fontenc}  

\usepackage{xcolor}
\usepackage{overpic}
\usepackage{appendix}

\newcommand\x{\bm{x}}

\newcommand\n{{\bm n}}

\allowdisplaybreaks

\newtheorem{thm}{Theorem}[section]

\newcommand{\be}{\begin{eqnarray}}
	\newcommand{\ee}{\end{eqnarray}}
\newcommand{\ben}{\begin{eqnarray*}}
	\newcommand{\een}{\end{eqnarray*}}

\newtheorem{theorem}{Theorem}[section]

\newtheorem{example}[theorem]{Example}

\newtheorem{remark}{Remark}[section]

\title{Phase-Field Modeling and Energy-Stable Schemes for Osmotic Flow through Semi-Permeable }
\author{Ruihan Guo, Xianmin Xu, Shixin Xu}
\author{ Ruihan Guo \thanks{School of Mathematics and Statistics, Zhengzhou University, Zhengzhou 450001, China}
	\and Jie Shen \thanks{School of Mathematical Science, Eastern Institute of Technology, Ningbo, 315200, China}
	\and Shixin Xu\thanks{Zu Chongzhi Center, Duke Kunshan University, 8 Duke Ave, Kunshan, Jiangsu, China. Corresponding Author} 
	\and Xianmin Xu \thanks{ LSEC, ICMSEC, NCMIS, Academy of Mathematics and Systems Science, Chinese Academy of
		Sciences, Beijing 100190, China}
}

\begin{document}
	\date{}
	\maketitle
	
	\begin{abstract}
		We present a thermodynamically consistent phase-field model for simulating fluid transport across semi-permeable membranes, with a particular focus on osmotic pressure effects. The model extends the classical Navier–Stokes–Cahn–Hilliard (NSCH) system by introducing an Allen–Cahn-type transmembrane flux governed by chemical potential imbalances, resulting in a strongly coupled system involving fluid motion, solute transport, and interface dynamics. To solve this system efficiently and accurately, we develop high-order, energy-stable numerical schemes. The local discontinuous Galerkin (LDG) method is employed for spatial discretization, offering high-order accuracy and geometric flexibility. For temporal integration, we first construct a first-order decoupled scheme with rigorous energy stability, and then improve temporal accuracy via a semi-implicit spectral deferred correction (SDC) method. Numerical experiments confirm the theoretical properties of the proposed scheme and demonstrate the influence of osmotic pressure and membrane permeability on droplet morphology at equilibrium. The framework offers a robust and versatile tool for modeling transmembrane fluid transport in both biological and industrial applications.
		
	\end{abstract}
	\paragraph{MSCcodes:}
	65M60, 65H10, 76T06, 92C10
	
	\section{Introduction}

	Semi-permeable membranes play a vital role in both biological and industrial systems by regulating the selective transport of substances across compartments \cite{gong2014immersed, jiang2013cellular}. In living cells, the plasma membrane is typically permeable to water but impermeable to solutes, enabling essential physiological processes such as hydration, nutrient uptake, and waste removal \cite{delpire2018water, tang2023phase}. Water transport across such membranes is driven by two competing forces: osmotic pressure, which results from solute concentration differences across the membrane, and hydrostatic pressure, which arises from fluid mechanical stress. Water tends to flow from regions of low to high osmotic pressure or from high to low hydrostatic pressure regions. The interplay between these forces determines the direction and magnitude of transmembrane flow. When imbalanced, this can lead to pathological effects such as cell swelling or shrinkage, impairing cellular function \cite{ho2006intracellular}.
	
	Modeling fluid transport through semi-permeable membranes under the influence of both osmotic and hydrostatic pressure presents significant challenges. Traditional sharp-interface methods, such as the immersed boundary method \cite{amiri2021immersed, layton2006modeling, wang2020immersed, yao2017numerical}, explicitly track the membrane as a moving interface between fluid domains. While these approaches can capture basic membrane dynamics, they typically require complex interface conditions to handle transmembrane flow and are limited when dealing with large deformations or topological changes.

	To overcome the limitations of sharp-interface models, phase-field methods offer a robust alternative by representing membranes as diffuse interfaces via continuous order parameters, naturally accommodating topological changes and large deformations. Phase-field approaches have been successfully used in vesicle dynamics, droplet interactions, and fluid–structure coupling problems \cite{AnMcFWh98, duqiang_2009_variational, lowengrub_quasiincompressible_1998, shen2022energy}. However, the modeling of osmotic-pressure-driven transmembrane water flow remains relatively unexplored. 
	Recent work has extended the framework to incorporate osmotic pressure and semi-permeable transport explicitly. For example, Rätz \cite{ratz2016diffuse} derived a thermodynamically consistent phase-field approximation that converges to a sharp-interface limit where only the concentrations of substances are considered.  
	Tang et al. \cite{tang2023phase} introduced a phase-field model for vesicle growth driven by osmotic pressure, incorporating an osmotic energy formulation.

	The Navier-Stokes–Cahn-Hilliard (NSCH) system offers a natural setting for modeling fluid–membrane interactions, particularly in two-phase flow systems. However, extending it to handle semi-permeable membranes—where water can independently traverse the interface due to osmotic pressure—poses new challenges. Specifically, it is nontrivial to capture the strong coupling between solute concentration, osmotic pressure, and transmembrane flow within a diffuse-interface framework. Several prior efforts have addressed related challenges: Camley et al. \cite{camley2013periodic} modeled intracellular confinement of active substances, Lowengrub et al. \cite{lowengrub2016numerical} introduced phase-field models with inextensibility constraints, and Garcke et al. \cite{garcke2013diffuse} proposed models for surfactants soluble in both phases but constrained at interfaces. However, these models do not account for osmotic-driven transport across semi-permeable boundaries.
	
	The primary objective of this work is to develop a thermodynamically consistent phase-field model that accounts for water transport across semi-permeable membranes driven by osmotic pressure differences. Our formulation extends the NSCH system to incorporate fluid–structure coupling and membrane permeability while preserving energy stability, thereby providing a more realistic framework for simulating transmembrane fluid transport in biological and industrial settings. 
	
	The energy variational approach \cite{eisenberg2010energy,shen2020energy,shen2022energy} leads to a highly nonlinear, coupled Navier–Stokes–Cahn– Hilliard–Allen–Cahn (NSCHAC) system, where the Allen–Cahn term captures the volume change induced by the transmembrane water flow. Simulating this system poses significant challenges due to its strong nonlinear couplings. Although many numerical schemes have been developed for Allen–Cahn or Cahn–Hilliard–Navier– Stokes systems \cite{yangxiaofeng_2015_decoupled, chen2016convergence, cheng2019energy, Diegel2019finite, guo_2014_midpoint, yan2018second, yang2020convergence, yang2017numerical}, our model exhibits a more intricate feedback structure: convection drives solute transport, the resulting solute gradient generates osmotic pressure, and osmotic flux in turn leads to volume changes that affect solute concentration. Consequently, the interplay between fluid motion, interfacial dynamics, and solute distribution is significantly stronger and more nonlinear than in traditional models.
	
	The second goal of this work is to develop energy-stable and decoupled numerical schemes for the proposed system. For spatial discretization, we adopt the local discontinuous Galerkin (LDG) method \cite{guo-xu,xu-shu}, which provides high-order accuracy, nonlinear stability, and adaptability to arbitrary mesh refinement (\(h\)- and \(p\)-adaptivity). For time integration, we first construct a first-order decoupled scheme and rigorously prove its energy stability under the LDG framework. Although efficient, this scheme is limited to first-order temporal accuracy. To address this, we incorporate the semi-implicit spectral deferred correction (SDC) method \cite{guo-xia, guo-xu-xia}, enabling high-order accuracy in both space and time for the thermodynamically consistent phase-field model of fluid transport through semi-permeable membranes.
	
	This paper is organized as follows. In Section~\ref{se:model}, we propose the phase-field model for semi-permeable membranes. Section~\ref{se:numerical-scheme} presents the numerical schemes: Subsection~\ref{subse1} introduces the first-order decoupled temporal scheme, Subsection~\ref{subse2} describes the LDG method for spatial discretization and proves its energy stability, and Subsection~\ref{subse3} applies the semi-implicit SDC method to enhance temporal accuracy. Section~\ref{se:numerical} provides numerical results validating the theoretical analysis and exploring the effects of interface permeability and osmotic pressure. Finally, concluding remarks are given in Section~\ref{se:concluding}.

	\section{Model derivation}
	\label{se:model} For simplicity, we consider the semi-permeable interfaces with constant surface tension.
	We start with the following conservation laws \cite{shen2022energy}
	\begin{equation}\label{kimematic assume oss}
		\left\{
		\begin{array}{ll}
			\rho \left(\frac{\partial \bm{u}}{\partial t} + (\bm{u}\cdot\nabla)\bm{u}\right) = \nabla\cdot \bm{\sigma}_{\eta} +\nabla\cdot \bm{\sigma}_{\phi}, &  \mbox{in} ~\Omega, \\
			\nabla\cdot\bm{u} = 0, &  \mbox{in} ~\Omega, \\
			\frac{\partial}{\partial t} (\zeta_{\pm} C_{\pm}) +\nabla\cdot(\zeta_{\pm} C_{\pm}\bm{u}) = -\nabla\cdot(\zeta_\pm \bm{J}_{\pm}),      & \mbox{in} ~\Omega ,\\
			\frac{\partial\phi}{\partial t} +\nabla\cdot(\phi\bm{u}) = -\nabla\cdot\bm{J}_{\phi} +S_{\phi}|\nabla\phi|, & \mbox{in} ~\Omega,
		\end{array}
		\right.
	\end{equation}
	where $\rho$ is the density,  $C_{\pm}$ is the concentration inside/outside of the droplet, $\bm{u}$ is the velocity and $\phi$ is the label function and $\zeta_+(\phi)$ is a smooth index function given by \cite{garcke2013diffuse}
	\begin{equation*}
		\zeta_+(\phi) = \left\{\begin{array}{cc}
			1&\phi \ge 1,  \\
			\frac{1}{2} (1+\frac 1 2\phi(3-\phi^2))& |\phi|<1, \\
			0 & \phi\le -1,
		\end{array}\right. ~\zeta_-(\phi) = 1- \zeta_+(\phi).
	\end{equation*}
	Here the first equation is the conservation law of momentum with two unknown stresses $\bm{\sigma}_{\eta}$ and $\bm{\sigma}_{\phi}$, the second is the fluid incompressibility, the third is the conservation of substance with unknown flux $\bm{J}_{\pm}$ and the last one is the Cahn-Hilliard type to track the movement of the interface  with unknown flux $\bm{J}_{\phi}$ and source term $S_{\phi}$ due to transmembrane water flow. 
	
	The total energy consists of kinetic, entropy and phase-mixing energy
	\begin{align}\label{energyfunction}
		E =&E_{kin}+E_{entropy} +E_{mix}\\
		=&\int_{\Omega}\rho \frac{|\bm{u}|^2}{2}d\x   +\!\!\! \int_{\Omega}RT \left( \zeta_+  C_+(\ln{\frac{C_+}{c_\infty}}-1)  +  \!\!  \zeta_- C_-(\ln{\frac{C_-}{c_\infty}}-1) \right) d\x\nonumber\\
		&+ \!\!\!\int_{\Omega} \lambda \left(\frac{l^2|\nabla\phi|^2 }{2}+F(\phi)\right) d\x,\nonumber
	\end{align}
	where $R$ is the  universal ideal gas constant, $T$ is temperature, $c_{\infty}$ is the characteristic concentration, $\rho$ is density, $\lambda$ is surface energy density, $l$ is surface thickness and $F(\phi)=\frac{1}{4}(\phi^2-1)^2$ is the double-well potential. 
	
	\begin{remark}
		More complicated cases with variable surface tensions, such as the bending modulus of vesicles, can be handled by adding more terms in the energetic function \cite{duqiang_2004_bending,shen2022energy}.      
	\end{remark}
	
	Corresponding to the total energy, we could define the chemical potentials
	\begin{align*}
		\tilde{\mu}_{\phi} = \lambda (-l^2\Delta \phi + F'(\phi)) -RT\left( \frac{\partial \zeta_+}{\partial \phi}  C_+ + \frac{\partial\zeta_-}{\partial\phi} C_-  \right),~~
		\mu_c^{\pm}=RT \ln\frac{C_{\pm}}{c_{\infty}}. 
	\end{align*}
	
	The dissipative functional consists of fluid friction, and irreversible mixing of the substance and two phases in the bulk
	\begin{align}\label{disspationfunctional}
		\Delta = \int_{\Omega}\left(2 \eta |D_{\eta}|^2 + \sum_{\pm}\zeta_{\pm}\frac{D_{\pm}C_{\pm}} {RT}|\nabla\mu_c^{\pm}|^2 +\mathcal{M}|\nabla\tilde{\mu}_{\phi}|^2+ K|\nabla\phi||\tilde{\mu}_{\phi}|^2\right) d\x,
	\end{align}
	where $\eta$ is the viscosity of the fluid, $D_{\eta} =\frac{1}{2}(\nabla \bm{u}+(\nabla\bm{u})^T )$ is the strain rate, $\mathcal{M}$ is the phenomenological mobility, $D_{\pm}$ is the intrinsic diffusion coefficient of substance in $\Omega^{\pm}$, $K$ is the interface permeability. 

	The energy dissipative law \cite{eisenberg2010energy,qin2022phase,shen2020energy} states that without external force acting on the system, the changing rate of total energy equals to  the dissipation
	\begin{equation*}
		\frac{dE}{dt} = \frac{dE_{kin}}{dt} +\frac{dE_{entropy}}{dt} +\frac{dE_{mix}}{dt} =-\Delta.
	\end{equation*}
	
	Using the definition of the total energy functional \eqref{energyfunction} and the conservation laws in \eqref{kimematic assume oss}, we can obtain the following energy dissipation relation (see Appendix for a detailed derivation):
	\begin{align*}
		\frac{dE}{dt} 
		= & -\int_{\Omega} \nabla \bm{u} : (\bm{\sigma}_{\eta} + \bm{\sigma}_{\phi}) \, d\bm{x}
		- \int_{\Omega} p \, \nabla \cdot \bm{u} \, d\bm{x}
		- \int_{\Omega} \lambda l^2 (\nabla \phi \otimes \nabla \phi) : \nabla \bm{u} \, d\bm{x} \\
		& + \int_{\Omega} \left( \zeta_+ \bm{J}_+ \cdot \nabla \mu_c^+ + \zeta_- \bm{J}_- \cdot \nabla \mu_c^- \right) \, d\bm{x}
		+ \int_{\Omega} \bm{J}_{\phi} \cdot \nabla \tilde{\mu}_{\phi} \, d\bm{x}
		+ \int_{\Omega} \tilde{\mu}_{\phi} S_{\phi} |\nabla \phi| \, d\bm{x}.
	\end{align*}

	By comparing the terms above with the dissipation functional given in equation \eqref{disspationfunctional}, we can identify the explicit expressions for the unknown quantities such as the fluxes and stress tensors.
	\begin{align*}
		&\bm{J}_{\phi}=-\mathcal{M}\nabla \tilde{\mu}_{\phi}, ~~
		\bm{J}_{\pm}=- \frac{D_{\pm}C_\pm}{RT}\nabla\mu_{c}^{\pm} = -D_{\pm} \nabla C_{\pm}, \\
		&\bm\sigma_{\eta}=2\eta D_{\eta}-p\textbf{I},~~ \bm\sigma_{\phi}=-\lambda l^{2}(\nabla\phi\otimes\nabla\phi),~~ S_{\phi}= -K\tilde{\mu}_{\phi}.
	\end{align*}
	Substituting the above into \eqref{kimematic assume oss} yields the phase-field model for the semipermeable interface 
	\begin{align*}
		&\frac{\textbf{D} \phi}{\textbf{D} t}=  \nabla\cdot(\mathcal{M}\nabla \tilde{\mu}_{\phi})-K\tilde{\mu}_{\phi}|\nabla\phi|, \\
		&\frac{\textbf{D}}{\textbf{D} t}( \zeta_{\pm}C_{\pm})=  \nabla\cdot (\zeta_{\pm}D_{\pm}\nabla C_{\pm}), \\
		&\rho\frac{\textbf{D} \bm{u}}{\textbf{D} t}= 
		\nabla\cdot(\eta(\nabla\bm{u}+(\nabla\bm{u})^T))-\nabla p
		-\lambda l^{2}\nabla\cdot(\nabla\phi\otimes\nabla\phi), \\
		&\nabla\cdot\bm{u}=0, \\
		& \tilde{\mu}_{\phi} = \lambda (-l^2\Delta \phi + F'(\phi))-RT\left( \frac{\partial \zeta_+}{\partial \phi}  C_+ + \frac{\partial\zeta_-}{\partial\phi} C_-  \right), 
	\end{align*}
	with boundary conditions  
	\begin{equation}\label{eqn: bdc}
		\nabla\phi\cdot\bm{n}|_{\partial\Omega}=0, \quad
		\nabla C_{\pm}\cdot\bm{n}|_{\partial\Omega}=0, \quad
		\nabla\tilde{\mu}_{\phi}\cdot\bm{n}|_{\partial\Omega}=0, \quad
		\bm{u}|_{\partial\Omega}=0.
	\end{equation}

	Let \( L^* \), \( U^* \), \( C^* = c_\infty \), and \( D^* \) denote the characteristic length, velocity, concentration, and diffusion coefficient, respectively. Define the characteristic time and chemical potential as \( t^* = \frac{L^*}{U^*} \) and \( \mu^* = \frac{\lambda l}{L^*} \). Using these scalings, the dimensionless form of the system is given by:
	
	\begin{align*}
		&\frac{\textbf{D} \phi}{\textbf{D} t}=  \nabla\cdot(\mathcal{M}\nabla \tilde{\mu}_{\phi})- \frac{K}{Ca} \tilde{\mu}_{\phi}|\nabla\phi|, \\
		&\frac{\textbf{D}}{\textbf{D} t}( \zeta_{\pm}C_{\pm})= \frac 1 {Pe} \nabla\cdot (\zeta_{\pm}D_{\pm}\nabla C_{\pm}), \\
		&Re\frac{\textbf{D} \bm{u}}{\textbf{D} t}= 
		\nabla\cdot(\eta(\nabla\bm{u}+(\nabla\bm{u})^T))-\nabla p
		- \frac{\epsilon}{Ca}\nabla\cdot(\nabla\phi\otimes\nabla\phi), \\
		&\nabla\cdot\bm{u}=0, \\
		& \tilde{\mu}_{\phi}  =   -\epsilon\Delta \phi + \frac 1 {\epsilon}F'(\phi)-\frac{\beta}{\epsilon}\left( \frac{\partial \zeta_+}{\partial \phi}  C_+ + \frac{\partial\zeta_-}{\partial\phi} C_-  \right), 
	\end{align*}
	where 
	$$Re= \frac{\rho U^* L^*}{\eta^*}, Pe = \frac{U^*L^*}{D^*}, Ca = \frac{\eta^* U^*}{\lambda l}, \beta = \frac{RTC^*}{\lambda},  K = \frac{K\eta^*}{ L^*}, \ \epsilon=\frac{l}{L^*}. $$
	Using the fact that
	\begin{align*}
		-\epsilon \nabla\cdot(\nabla\phi\otimes\nabla\phi)&=-\epsilon \nabla\phi\Delta\phi   - \frac{\epsilon}{2}\nabla|\nabla\phi|^2\\
		&=\tilde{\mu}_{\phi}\nabla\phi+\frac{\beta}{\epsilon}(C_+\nabla\zeta_++C_-\nabla \zeta_{-} ) -\nabla(\frac 1 {\epsilon} F(\phi) +\frac{\epsilon}{2}|\nabla\phi|^2),
	\end{align*}
	the above system could be written as 
	\begin{subequations} \label{eqn: maindimensionless}
		\begin{align}
			&\frac{\partial \phi}{\partial t}+\bm{u}\cdot\nabla\phi=  \nabla\cdot(\mathcal{M}\nabla \tilde{\mu}_{\phi})- \frac{K}{Ca} \tilde{\mu}_{\phi}|\nabla\phi|, 
			\label{eqn: phi}\\
			&\frac{\partial ( \zeta_{\pm}C_{\pm})}{\partial t}+\nabla\cdot ( \bm{u}\zeta_{\pm}C_{\pm})= \frac 1 {Pe} \nabla\cdot (\zeta_{\pm}D_{\pm}\nabla C_{\pm}), 
			\label{eqn: c}\\
			&Re(\frac{\partial \bm{u}}{\partial t}+(\bm{u}\cdot\nabla)\bm{u})= 
			\nabla\cdot(\eta(\nabla\bm{u}+(\nabla\bm{u})^T))-\nabla p
			+\frac {1}{Ca} \tilde{\mu}_{\phi}\nabla\phi  +\frac{\beta}{Ca\epsilon}(C_+\nabla\zeta_++C_-\nabla\zeta_-) ,  \label{eqn: u}\\
			&\nabla\cdot\bm{u}=0, 
			\label{eqn: nabla_u}\\
			& \tilde{\mu}_{\phi}  =   -\epsilon\Delta \phi + \frac 1 {\epsilon}F'(\phi)-\frac{\beta}{\epsilon}\left( \frac{\partial \zeta_+}{\partial \phi}  C_+ + \frac{\partial\zeta_-}{\partial\phi} C_-  \right), \label{eqn: muphi} 
		\end{align}
	\end{subequations}
	with boundary conditions \eqref{eqn: bdc}.

	\begin{remark}
		If we use the identities
		\[
		\bm{n} = -\frac{\nabla\phi}{|\nabla\phi|}, \quad \text{and} \quad |\nabla\phi| = \frac{\nabla\phi}{|\nabla\phi|} \cdot \nabla\phi = -\bm{n} \cdot \nabla\phi,
		\]
		then the equation \eqref{eqn: phi} can be rewritten as
		\[
		\frac{\partial \phi}{\partial t} + \bm{u} \cdot \nabla\phi - \frac{K}{Ca} \tilde{\mu}_{\phi} \, \bm{n} \cdot \nabla\phi = \frac{\partial \phi}{\partial t} + \bm{v} \cdot \nabla\phi = \nabla \cdot (\mathcal{M} \nabla \tilde{\mu}_{\phi}),
		\]
		where  the interface velocity is defined by
		$\bm{v} = \bm{u} - \frac{K}{Ca} \tilde{\mu}_{\phi} \bm{n}.$
		As a result, the transmembrane water velocity relative to the interface is given by
		$U \bm{n} = \bm{u} - \bm{v} = \frac{K}{Ca} \tilde{\mu}_{\phi} \bm{n}.$
		
	\end{remark}

	\begin{theorem} \label{energy law theorem}
		The above system \eqref{eqn: maindimensionless} with boundary conditions \eqref{eqn: bdc} satisfies the following energy dissipation law 
		
		\begin{align*}
			& \frac {d}{dt} \left(\frac{Re}2\int_{\Omega} |\bm{u}|^2d\x +   \frac 1 {Ca}\int_{\Omega}  (\frac{\epsilon}{2}|\nabla\phi|^2 +\frac 1 {\epsilon}F(\phi))d\x + \sum_{\pm}\frac{\beta}{Ca\epsilon}\int_{\Omega}\zeta_{\pm}C_{\pm}(\ln{C_{\pm}}-1) d\x\right) \nonumber\\
			=& -\int_{\Omega}\eta|\nabla\bm{u}|^2d\x-\frac 1 {Ca}\int_{\Omega}\mathcal{M}|\nabla\tilde\mu_{\phi}|^2 d\x -\frac 1 {Ca}\int_{\Omega}\frac{K}{Ca}|\tilde\mu_{\phi}|^2|\nabla\phi|  d\x-\frac{\beta}{Ca\epsilon}\frac{1}{Pe}\sum_{\pm}\int_{\Omega}\zeta_\pm D_\pm  C_\pm|\nabla\mu_c^\pm|^2 d\x,
		\end{align*}
		where $\mu_c^{\pm}=\ln C_\pm$.
	\end{theorem}
	
	\begin{proof} Multiplying equation \eqref{eqn: phi} by $\frac{\tilde\mu_{\phi}}{Ca}$ and integration by parts yields
		\begin{align}
			\frac 1 {Ca}\int_{\Omega} \frac{\partial\phi}{\partial t}\tilde\mu_{\phi} d\x+\frac 1 {Ca}\int_{\Omega}\bm{u}\cdot\nabla\phi \tilde\mu_{\phi} d\x 
			=-\frac 1 {Ca}\int_{\Omega}\mathcal{M}|\nabla\tilde\mu_{\phi}|^2 d\x -\frac 1 {Ca}\int_{\Omega}\frac{K}{Ca}|\tilde\mu_{\phi}|^2|\nabla\phi|  d\x.
		\end{align}
		Multiplying  equation \eqref{eqn: muphi}  by $\frac{1}{Ca}\frac{\partial \phi}{\partial t}$ and integration by parts, we have 
		\begin{align}
			\frac{1}{Ca}\int_{\Omega}\tilde{\mu}_{\phi}\frac{\partial \phi}{\partial t} d\x=\frac 1 {Ca}\frac{d}{dt} \int_{\Omega}  (\frac{\epsilon}{2}|\nabla\phi|^2 +\frac 1 {\epsilon}F(\phi))d\x - \frac{\beta}{\epsilon Ca}\int_{\Omega}   \frac{\partial\phi}{\partial t}\left( \frac{\partial \zeta_+}{\partial \phi}  C_+ + \frac{\partial\zeta_-}{\partial\phi} C_-  \right) d\x.
		\end{align}
		Then, the above two equations give 
		\begin{align}
			&\frac 1 {Ca}\frac{d}{dt} \int_{\Omega}  (\frac{\epsilon}{2}|\nabla\phi|^2 +\frac 1 {\epsilon}F(\phi))d\x - \frac{\beta}{\epsilon Ca}\int_{\Omega}   \frac{\partial\phi}{\partial t}\left( \frac{\partial \zeta_+}{\partial \phi}  C_+ + \frac{\partial\zeta_-}{\partial\phi} C_-  \right) d\x +\frac 1 {Ca}\int_{\Omega}\bm{u}\cdot\nabla\phi \tilde\mu_{\phi} d\x\nonumber\\
			=&-\frac 1 {Ca}\int_{\Omega}\mathcal{M}|\nabla\tilde\mu_{\phi}|^2 d\x -\frac 1 {Ca}\int_{\Omega}\frac{K}{Ca}|\tilde\mu_{\phi}|^2|\nabla\phi|  d\x.
		\end{align}
		For equation \eqref{eqn: c}, using the definition  $\mu_c^\pm =\ln C_\pm$, it yields 
		\begin{align}
			&\frac{\beta}{Ca\epsilon}\int_{\Omega} \frac{\partial ( \zeta_{\pm}C_{\pm})}{\partial t}\mu_c^\pm d\x +\frac{\beta}{Ca\epsilon}\int_{\Omega}\bm{u}\cdot \nabla(\zeta_{\pm}C_\pm)\mu_c^{\pm}d\x \nonumber\\
			=& \frac{\beta}{Ca\epsilon}\int_{\Omega} \frac{\partial ( \zeta_{\pm}C_{\pm})}{\partial t}\mu_c^\pm d\x -\frac{\beta}{Ca\epsilon}\int_{\Omega}\bm{u}\cdot (C_\pm \nabla\mu_c^{\pm})\zeta_{\pm}d\x \nonumber\\
			=&\frac{\beta}{Ca\epsilon} \int_{\Omega} \frac{\partial ( \zeta_{\pm}C_{\pm})}{\partial t}\mu_c^\pm d\x -\frac{\beta}{Ca\epsilon}\int_{\Omega}\bm{u}\cdot  \nabla C_{\pm}\zeta_{\pm}d\x \nonumber\\
			=& \frac{\beta}{Ca\epsilon}\int_{\Omega} \frac{\partial ( \zeta_{\pm}C_{\pm})}{\partial t}\mu_c^\pm d\x +\frac{\beta}{Ca\epsilon}\int_{\Omega}\bm{u}\cdot  \nabla\zeta_{\pm} C_{\pm}d\x \nonumber\\    
			=&-\frac{\beta}{Ca\epsilon}\frac{1}{Pe}\int_{\Omega}\zeta_\pm D_{\pm} C_\pm|\nabla\mu_c^\pm|^2 d\x.
		\end{align}
		Multiplying equation \eqref{eqn: u} with $\bm{u}$ and by the incompressibility constraint $\nabla \cdot \bm{u}=0$,  we have 
		\begin{align}
			\frac {d}{dt} \frac{Re}2\int_{\Omega} |\bm{u}|^2d\x = -\int_{\Omega}\eta|\nabla\bm{u}|^2d\x+\frac 1 {Ca}\int_{\Omega}\tilde{\mu}_{\phi}\nabla\phi\cdot\bm{u}d\x +\frac{\beta}{Ca\epsilon}\int_{\Omega}(C_+\nabla\zeta_++C_-\nabla\zeta_-)\cdot \bm{u}d\x.
		\end{align}
		Adding the above three equations yields
		\begin{align}
			& \frac {d}{dt} \frac{Re}2\int_{\Omega} |\bm{u}|^2d\x +\frac{d}{dt}  \frac 1 {Ca}\int_{\Omega}  (\frac{\epsilon}{2}|\nabla\phi|^2 +\frac 1 {\epsilon}F(\phi))d\x\nonumber\\
			&  - \frac{\beta}{\epsilon Ca}\int_{\Omega}   \frac{\partial\phi}{\partial t}\left( \frac{\partial \zeta_+}{\partial \phi}  C_+ + \frac{\partial\zeta_-}{\partial\phi} C_-  \right) d\x+\frac{\beta}{Ca\epsilon} \int_{\Omega}\left( \frac{\partial ( \zeta_{+}C_{+})}{\partial t}\mu_c^+ + \frac{\partial ( \zeta_{-}C_{-})}{\partial t}\mu_c^- \right) d\x\nonumber\\
			=& \frac {d}{dt} \frac{Re}2\int_{\Omega} |\bm{u}|^2d\x +\frac{d}{dt}  \frac 1 {Ca}\int_{\Omega}  (\frac{\epsilon}{2}|\nabla\phi|^2 +\frac 1 {\epsilon}F(\phi))d\x\nonumber\\
			&  +\frac{\beta}{Ca\epsilon} \int_{\Omega}\left( \frac{\partial ( \zeta_{+}C_{+})}{\partial t}\mu_c^+ - \frac{\partial \zeta_+}{\partial t}  C_+ + \frac{\partial ( \zeta_{-}C_{-})}{\partial t}\mu_c^--\frac{\partial \zeta_-}{\partial t}  C_- \right)d\x\nonumber\\
			=& \frac {d}{dt} \frac{Re}2\int_{\Omega} |\bm{u}|^2dx +\frac{d}{dt}  \frac 1 {Ca}\int_{\Omega}  (\frac{\epsilon}{2}|\nabla\phi|^2 +\frac 1 {\epsilon}F(\phi))d\x \nonumber\\
			&+\frac{d }{d t}\frac{\beta}{Ca\epsilon}\int_{\Omega}( \zeta_{+}C_+(\ln{C_+}-1)+ \zeta_{-}C_-(\ln{C_-}-1)) d\x \nonumber\\
			=& -\int_{\Omega}\eta|\nabla\bm{u}|^2d\x-\frac 1 {Ca}\int_{\Omega}\mathcal{M}|\nabla\tilde\mu_{\phi}|^2 d\x -\frac 1 {Ca}\int_{\Omega}\frac{K}{Ca}|\tilde\mu_{\phi}|^2|\nabla\phi|  d\x-\frac{\beta}{Ca\epsilon}\frac{1}{Pe}\sum_{\pm}\int_{\Omega}\zeta_\pm D_\pm C_\pm|\nabla\mu_c^\pm|^2 d\x. \nonumber
		\end{align}
		
	\end{proof} 
	
	\section{Numerical scheme}
	\label{se:numerical-scheme}
	The purpose of this section is to develop high-order, decoupled, and energy-stable numerical schemes to solve the model \eqref{eqn: maindimensionless}.
	\subsection{Semi-discrete scheme in time}
	\label{subse1}
	A first-order and decoupled energy stable numerical scheme is as follows: assuming that $(\phi^n,\bm{u}^n,C_{\pm}^{n})$ are already known, we then compute $(\phi^{n+1},\bm{u}^{n+1},C_{\pm}^{n+1})$ from the following temporal discrete system.
	
	\noindent {\bf Step 1:}
	
	\begin{equation}\label{eq:semi-phi}
		\left\{
		\begin{aligned}
			& \frac{\phi^{n+1}-\phi^n}{\Delta t}=\mathcal{M} \Delta \tilde\mu_\phi^{n+1}-\nabla \cdot (\phi^n \bm{u}^n_{*})-\frac{K}{Ca}\tilde\mu_{\phi}^{n+1} |\nabla \phi^n|,\\
			& \tilde\mu_{\phi}^{n+1}=-\epsilon \Delta \phi^{n+1}+\frac{S}{\epsilon}(\phi^{n+1}-\phi^n)+\frac{1}{\epsilon}f(\phi^n)-\frac{\beta}{\epsilon}\sum_{\pm}\frac{\zeta_{\pm}^{n+1}-\zeta_{\pm}^{n}}{\phi^{n+1}-\phi^n}C_{\pm}^{n+1}, \\
			&\frac{\zeta_{\pm}^{n+1} C_{\pm}^{n+1}-\zeta_{\pm}^{n}C_{\pm}^{n}}{\Delta t}=\frac{1}{Pe}\nabla \cdot (\zeta_{\pm}^n D_{\pm}^n   C_{\pm}^{n+1}\nabla \mu_{c_{\pm}}^{n+1})-\nabla \cdot(\bm{u}_{*}^n \zeta_{\pm}^n C_{\pm}^{n}),\\
			& \mu_{c_{\pm}}^{n+1}=\ln C_{\pm}^{n+1},
		\end{aligned}
		\right .
	\end{equation}
	with
	\begin{equation}\label{eq:semi-ustar}
		\bm{u}_{*}^n=\bm{u}^n-\frac{\Delta t}{Ca Re} \phi^n \nabla \tilde\mu_\phi^{n+1}-\frac{\beta \Delta t}{Ca \epsilon Re}\sum_{\pm}\zeta_{\pm}^n C_{\pm}^{n} \nabla \mu^{n+1}_{c_{\pm}}.
	\end{equation}
	{\bf Step 2:}
	\begin{equation}\label{eq:semi-u}
		Re \left(\frac{\widetilde{\bm{u}}^{n+1}-\bm{u}^n}{\Delta t}+(\bm{u}^n \cdot \nabla) \widetilde{\bm{u}}^{n+1}\right)=\eta \Delta \widetilde{\bm{u}}^{n+1}-\nabla p^n-\frac{1}{Ca} \phi^n \nabla \tilde\mu_{\phi}^{n+1}-\frac{\beta}{Ca \epsilon}\sum_{\pm}\zeta_{\pm}^n C_{\pm}^{n} \nabla \mu^{n+1}_{c_{\pm}}.
	\end{equation}
	{\bf Step 3:}
	\begin{equation}\label{eq:projection}
		\left\{
		\begin{aligned}
			&  Re \frac{\bm{u}^{n+1}-\widetilde{\bm{u}}^{n+1}}{\Delta t}+\nabla (p^{n+1}-p^n)=0, \\
			& \nabla \cdot \bm{u}^{n+1}=0.
		\end{aligned}
		\right .
	\end{equation}
	Taking divergence of the first equation in \eqref{eq:projection} yields
	\begin{equation}
		-\Delta (p^{n+1}-p^n)=-\frac{Re}{\Delta t} \nabla \cdot \widetilde{\bm{u}}^{n+1},
	\end{equation}
	due to the condition $\nabla \cdot \bm{u}^{n+1}=0$.

	Actually, the time discretization technique for the double-well potential $F(\phi)$ is the stabilized semi-implicit method. We assume that $F(\phi)$ satisfies the following condition:
	there exists a constant $L$ such that 
	\begin{equation}\label{truncation}
		\max_{\phi} |F''(\phi)| \leq L.
	\end{equation}
	It is known that $F(\phi)$ does not satisfy \eqref{truncation}. However, its has been a common practice \cite{caff} to consider the Cahn-Hilliard equation with a truncated double well potential $\tilde F(\phi)$. It is then obvious that there exists a constant $L$ such that \eqref{truncation} is satisfied with $F(\phi)$ replaced by $\tilde F(\phi)$.
	
	\begin{thm}
		Assuming that the condition \eqref{truncation} is satisfied and $S \geq L/2$. Then the  scheme \eqref{eq:semi-phi}-\eqref{eq:projection} is energy stable and satisfies the following discrete energy law
		\begin{align*}
			& \mathcal{E}^{n+1}-\mathcal{E}^n \\
			\leq &  -\Delta t \eta  ||\nabla \widetilde{\bm{u}}^{n+1}||^2-\frac{\Delta t \mathcal{M}}{Ca}||\nabla \tilde\mu_{\phi}^{n+1}||^2-\frac{\Delta t}{Ca}\frac{ K}{Ca}\int_{\Omega}|\tilde\mu_{\phi}^{n+1}|^2|\nabla \phi^n| d \bm{x}-\frac{\Delta t \beta}{Ca \epsilon Pe} \sum_{\pm}\int_{\Omega}\zeta_{\pm}^{n}D_{\pm}^{n} C_{\pm}^{n+1} |\nabla \mu^{n+1}_{c_{\pm}}|^2 d \bm{x}.
		\end{align*}
		where $||\cdot||$ denotes the discrete $L^2$ norm in domain $\Omega$ and 
		\begin{equation*}
			\mathcal{E}^n=\frac{Re}{2}||\bm{u}^n||^2+\frac{1}{Ca}\left(\frac{\epsilon}{2}||\nabla \phi^n||^2+\frac{\Delta t^2}{2Re}||\nabla p^n||^2+\frac{1}{\epsilon}(F(\phi^n),1)\right)+\frac{\beta}{Ca \epsilon} \sum_{\pm} \left( \zeta_{\pm}^nC_{\pm}^n(\ln C_{\pm}^n-1),1\right).
		\end{equation*}
	\end{thm}
	\begin{proof}
		From \eqref{eq:semi-ustar}, we have
		\begin{equation*}
			Re \frac{\widetilde{\bm{u}}^{n+1}-\bm{u}^n}{\Delta t}+\frac{1}{Ca} \phi^n \nabla \tilde\mu_{\phi}^{n+1}+\frac{\beta}{Ca \epsilon}\sum_{\pm}\zeta_{\pm}^n C_{\pm}^{n} \nabla \mu^{n+1}_{c_{\pm}}=Re \frac{\widetilde{\bm{u}}^{n+1}-\bm{u}^n_{*}}{\Delta t}.
		\end{equation*}
		Taking the inner product of \eqref{eq:semi-u} with $\Delta t \widetilde{\bm{u}}^{n+1}$, using the above relation, we derive
		\begin{equation}\label{semi-pf1}
			\frac{Re}{2}||\widetilde{\bm{u}}^{n+1}||^2-\frac{Re}{2}||\bm{u}_{*}^n||^2+\frac{Re}{2}||\widetilde{\bm{u}}^{n+1}-\bm{u}_{*}^n||^2+\eta \Delta t ||\nabla \widetilde{\bm{u}}^{n+1}||^2+\Delta t(\nabla p^n, \widetilde{\bm{u}}^{n+1})=0.
		\end{equation}
		To deal with the last term in the above, we first take the inner product of \eqref{eq:projection} with $\Delta t \nabla p^n$ to obtain
		\begin{equation}\label{semi-pf2}
			\frac{\Delta t^2}{2Re}(||\nabla p^{n+1}||^2-||\nabla p^{n}||^2-||\nabla p^{n+1}-\nabla p^{n}||^2)=\Delta t (\widetilde{\bm{u}}^{n+1}, \nabla p^n).
		\end{equation}
		We also derive from  \eqref{eq:projection} that
		\begin{equation}\label{semi-pf3}
			\frac{\Delta t^2}{2Re} ||\nabla p^{n+1}-\nabla p^{n}||^2=\frac{Re}{2}||\widetilde{\bm{u}}^{n+1}-\bm{u}^{n+1}||^2.
		\end{equation}
		We then take the inner product of \eqref{eq:projection} with $\bm{u}^{n+1}$ to get
		\begin{equation}\label{semi-pf4}
			\frac{Re}{2}||\bm{u}^{n+1}||^2+\frac{Re}{2}||\bm{u}^{n+1}-\widetilde{\bm{u}}^{n+1}||^2=\frac{Re}{2}||\widetilde{\bm{u}}^{n+1}||^2.
		\end{equation}
		Combining equations \eqref{semi-pf1}-\eqref{semi-pf4}, we find
		\begin{equation}\label{semi-pf5}
			\frac{Re}{2}||\bm{u}^{n+1}||^2-\frac{Re}{2}||\bm{u}_{*}^{n}||^2+\frac{Re}{2}||\widetilde{\bm{u}}^{n+1}-\bm{u}_{*}^{n}||^2+\eta \Delta t||\nabla \widetilde{\bm{u}}^{n+1}||^2+\frac{\Delta t^2}{2Re}(||\nabla p^{n+1}||^2-||\nabla p^{n}||^2)=0.
		\end{equation}
		Next, we use the relation \eqref{eq:semi-ustar} to deal with $||\bm{u}_{*}^{n}||^2$ in \eqref{semi-pf5}. Taking the inner product of \eqref{eq:semi-ustar} with $Re \bm{u}_{*}^{n}$, we obtain
		\begin{equation}\label{semi-pf6}
			\frac{Re}{2}||\bm{u}_{*}^{n}||^2-\frac{Re}{2}||\bm{u}^{n}||^2+\frac{Re}{2}||\bm{u}_{*}^{n}-\bm{u}^{n}||^2=-\frac{\Delta t}{Ca}(\phi^n \nabla \tilde\mu_{\phi}^{n+1},\bm{u}_{*}^{n})-\frac{\beta \Delta t}{Ca \epsilon}\sum_{\pm}(\zeta_{\pm}^n C_{\pm}^{n} \nabla \mu^{n+1}_{c_{\pm}},\bm{u}_{*}^{n}).
		\end{equation}
		Adding \eqref{semi-pf5} and \eqref{semi-pf6}, we have
		\begin{align}\label{semi-pf7}
			&\frac{Re}{2}||\bm{u}^{n+1}||^2-\frac{Re}{2}||\bm{u}^{n}||^2+\frac{Re}{2}||\widetilde{\bm{u}}^{n+1}-\bm{u}_{*}^{n}||^2+\frac{Re}{2}||\bm{u}_{*}^{n}-\bm{u}^n||^2+\eta \Delta t||\nabla \widetilde{\bm{u}}^{n+1}||^2 \nonumber \\
			&+\frac{\Delta t^2}{2Re}(||\nabla p^{n+1}||^2-||\nabla p^{n}||^2)\\
			=&-\frac{\Delta t}{Ca}(\phi^n \nabla \tilde\mu_{\phi}^{n+1},\bm{u}_{*}^{n})-\frac{\beta \Delta t}{Ca \epsilon}\sum_{\pm}(\zeta_{\pm}^n C_{\pm}^{n} \nabla \mu^{n+1}_{c_{\pm}},\bm{u}_{*}^{n}). \nonumber
		\end{align}
		It now remains to deal with the last two terms in the above.
		
		Taking the inner product of the first equation of \eqref{eq:semi-phi} with $\frac{\Delta t}{Ca} \tilde\mu_{\phi}^{n+1}$, we get
		\begin{equation}\label{semi-pf8}
			\frac{1}{Ca}(\phi^{n+1}-\phi^n,\tilde\mu_{\phi}^{n+1})=-\frac{\mathcal{M} \Delta t}{Ca}||\nabla \tilde\mu_{\phi}^{n+1}||^2-\frac{\Delta t}{Ca} (\nabla \cdot(\bm{u}_{*}^n \phi^n),\tilde\mu_{\phi}^{n+1})-\frac{\Delta t  }{Ca} \frac{K}{Ca} \int_{\Omega}|\tilde\mu_{\phi}^{n+1}|^2|\nabla \phi^n| d \bm{x},
		\end{equation}
		and taking the inner product of the second equation of \eqref{eq:semi-phi} with $-\frac{1}{Ca} (\phi^{n+1}-\phi^n)$, we obtain
		\begin{align}\label{semi-pf9}
			-\frac{1}{Ca}(\tilde\mu_{\phi}^{n+1},\phi^{n+1}-\phi^n)=&-\frac{\epsilon}{2Ca}(||\nabla \phi^{n+1}||^2-||\nabla \phi^n||^2+||\nabla \phi^{n+1}-\nabla \phi^n||^2) \nonumber \\
			&+\frac{\beta}{Ca \epsilon} \sum_{\pm}((\zeta_{\pm}^{n+1}-\zeta_{\pm}^{n})C_{\pm}^{n+1},1) \\
			&-\frac{1}{Ca \epsilon}(S(\phi^{n+1}-\phi^n)+f(\phi^n),\phi^{n+1}-\phi^n). \nonumber
		\end{align}
		For the last term in \eqref{semi-pf9}, we use the Taylor expansion
		\begin{equation*}
			F(\phi^{n+1})-F(\phi^n)=f(\phi^n)(\phi^{n+1}-\phi^n)+\frac{f'(\xi_n)}{2}(\phi^{n+1}-\phi^n)^2.
		\end{equation*}
		Taking the inner product of the third equation of \eqref{eq:semi-phi} with $\frac{\beta \Delta t}{Ca \epsilon} \mu_{c_{\pm}}^{n+1}$, we have
		\begin{equation}\label{semi-pf10}
			\frac{\beta}{Ca \epsilon}(\zeta_{\pm}^{n+1}C_{\pm}^{n+1}-\zeta_{\pm}^{n}C_{\pm}^{n},\mu_{c_{\pm}}^{n+1})=-\frac{\beta \Delta t}{Pe Ca \epsilon} \int_{\Omega}\zeta_{\pm}^{n}D_{\pm}^{n} C_{\pm}^{n+1} |\nabla \mu^{n+1}_{c_{\pm}}|^2 d \bm{x}-\frac{\beta \Delta t}{Ca \epsilon} (\nabla \cdot(\bm{u}_{*}^n \zeta_{\pm}^n C_{\pm}^{n}),\mu_{c_{\pm}}^{n+1}).
		\end{equation}
		Combining \eqref{semi-pf7}, \eqref{semi-pf8}, \eqref{semi-pf9} and \eqref{semi-pf10}, we have
		\begin{align}\label{semi-pf11}
			&\frac{Re}{2}||\bm{u}^{n+1}||^2-\frac{Re}{2}||\bm{u}^{n}||^2+\frac{Re}{2}||\widetilde{\bm{u}}^{n+1}-\bm{u}_{*}^{n}||^2+\frac{Re}{2}||\bm{u}_{*}^{n}-\bm{u}^n||^2+\eta \Delta t||\nabla \widetilde{\bm{u}}^{n+1}||^2+\frac{\Delta t^2}{2Re}(||\nabla p^{n+1}||^2-||\nabla p^{n}||^2) \nonumber\\
			&+\frac{\mathcal{M} \Delta t}{Ca}||\nabla \tilde\mu_{\phi}^{n+1}||^2+\frac{\Delta t  }{Ca} \frac{K}{Ca} \int_{\Omega}|\tilde\mu_{\phi}^{n+1}|^2|\nabla \phi^n| d \bm{x}
			+\frac{\epsilon}{2Ca}(||\nabla \phi^{n+1}||^2-||\nabla \phi^n||^2+||\nabla \phi^{n+1}-\nabla \phi^n||^2)\\
			&+\frac{\beta \Delta t}{Pe Ca \epsilon} \sum_{\pm}\int_{\Omega}\zeta_{\pm}^{n}D_{\pm}^{n} C_{\pm}^{n+1} |\nabla \mu^{n+1}_{c_{\pm}}|^2 d \bm{x}+\frac{1}{Ca \epsilon}(F(\phi^{n+1})-F(\phi^n),1)+\frac{1}{Ca \epsilon}((S-\frac{f'(\xi_n)}{2})(\phi^{n+1}-\phi^n),\phi^{n+1}-\phi^n) \nonumber\\
			&+\frac{\beta}{Ca \epsilon}\sum_{\pm}(\zeta_{\pm}^{n+1}C_{\pm}^{n+1}-\zeta_{\pm}^{n}C_{\pm}^{n},\mu_{c_{\pm}}^{n+1})+\frac{\beta}{Ca \epsilon}\sum_{\pm}((\zeta_{\pm}^{n+1}-\zeta_{\pm}^n)C_{\pm}^{n+1},1)=0. \nonumber
		\end{align}
		For the last two terms in the above, we use the following relation
		\begin{align*}
			&(\zeta_{\pm}^{n+1}C_{\pm}^{n+1}-\zeta_{\pm}^n C_{\pm}^n)\ln C_{\pm}^{n+1}-(\zeta_{\pm}^{n+1}-\zeta_{\pm}^n)C_{\pm}^{n+1} \\
			=&[(\zeta_{\pm}^{n+1}-\zeta_{\pm}^n)C_{\pm}^{n+1}+\zeta_{\pm}^n(C_{\pm}^{n+1}-C_{\pm}^n)]\ln C_{\pm}^{n+1}-(\zeta_{\pm}^{n+1}-\zeta_{\pm}^n)C_{\pm}^{n+1} \\
			=&(\zeta_{\pm}^{n+1}-\zeta_{\pm}^n)C_{\pm}^{n+1}(\ln C_{\pm}^{n+1}-1)+\zeta_{\pm}^n(C_{\pm}^{n+1}-C_{\pm}^n)\ln C_{\pm}^{n+1} \\
			=&(\zeta_{\pm}^{n+1}-\zeta_{\pm}^n)C_{\pm}^{n+1}(\ln C_{\pm}^{n+1}-1)+\zeta_{\pm}^n(C_{\pm}^{n+1}-C_{\pm}^n)(\ln C_{\pm}^{n+1}-1)+\zeta_{\pm}^n(C_{\pm}^{n+1}-C_{\pm}^n).
		\end{align*}
		Using the Taylor expansion, it is easy to get
		\begin{equation*}
			\ln C_{\pm}^{n+1}-\ln C_{\pm}^n=\frac{1}{C_{\pm}^n}(C_{\pm}^{n+1}-C_{\pm}^n)-\frac{1}{2\xi^2}(C_{\pm}^{n+1}-C_{\pm}^n)^2.
		\end{equation*}
		Combining the above two 
		equalities, we find
		\begin{align}\label{semi-pf12}
			&(\zeta_{\pm}^{n+1}C_{\pm}^{n+1}-\zeta_{\pm}^n C_{\pm}^n)\ln C_{\pm}^{n+1}-(\zeta_{\pm}^{n+1}-\zeta_{\pm}^n)C_{\pm}^{n+1} \\
			=&\zeta_{\pm}^{n+1}C_{\pm}^{n+1}(\ln C_{\pm}^{n+1}-1)-\zeta_{\pm}^n C_{\pm}^n (\ln C_{\pm}^n-1)+\frac{\zeta_{\pm}^n C_{\pm}^n}{2 \xi^2}(C_{\pm}^{n+1}-C_{\pm}^n)^2.\nonumber
		\end{align}
		Therefore, combining \eqref{semi-pf11} and \eqref{semi-pf12}, we have the energy stability result.
	\end{proof}

	\subsection{The fully-discrete LDG scheme and its energy stability}
	\label{subse2}
	In this subsection,   the LDG scheme to solve the scheme \eqref{eq:semi-phi}-\eqref{eq:projection} is proposed. 
	
	Let $\mathcal{T}_h=\{K\}$ be a shape-regular subdivision of $\Omega$. $\mathcal{E}_h$ denotes the union of the boundary faces of elements $K \in \mathcal{T}_h$, and $\mathcal{E}_0=\mathcal{E}_h \setminus  \partial \Omega$. Let $\mathcal{P}^k(K)$ be the
	space of polynomials of degree at most $k\geq 0$ on $K \in
	\mathcal{T}_h$. The DG finite element spaces are denoted by
	\begin{align*}
		V_{h}=&\Bigl\{\varphi:\;\;\;\varphi\vert_K\in
		\mathcal{P}^k(K),\;\;\;\; \; \forall K\in \mathcal{T}_{h}\Bigr\},\\
		\bm{W}_h = & \Bigl\{\Phi=(\phi_1,\cdots,\phi_d)^T :\;\;\;
		\phi_l\vert_{ K }\in \mathcal{P}^k(K), \;\;l=1\cdots d,\;\;\;
		\forall K \in \mathcal{T}_{h}\Bigr\},\\
		\bm{\Pi}_h=&\Bigl\{\bm{\Theta}=(\bm{\theta}_1,\cdots,\bm{\theta}_d)^T :\;\;\;
		\bm{\theta}_l\vert_{ K }\in (\mathcal{P}^k(K))^d, \;\;l=1\cdots d,\;\;\;
		\forall K \in \mathcal{T}_{h}\Bigr\}.
	\end{align*}
	
	To develop the LDG scheme, we first rewrite \eqref{eq:semi-phi}-\eqref{eq:projection} as a first-order system
	\begin{subequations}\label{first_system}
		\begin{align}
			\frac{\phi^{n+1}-\phi^n}{\Delta t}=&\mathcal{M} \nabla \cdot \bm{w}^{n+1}-\nabla \cdot \bm{q}^{n+1}-\frac{K}{Ca} \tilde\mu_{\phi}^{n+1} |\bm{v}^n|,\\
			\bm{w}^{n+1}=&\nabla \tilde\mu_{\phi}^{n+1},\\
			\bm{q}^{n+1} =&\phi^n \bm{u}_{*}^n,\\
			\tilde\mu_{\phi}^{n+1}=&-\epsilon \nabla \cdot \bm{v}^{n+1}+\frac{S}{\epsilon}(\phi^{n+1}-\phi^n)+\frac{1}{\epsilon}f(\phi^n)-\frac{\beta}{\epsilon}\sum_{\pm} \frac{\zeta_{\pm}^{n+1}-\zeta_{\pm}^n}{\phi^{n+1}-\phi^n} C_{\pm}^{n+1},\\
			\bm{v}^{n+1} =& \nabla \phi^{n+1},\\
			\frac{\zeta_{\pm}^{n+1}C_{\pm}^{n+1}-\zeta_{\pm}^nC_{\pm}^n}{\Delta t} =&\frac{1}{Pe} \nabla \cdot \bm{w}_{\pm}^{n+1}-\nabla \cdot \bm{q}_{\pm}^{n+1},\\
			\bm{w}_{\pm}^{n+1} =&\zeta_{\pm}^n D_{\pm}^n   C_{\pm}^{n+1} \bm{s}_{\pm}^{n+1},\\
			\bm{s}_{\pm}^{n+1} =&\nabla \mu_{c_{\pm}}^{n+1},\\
			\mu_{c_{\pm}}^{n+1} =&\ln C_{\pm}^{n+1},\\
			\bm{q}_{\pm}^{n+1}=&\bm{u}_{*}^n \zeta_{\pm}^n C_{\pm}^{n},\\
			Re \frac{\widetilde{\bm{u}}^{n+1}-\bm{u}^n}{\Delta t}=&\eta \nabla \cdot \bm{Q}^{n+1}-Re (\bm{u}^n \cdot \nabla)\widetilde{\bm{u}}^{n+1}-\bm{r}^n-\frac{1}{Ca} \phi^n \bm{w}^{n+1}-\frac{\beta}{Ca \epsilon} \sum_{\pm} \zeta^n_{\pm}C_{\pm}^{n}\bm{s}_{\pm}^{n+1},\\
			\bm{Q}^{n+1}=&\nabla \widetilde{\bm{u}}^{n+1},\\
			\bm{r}^{n+1} =&\nabla p^{n+1},\\
			Re \frac{\bm{u}^{n+1}-\widetilde{\bm{u}}^{n+1}}{\Delta t} =&-(\bm{r}^{n+1}-\bm{r}^{n}),\\
			\nabla \cdot \bm{u}^{n+1} =&0.
		\end{align}
	\end{subequations}
	Then the LDG scheme to solve \eqref{first_system} is: find $\phi^{n+1}$, $\tilde\mu_{\phi}^{n+1}$, $C_{\pm}^{n+1}$, $\mu_{c_{\pm}}^{n+1}$, $p^{n+1} \in V_h$, $\bm{w}^{n+1}$, $\bm{q}^{n+1}$, $\bm{v}^{n+1}$, $\bm{w}_{\pm}^{n+1}$, $\bm{s}_{\pm}^{n+1}$, $\bm{q}_{\pm}^{n+1}$, $\widetilde{\bm{u}}^{n+1}$, $\bm{u}^{n+1}$, $\bm{r}^{n+1} \in \bm{W}_h$, $\bm{Q}^{n+1} \in \bm{\Pi}_h$, such that, for all test functions $\varphi_1$, $\varphi_2$, $\varphi_3$, $\psi_{\pm}$, $\chi_{\pm} \in V_h$, $\bm{\theta}_1$, $\bm{\theta}_2$, $\bm{\theta}_3$, $\bm{\theta}_4$, $\bm{\theta}_5$, $\bm{\theta}_6$, $\bm{\varrho}_{\pm}$, $\bm{\varpi}_{\pm}$,  $\bm{\vartheta}_{\pm} \in \bm{W}_h$, and $\bm{\Theta} \in \bm{\Pi}_h$, we have
	\begin{subequations}\label{LDG_scheme}
		\begin{align}
			\int_K \frac{\phi^{n+1}-\phi^n}{\Delta t} \varphi_1 dK=&-\int_K (\mathcal{M}\bm{w}^{n+1}-\bm{q}^{n+1}) \cdot \nabla \varphi_1 dK+\int_{\partial K} (\mathcal{M}\widehat{\bm{w}}^{n+1}\nonumber \\
			&-\widehat{\bm{q}}^{n+1}) \cdot \n \varphi_1 ds-\int_K \frac{K}{Ca} \tilde\mu_{\phi}^{n+1} |\bm{v}^n|\varphi_1 dK, 
			\label{LDG_scheme1}\\
			\int_K \bm{w}^{n+1} \cdot \bm{\theta}_1 dK=&-\int_K \tilde\mu_{\phi}^{n+1} \nabla \cdot \bm{\theta}_1 dK+\int_{\partial K} \widehat{\tilde\mu}_{\phi}^{n+1} \bm{\theta}_1 \cdot \n ds,\label{LDG_scheme2} \\
			\int_K \bm{q}^{n+1} \cdot \bm{\theta}_2 dK =& \int_K \phi^n \bm{u}_{*}^n \cdot \bm{\theta}_2 dK,\label{LDG_scheme3}\\
			\int_K \tilde\mu_{\phi}^{n+1} \varphi_2 dK=& \epsilon \int_K \bm{v}^{n+1} \cdot \nabla \varphi_2 dK-\epsilon \int_{\partial K} \widehat{\bm{v}}^{n+1} \cdot \n \varphi_2 ds+\frac{1}{\epsilon}\int_K f(\phi^n)\varphi_2 dK\nonumber \\
			&+ \int_K \left(\frac{S}{\epsilon}(\phi^{n+1}-\phi^n)-\frac{\beta}{\epsilon}\sum_{\pm} \frac{\zeta_{\pm}^{n+1}-\zeta_{\pm}^n}{\phi^{n+1}-\phi^n}C_{\pm}^{n+1}\right) \varphi_2 dK, \label{LDG_scheme4}\\
			\int_K \bm{v}^{n+1} \cdot \bm{\theta}_3 dK =& -\int_K \phi^{n+1} \nabla \cdot \bm{\theta}_3 dK+\int_{\partial K} \widehat{\phi}^{n+1} \bm{\theta}_3 \cdot \n ds,\label{LDG_scheme5}\\
			\int_K  \frac{\zeta_{\pm}^{n+1}C_{\pm}^{n+1}-\zeta_{\pm}^nC_{\pm}^n}{\Delta t}  \psi_{\pm} dK =&-\int_K (\frac{1}{Pe}\bm{w}_{\pm}^{n+1}-\bm{q}_{\pm}^{n+1}) \cdot \nabla \psi_{\pm} dK \nonumber \\
			&+\int_{\partial K}(\frac{1}{Pe} \widehat{\bm{w}}^{n+1}_{\pm}-\widehat{\bm{q}}^{n+1}_{\pm}) \cdot \n \psi_{\pm}ds,\label{LDG_scheme6}\\
			\int_K \bm{w}^{n+1}_{\pm} \cdot \bm{\varrho}_{\pm}dK=&\int_K \zeta_{\pm}^n D_{\pm}^n   C_{\pm}^{n+1} \bm{s}_{\pm}^{n+1} \cdot \bm{\varrho}_{\pm} dK,\label{LDG_scheme7}\\
			\int_K \bm{s}_{\pm}^{n+1} \cdot \bm{\varpi}_{\pm} dK =&-\int_K \mu_{c_{\pm}}^{n+1} \nabla \cdot \bm{\varpi}_{\pm} dK+\int_{\partial K} \widehat{\mu}_{c_{\pm}}^{n+1} \bm{\varpi}_{\pm} \cdot \n ds,\label{LDG_scheme8}\\
			\int_K \mu_{c_{\pm}}^{n+1} \chi_{\pm} dK =& \int_K \ln C_{\pm}^{n+1} \chi_{\pm} dK,\label{LDG_scheme9}\\
			\int_K \bm{q}_{\pm}^{n+1} \cdot \bm{\vartheta}_{\pm} dK =& \int_K \bm{u}_{*}^n \zeta^n_{\pm} C_{\pm}^{n} \cdot \bm{\vartheta}_{\pm} dK, \label{LDG_scheme10}\\
			Re \int_K \frac{\widetilde{\bm{u}}^{n+1}-\bm{u}^n}{\Delta t} \cdot \bm{\theta}_4 dK =& -\eta \int_K \bm{Q}^{n+1} \cdot \nabla  \bm{\theta}_4 dK+\eta \int_{\partial K} (\widehat{\bm{Q}}^{n+1} \cdot \n) \cdot \bm{\theta}_4 ds\nonumber\\
			&- \int_K (Re(\bm{u}^n \cdot \nabla) \widetilde{\bm{u}}^{n+1} +\bm{r}^n+\frac{1}{Ca} \phi^n \bm{w}^{n+1} \nonumber\\
			&+\frac{\beta}{Ca \epsilon} \sum_{\pm} \zeta_{\pm}^n C_{\pm}^{n} \bm{s}_{\pm}^{n+1}) \cdot \bm{\theta}_4 dK,\label{LDG_scheme11}\\
			\int_K \bm{Q}^{n+1} \cdot \bm{\Theta} dK =& -\int_K \widetilde{\bm{u}}^{n+1} \nabla \cdot \bm{\Theta} dK+\int_{\partial K} \widehat{\widetilde{\bm{u}}}^{n+1} \bm{\Theta} \cdot \n ds,\label{LDG_scheme12}\\
			\int_K \bm{r}^{n+1} \cdot \bm{\theta}_5 dK =&-\int_K p^{n+1} \nabla \cdot \bm{\theta}_5 dK+\int_{\partial K} \widehat{p}^{n+1} \bm{\theta}_5 \cdot \n ds,\label{LDG_scheme13}\\
			Re \int_K \frac{\bm{u}^{n+1}-\widetilde{\bm{u}}^{n+1}}{\Delta t} \cdot \bm{\theta}_6 dK=&-\int_K (\bm{r}^{n+1}-\bm{r}^n) \cdot \bm{\theta}_6 dK,\label{LDG_scheme14}\\ 
			0 =& -\int_K \bm{u}^{n+1} \cdot \nabla \varphi_3 dK +\int_{\partial K} \widehat{\bm{u}}_p^{n+1} \cdot \n \varphi_3 ds, \label{LDG_scheme15} 
		\end{align}
	\end{subequations}
	with 
	\begin{equation}\label{LDG_scheme_star}
		\bm{u}_{*}^n=\bm{u}^n-\frac{\Delta t}{Ca Re} \phi^n \bm{w}^{n+1}-\frac{\beta \Delta t}{Ca \epsilon Re} \sum_{\pm} \zeta_{\pm}^n C_{\pm}^{n} \bm{s}_{\pm}^{n+1}.
	\end{equation}
	Here, the ``hat'' terms are the so-called numerical fluxes. To properly define the numerical fluxes, we need to introduce the following notations.
	
	Let $e$ be an interior  face shared by the ``left'' and ``right'' elements $K_L$ and $K_R$ and define the normal vectors $\n_L$ and  $\n_R$ on $e$ pointing exterior to $K_L$ and $K_R$, respectively. For our purpose, ``left'' and ``right'' can be uniquely defined for each face according to any fixed rule. For example, we choose $\n_0$ as a constant vector. The left element $K_L$ to the face requires that $\n_L \cdot \n_0<0$, and the right one $K_R$ requires $\n_R \cdot \n_0 \geq 0$. If $\psi$ is a function on $K_L$ and $K_R$, but possibly discontinuous across $e$ , let $\psi_L$ denote $(\psi|_{K_L})|_e$ and $\psi_R$ denote $(\psi|_{K_R})|_e$, the left and right trace, respectively.
	
	In the following proof of the energy stability, we can see the alternating numerical fluxes can guarantee the energy stability, such as
	\begin{align}\label{flux}
		& \widehat{\bm{w}}^{n+1}=\bm{w}^{n+1}_{L}, \;\;\;\; \widehat{\bm{q}}^{n+1}=\bm{q}^{n+1}_L, \;\;\;\; 
		\widehat{\tilde\mu}_{\phi}^{n+1}=\tilde\mu^{n+1}_{\phi,R},\;\;\;\; 
		\widehat{\bm{v}}^{n+1}=\bm{v}^{n+1}_L, 
		\nonumber \\
		& 
		\widehat{\phi}^{n+1}=\phi^{n+1}_R,\;\;\;\;\widehat{\bm{q}}_{\pm}^{n+1}=\bm{q}^{n+1}_{\pm,L},\;\;\;\;\widehat{\bm{w}}_{\pm}^{n+1}=\bm{w}^{n+1}_{\pm,L},\;\;\;\;
		\widehat{\mu}^{n+1}_{c_{\pm}}=\mu^{n+1}_{c_{\pm},R},\\
		&\widehat{\bm{Q}}^{n+1}=\bm{Q}^{n+1}_R,\;\;\;\;  \widehat{\widetilde{\bm{u}}}^{n+1}=\widetilde{\bm{u}}^{n+1}_{L},\;\;\;\;\widehat{p}^{n+1}=p^{n+1}_R, \;\;\;\; \widehat{\bm{u}}_p^{n+1}=\bm{u}^{n+1}_L. \nonumber
	\end{align}
	Under the boundary condition \eqref{eqn: bdc}, the boundary numerical fluxes are
	\begin{equation}\label{flux_bc}
		\widehat{\bm{w}}^{n+1}\cdot \n=0, ~\widehat{\bm{q}}^{n+1}=\bm{0},  ~
		\widehat{\bm{v}}^{n+1}\cdot\n = 0,~
		\widehat{\bm{w}}^{n+1}_{\pm} \cdot \n=0,~
		\widehat{\bm{q}}^{n+1}_{\pm} =\bm{0},~
		\widehat{\bm{u}}_p^{n+1}=\bm{0},    
	\end{equation}
	and the rest boundary fluxes come from the interior of the domain.
	
	For the convenience of presentation, we introduce some LDG operators. Denote
	\begin{align*}
		&H_K^{+}(r,\bm{v})=\int_K r \nabla \cdot \bm{v} dK\!-\!\int_{\partial K} r_R \bm{v} \cdot \n ds,~
		H_K^{-}(\bm{v},r)=\int_K \bm{v} \cdot \nabla r dK\!-\!\int_{\partial K} \bm{v}_L \cdot \n r ds.
	\end{align*}
	Summing up over $K$, we define
	\begin{equation*}
		H^{+}(r,\bm{v})=\sum_{K}  H_K^{+}(r,\bm{v}),\;\;\;\;\; H^{-}(\bm{v},r)=\sum_{K}  H_K^{-}(\bm{v},r).
	\end{equation*}
	With the special choice of the flux \eqref{flux} and the boundary conditions \eqref{eqn: bdc}, we have the property 
	\begin{equation}\label{H_propeety}
		H^{+}(r,\bm{v})=-H^{-}(\bm{v},r).
	\end{equation}
	
	Next, we will prove its energy stability.
	
	\begin{thm}
		Assuming that the condition \eqref{truncation} is satisfied and $S \geq L/2$. Then the solution to the fully-discrete LDG scheme \eqref{LDG_scheme} with the numerical fluxes \eqref{flux} and the boundary conditions \eqref{flux_bc} satisfies the following discrete energy dissipation
		\begin{equation*}
			\mathcal{E}_h^{n+1}-\mathcal{E}_h^n \leq -\Delta t \int_{\Omega} \left(\eta |\bm{Q}^{n+1}|^2+\frac{\mathcal{M}}{Ca} |\bm{w}^{n+1}|^2+\frac{1}{Ca}\frac{K}{Ca}|\bm{v}^n||\tilde\mu_{\phi}^{n+1}|^2+\sum_{\pm} \frac{\beta}{Ca \epsilon Pe}\zeta_{\pm}^nD_{\pm}^n   C_{\pm}^{n+1} |\bm{s}_{\pm}^{n+1}|^2\right)d\x,
		\end{equation*}
		where
		\begin{align*}
			\mathcal{E}_h^n=&\frac{Re}{2} \int_{\Omega}|\bm{u}^n|^2 d\x+\frac{1}{Ca}\int_{\Omega} \left(\frac{\epsilon}{2}|\bm{v}^n|^2+\frac{1}{\epsilon}F(\phi^n)\right)d\x+\frac{\Delta t^2}{2Re}\int_{\Omega} |\bm{r}^n|^2 d\x+\frac{\beta}{Ca \epsilon} \sum_{\pm} \int_{\Omega} \zeta_{\pm}^n C_{\pm}^n (\ln C_{\pm}^n-1)d\x.
		\end{align*}
	\end{thm}
	\begin{proof}
		Taking the inner product of \eqref{LDG_scheme_star} with $\bm{u}_{*}^n$, we obtain
		\begin{equation}\label{LDG-pf1}
			\frac{Re}{\Delta t} \int_K \bm{u}_{*}^n\cdot (\bm{u}_{*}^n-\bm{u}^n)dK=-\int_K \left(\frac{1}{Ca} \phi^n \bm{w}^{n+1}+\frac{\beta}{Ca \epsilon} \sum_{\pm} \zeta_{\pm}^n C_{\pm}^{n} \bm{s}_{\pm}^{n+1}\right) \cdot  \bm{u}_{*}^n dK.   
		\end{equation}
		From \eqref{LDG_scheme_star}, we have
		\begin{equation}\label{LDG-pf2}
			Re \frac{\widetilde{\bm{u}}^{n+1}-\bm{u}^n}{\Delta t}+\frac{1}{Ca} \phi^n \bm{w}^{n+1}+\frac{\beta}{Ca \epsilon}  \sum_{\pm} \zeta_{\pm}^n C_{\pm}^{n} \bm{s}_{\pm}^{n+1} =Re  \frac{\widetilde{\bm{u}}^{n+1}-\bm{u}_{*}^n}{\Delta t}.
		\end{equation}
		For \eqref{LDG_scheme11} of the LDG scheme, taking the test function $\bm{\theta}_4=\widetilde{\bm{u}}^{n+1}$ and using the relation \eqref{LDG-pf2}, we derive
		\begin{align}\label{LDG-pf3}
			Re \int_K \frac{\widetilde{\bm{u}}^{n+1}-\bm{u}_{*}^n}{\Delta t} \cdot  \widetilde{\bm{u}}^{n+1} dK =& -\eta \int_K \bm{Q}^{n+1} \cdot \nabla  \widetilde{\bm{u}}^{n+1} dK+\eta \int_{\partial K} (\widehat{\bm{Q}}^{n+1} \cdot \n) \cdot \widetilde{\bm{u}}^{n+1} ds \nonumber\\
			&- \int_K (Re(\bm{u}^n \cdot \nabla) \widetilde{\bm{u}}^{n+1}+\bm{r}^n) \cdot \widetilde{\bm{u}}^{n+1} dK.
		\end{align}
		For the convenience of presentation, for any $w$ we denote $\mathcal{D}w=w^{n+1}-w^n$. For \eqref{LDG_scheme5} and \eqref{LDG_scheme13}, subtracting the equation at time level $t^n$ from the equation at time level $t^{n+1}$, and choosing test functions $\bm{\theta}_3=\frac{\epsilon}{Ca \Delta t} \bm{v}^{n+1}$, $\bm{\theta}_5=\bm{u}^{n+1}$, we get
		\begin{align}
			\frac{\epsilon}{Ca \Delta t} \int_K \mathcal{D} \bm{v} \cdot \bm{v}^{n+1} dK &=-\frac{\epsilon}{Ca \Delta t} H_K^{+}(\mathcal{D}\phi, \bm{v}^{n+1}),\label{LDG-pf4}\\
			\int_K \mathcal{D} \bm{r} \cdot \bm{u}^{n+1} dK &=-H_K^{+}(\mathcal{D}p,\bm{u}^{n+1}).\label{LDG-pf5}
		\end{align}
		For \eqref{LDG_scheme1}, \eqref{LDG_scheme2}, \eqref{LDG_scheme3}, \eqref{LDG_scheme4}, \eqref{LDG_scheme6}, \eqref{LDG_scheme7}, \eqref{LDG_scheme8}, \eqref{LDG_scheme9}, \eqref{LDG_scheme10}, \eqref{LDG_scheme12}, \eqref{LDG_scheme14} and \eqref{LDG_scheme15} in the fully-discrete scheme, taking the test functions as
		\begin{align*}
			& \varphi_1=\frac{1}{Ca} \tilde\mu_{\phi}^{n+1},\;\; \bm{\theta}_1=\frac{1}{Ca}(\mathcal{M} \bm{w}^{n+1}-\bm{q}^{n+1}),\;\;\bm{\theta}_2=\frac{1}{Ca} \bm{w}^{n+1},\;\; \varphi_2=-\frac{1}{Ca \Delta t} \mathcal{D} \phi,\\
			&\psi_{\pm}=\frac{\beta }{Ca \epsilon} \mu_{c_{\pm}}^{n+1},\;\; \bm{\varrho}_{\pm}=-\frac{\beta }{Ca \epsilon Pe} \bm{s}_{\pm}^{n+1},\;\;\bm{\varpi}_{\pm}=\frac{\beta}{Ca \epsilon} (\frac{1}{Pe} \bm{w}_{\pm}^{n+1}-\bm{q}_{\pm}^{n+1}),\;\;\bm{\theta}_6=\bm{u}^{n+1},\\
			& \chi_{\pm}=-\frac{\beta}{Ca \epsilon \Delta t}(\zeta_{\pm}^{n+1}C_{\pm}^{n+1}-\zeta_{\pm}^{n}C_{\pm}^{n}),\;\;\bm{\vartheta}_{\pm}=\frac{\beta }{Ca \epsilon} \bm{s}_{\pm}^{n+1},\;\;\bm{\Theta}=\eta \bm{Q}^{n+1}, \;\;\varphi_3=\mathcal{D} p.
		\end{align*}
		We have
		\begin{align}
			\frac{1}{Ca \Delta t}\int_K \mathcal{D} \phi \tilde\mu_{\phi}^{n+1} dK =&-\frac{1}{Ca}H_K^{-}(\mathcal{M} \bm{w}^{n+1}-\bm{q}^{n+1},\tilde\mu_{\phi}^{n+1}) \nonumber\\
			&-\frac{1}{Ca}\frac{K}{Ca}\int_K |\bm{v}^n| |\tilde\mu_{\phi}^{n+1}|^2 dK,\\
			\frac{1}{Ca}\int_K \bm{w}^{n+1} \cdot (\mathcal{M} \bm{w}^{n+1}-\bm{q}^{n+1}) dK =&-\frac{1}{Ca} H_K^{+}(\tilde\mu_{\phi}^{n+1},\mathcal{M} \bm{w}^{n+1}-\bm{q}^{n+1}),\\
			\frac{1}{Ca}\int_K \bm{q}^{n+1} \cdot \bm{w}^{n+1} dK &=\frac{1}{Ca} \int_K \phi^n \bm{u}_{*}^n \cdot \bm{w}^{n+1}dK,\\
			-\frac{1}{Ca \Delta t}\int_K \tilde\mu_{\phi}^{n+1} \mathcal{D} \phi dK =&-\frac{\epsilon}{Ca \Delta t} H_K^{-}(\bm{v}^{n+1},\mathcal{D}\phi)+\frac{\beta}{Ca \Delta t \epsilon}\int_K \sum_{\pm} \mathcal{D} \zeta_{\pm} C_{\pm}^{n+1}dK\nonumber\\
			&-\frac{1}{Ca \Delta t}\int_K \left(\frac{S}{\epsilon}(\mathcal{D}\phi)^2+\frac{1}{\epsilon} f(\phi^n) \mathcal{D}\phi\right)dK,\\
			\frac{\beta}{Ca \epsilon \Delta t}\int_K (\zeta_{\pm}^{n+1}C_{\pm}^{n+1}-\zeta_{\pm}^{n}C_{\pm}^{n})\mu_{c_{\pm}}^{n+1}dK=& -\frac{\beta}{Ca \epsilon} H_K^{-}(\frac{1}{Pe} \bm{w}_{\pm}^{n+1}-\bm{q}_{\pm}^{n+1},\mu_{c_{\pm}}^{n+1}),\\
			-\frac{\beta}{Ca \epsilon Pe}\int_K \bm{w}^{n+1}_{\pm} \cdot \bm{s}_{\pm}^{n+1}dK=&-\frac{\beta}{Ca \epsilon Pe}\int_K \zeta_{\pm}^n D_{\pm}^n   C_{\pm}^{n+1} \bm{s}_{\pm}^{n+1} \cdot \bm{s}_{\pm}^{n+1} dK,\\
			\frac{\beta}{Ca \epsilon}\int_K \bm{s}_{\pm}^{n+1} \cdot (\frac{1}{Pe}\bm{w}^{n+1}_{\pm}-\bm{q}^{n+1}_{\pm}) dK =&-\frac{\beta }{Ca \epsilon}H_K^{+}(\mu_{c_{\pm}}^{n+1},\frac{1}{Pe}\bm{w}^{n+1}_{\pm}-\bm{q}^{n+1}_{\pm}),\\
			-\frac{\beta}{Ca \epsilon \Delta t}\int_K \mu_{c_{\pm}}^{n+1} (\zeta_{\pm}^{n+1}C_{\pm}^{n+1}-\zeta_{\pm}^{n}C_{\pm}^{n}) dK =&-\frac{\beta}{Ca \epsilon \Delta t} \int_K \ln C_{\pm}^{n+1} (\zeta_{\pm}^{n+1}C_{\pm}^{n+1}-\zeta_{\pm}^{n}C_{\pm}^{n}) dK,\\
			\frac{\beta}{Ca \epsilon}\int_K \bm{q}_{\pm}^{n+1} \cdot \bm{s}_{\pm}^{n+1}dK =& \frac{\beta}{Ca \epsilon}\int_K \bm{u}_{*}^n \zeta^n_{\pm} C_{\pm}^{n} \cdot \bm{s}_{\pm}^{n+1} dK, \\
			\eta \int_K \bm{Q}^{n+1} \cdot \bm{Q}^{n+1} dK =& -\eta \int_K \widetilde{\bm{u}}^{n+1} \nabla \cdot \bm{Q}^{n+1} dK+\eta \int_{\partial K} \widehat{\widetilde{\bm{u}}}^{n+1} \bm{Q}^{n+1} \cdot \n ds,\\
			Re \int_K \frac{\bm{u}^{n+1}-\widetilde{\bm{u}}^{n+1}}{\Delta t} \cdot \bm{u}^{n+1} dK=&-\int_K (\bm{r}^{n+1}-\bm{r}^n) \cdot \bm{u}^{n+1} dK,\\
			0=&-H_K^{-}(\bm{u}^{n+1},\mathcal{D}p).\label{LDG-pf6}
		\end{align}
		Now combining the equations \eqref{LDG-pf1}, \eqref{LDG-pf3}-\eqref{LDG-pf6}, summing over all elements $K$ and by the property \eqref{H_propeety}, we have
		\begin{align}\label{LDG-pf7}
			& \frac{\epsilon}{Ca \Delta t} \int_{\Omega}  \bm{v}^{n+1} \cdot(\bm{v}^{n+1}-\bm{v}^n)d\x +\frac{Re}{\Delta t} \int_{\Omega} \bm{u}^{n+1} \cdot (\bm{u}^{n+1}-\widetilde{\bm{u}}^{n+1})d\x \\
			&+\frac{Re}{\Delta t} \int_{\Omega} \bm{u}_{*}^n \cdot (\bm{u}_{*}^n-\bm{u}^n) d\x\!\! +\!\!\frac{Re}{\Delta t} \int_{\Omega} \widetilde{\bm{u}}^{n+1} \cdot (\widetilde{\bm{u}}^{n+1}-\bm{u}_{*}^n)d\x \!\!+\!\!\frac{1}{Ca} \int_{\Omega} \mathcal{M} \bm{w}^{n+1} \cdot  \bm{w}^{n+1} d\x\nonumber \\
			&+\frac{K}{Ca} \int_{\Omega} |\bm{v}^n| |\tilde\mu_{\phi}^{n+1}|^2 d\x +\frac{\beta}{Ca \epsilon Pe} \sum_{\pm} \int_{\Omega} \zeta_{\pm}^n D_{\pm}^n   C_{\pm}^{n+1} \bm{s}_{\pm}^{n+1} \cdot  \bm{s}_{\pm}^{n+1} d\x\nonumber \\
			&+\eta \int_{\Omega} \bm{Q}^{n+1} \cdot \bm{Q}^{n+1} d\x+\int_{\Omega} \bm{r}^n \cdot \widetilde{\bm{u}}^{n+1}d\x +\frac{1}{Ca \Delta t}\int_{\Omega} \left(\frac{S}{\epsilon}\mathcal{D}\phi+\frac{f(\phi^n)}{\epsilon}\right)\mathcal{D} \phi d\x\nonumber\\
			&+\frac{\beta}{Ca \epsilon \Delta t}\sum_{\pm}\int_{\Omega} \left(\ln C_{\pm}^{n+1}(\zeta_{\pm}^{n+1} C_{\pm}^{n+1}-\zeta_{\pm}^{n} C_{\pm}^{n})-\mathcal{D} \zeta_{\pm} C_{\pm}^{n+1}\right) d\x =0.\nonumber
		\end{align}

		For the last two terms in \eqref{LDG-pf7}, we have
		\begin{align}\label{LDG-pf8}
			&\frac{1}{Ca}\int_{\Omega} \left(\frac{S}{\epsilon}\mathcal{D}\phi+\frac{f(\phi^n)}{\epsilon}\right)\mathcal{D} \phi d\x+\frac{\beta}{Ca \epsilon}\sum_{\pm}\int_{\Omega} \left(\ln C_{\pm}^{n+1}(\zeta_{\pm}^{n+1} C_{\pm}^{n+1}-\zeta_{\pm}^{n} C_{\pm}^{n})-\mathcal{D} \zeta_{\pm} C_{\pm}^{n+1}\right) d\x  \\
			=&\frac{1}{Ca \epsilon} \int_{\Omega} \left(F(\phi^{n+1})-F(\phi^n)\right) d\x+\frac{1}{Ca \epsilon} \int_{\Omega} \left(S-\frac{f'(\xi_n)}{2}\right) (\mathcal{D}\phi)^2 d\x\nonumber\\
			&+\frac{\beta}{Ca \epsilon} \sum_{\pm} \int_{\Omega}\left(\zeta_{\pm}^{n+1}C_{\pm}^{n+1}(\ln C_{\pm}^{n+1}-1)-\zeta_{\pm}^{n}C_{\pm}^{n}(\ln C_{\pm}^{n}-1)\right) d\x+\frac{\beta}{Ca \epsilon} \sum_{\pm} \int_{\Omega} \frac{\zeta_{\pm}^n C_{\pm}^n}{2\xi^2}(\mathcal{D} C_{\pm})^2d\x.\nonumber 
		\end{align}
		To deal with the term $\int_{\Omega} \bm{r}^n \cdot \widetilde {\bm{u}}^{n+1}$, we take $\bm{\theta}_6=\bm{r}^n$  and $\bm{r}^{n+1}$ in \eqref{LDG_scheme14}, sum over all elements,
		\begin{align}\label{LDG-pf9}
			\frac{Re}{\Delta t} \int_{\Omega} \widetilde{\bm{u}}^{n+1} \cdot \bm{r}^n d\x=\frac{1}{2} \int_{\Omega}\left(|\bm{r}^{n+1}|^2-|\bm{r}^{n}|^2-|\bm{r}^{n+1}-\bm{r}^{n}|^2\right) d\x,\\
			\frac{Re}{\Delta t} \int_{\Omega} \widetilde{\bm{u}}^{n+1} \cdot \bm{r}^{n+1} d\x=\frac{1}{2} \int_{\Omega}\left(|\bm{r}^{n+1}|^2-|\bm{r}^{n}|^2+|\bm{r}^{n+1}-\bm{r}^{n}|^2\right) d\x.
		\end{align}
		where the term $\int_{\Omega} \bm{u}^{n+1} \cdot \bm{r}^n d\x$ is zero because of the divergence-free condition.
		
		Using the above two equations and  \eqref{LDG_scheme14}  yield
		\begin{equation}\label{LDG-pf10}
			\frac{Re^2 }{\Delta t^2} \int_{\Omega}|\bm{u}^{n+1}-\widetilde{\bm{u}}^{n+1}|^2 d\x=\int_{\Omega} |\bm{r}^{n+1}-\bm{r}^n|^2 d\x.  
		\end{equation}
		Combining \eqref{LDG-pf9} and \eqref{LDG-pf10}, we find
		\begin{equation}\label{LDG-pf11}
			\int_{\Omega} \bm{r}^n \cdot \widetilde{\bm{u}}^{n+1}d\x=\frac{\Delta t}{2Re}\int_{\Omega} (|\bm{r}^{n+1}|^2-|\bm{r}^{n}|^2)d\x-\frac{Re}{2 \Delta t} \int_{\Omega} |\bm{u}^{n+1}-\widetilde{\bm{u}}^{n+1}|^2d\x.
		\end{equation}
		Combining \eqref{LDG-pf7}, \eqref{LDG-pf8}, and \eqref{LDG-pf11}, we obtain
		\begin{align*}
			&\frac{\epsilon}{2Ca \Delta t}\int_{\Omega} \left(|\bm{v}^{n+1}|^2-|\bm{v}^n|^2\right)d\x+\frac{Re}{2\Delta t} \int_{\Omega} \left(|\bm{u}^{n+1}|^2-|\bm{u}^n|^2\right)d\x+\frac{\Delta t}{2Re}\int_{\Omega} \left(|\bm{r}^{n+1}|^2-|\bm{r}^n|^2\right)d\x\\
			& +\frac{1}{Ca \epsilon \Delta t} \int_{\Omega} \left(F(\phi^{n+1})-F(\phi^n)\right) d\x+\frac{\beta}{Ca \epsilon \Delta t} \sum_{\pm} \int_{\Omega} \left(\zeta_{\pm}^{n+1}C_{\pm}^{n+1}(\ln C_{\pm}^{n+1}-1)-\zeta_{\pm}^{n}C_{\pm}^{n}(\ln C_{\pm}^{n}-1)\right)d\x \\
			\leq &-\left(\frac{1}{Ca} \int_{\Omega} \mathcal{M} |\bm{Q}^{n+1}|^2 d\x+\frac{1}{Ca}\frac{K}{Ca} \int_{\Omega} |\tilde\mu_{\phi}^{n+1}|^2 |\bm{v}^n|d\x+\eta\int_{\Omega} |\bm{Q}^{n+1}|^2 d\x+\frac{\beta}{Ca \epsilon Pe} \sum_{\pm} \int_{\Omega} \zeta_{\pm}^n D_{\pm}^n   C_{\pm}^{n+1} |\bm{s}_{\pm}^{n+1}|^2 d\x\right),
		\end{align*}
		where we use the identity
		$ a(a-b)=\frac{1}{2}a^2-\frac{1}{2}b^2+\frac{1}{2}(a-b)^2 $
		for $a=\bm{u}^{n+1},\bm{u}_{*}^n, \widetilde{\bm{u}}^{n+1}, \bm{v}^{n+1}$, and $b=\widetilde{\bm{u}}^{n+1}, \bm{u}^{n}, \bm{u}_{*}^n, \bm{v}^n$, respectively. 
		Therefore, we have the energy stability result.
	\end{proof}

	\subsection{The semi-implicit SDC method}
	\label{subse3}
	The scheme proposed above is energy stable. However, it is limited to first-order temporal accuracy, while the LDG method is high-order accurate in space. To achieve high-order accuracy in time and space, we iteratively incorporate the semi-implicit spectral deferred correction (SDC) method \cite{guo-xia, guo-xu-xia}, based on the energy-stabilized first-order scheme. 
	
	We now perform the SDC procedure over each time interval \([t^n, t^{n+1}]\). To do this, we divide the interval \([t^n, t^{n+1}]\) into \(P\) subintervals by selecting interpolation points \(\{t^{n,m}\}_{m=0}^P\) such that
	\[
	t^n = t^{n,0} < t^{n,1} < \cdots < t^{n,m} < \cdots < t^{n,P} = t^{n+1}.
	\]
	Let \(\Delta t^{n,m} = t^{n,m+1} - t^{n,m}\) and denote by \(\phi_L^{n,m}\) the \(L\)th-order approximation of \(\phi(t^{n,m})\). To ensure high-order accuracy and avoid the instability associated with equispaced nodes, the time nodes \(\{t^{n,m}\}_{m=0}^P\) can be chosen as the Chebyshev–Gauss–Lobatto points on \([t^n, t^{n+1}]\).
	
	Starting from the known solution \((\phi^n, \bm{u}^n, C_{\pm}^n)\) at time \(t^n\), we outline below the algorithm used to compute the updated solution \((\phi^{n+1}, \bm{u}^{n+1}, C_{\pm}^{n+1})\).
	
	{\bf Compute the initial approximation:}
	
	Let $\phi_1^{n,0}=\phi^n$, $\bm{u}_1^{n,0}=\bm{u}^n$ and $C_{\pm,1}^{n,0}=C_{\pm}^n$.
	
	Use the first-order scheme  \eqref{eq:semi-phi}-\eqref{eq:projection}  to compute a first-order accurate approximate solution $\phi_1$, $\bm{u}_1$ and $C_{\pm,1}$ at the nodes $\{t^{n,m}\}_{m=1}^P$, i.e. 
	
	For $m=0,\ldots,P-1$
	
	\noindent {\bf Step 1:}
	\begin{equation}
		\left\{
		\begin{aligned}
			& \phi^{n,m+1}_1=\phi^{n,m}_1+\Delta t^{n,m} (\mathcal{M} \Delta \tilde \mu_{\phi,1}^{n,m+1}-\nabla \cdot (\phi_1^{n,m}\bm{u}^{n,m}_{*,1})-\frac{K}{Ca}\tilde \mu_{\phi,1}^{n,m+1} |\nabla \phi_{1}^{n,m}|),\\
			& \tilde \mu_{\phi,1}^{n,m+1}=-\epsilon \Delta \phi_1^{n,m+1}+\frac{S}{\epsilon}(\phi_1^{n,m+1}-\phi_1^{n,m})+\frac{1}{\epsilon} f(\phi_1^{n,m})-\frac{\beta}{\epsilon}\sum_{\pm} \frac{\zeta_{\pm,1}^{n,m+1}-\zeta_{\pm,1}^{n,m}}{\phi_1^{n,m+1}-\phi_1^{n,m}} C_{\pm,1}^{n,m+1}, \\
			&\zeta_{\pm,1}^{n,m+1}C_{\pm,1}^{n,m+1}=\zeta_{\pm,1}^{n,m}C_{\pm,1}^{n,m}+\Delta t^{n,m}(\frac{1}{Pe} \nabla \cdot (\zeta_{\pm,1}^{n,m}D_{\pm,1}^{n,m}C_{\pm,1}^{n,m+1}\nabla \mu_{{c}_{\pm,1}}^{n,m+1})-\nabla \cdot(\bm{u}^{n,m}_{*,1}\zeta_{\pm,1}^{n,m}C_{\pm,1}^{n,m} )),\\
			&\mu_{{c}_{\pm,1}}^{n,m+1}=\ln C_{\pm,1}^{n,m+1},
		\end{aligned}
		\right .
	\end{equation}
	with
	\begin{equation}
		\bm{u}_{*,1}^{n,m}=\bm{u}_1^{n,m}-\frac{\Delta t^{n,m}}{Ca Re} \phi_1^{n,m} \nabla \tilde \mu_{\phi,1}^{n,m+1} -\frac{\beta \Delta t^{n,m}}{Ca \epsilon Re}\sum_{\pm}\zeta_{\pm,1}^{n,m} C_{\pm,1}^{n,m} \nabla \mu_{c_{\pm,1}}^{n,m+1};
	\end{equation}
	\noindent {\bf Step 2:}
	\begin{align}
		\widetilde{\bm{u}}_1^{n,m+1}=&\bm{u}_1^{n,m}-\Delta t^{n,m} (\bm{u}_1^{n,m} \cdot \nabla) \widetilde{\bm{u}}_1^{n,m+1} \nonumber\\
		&+\frac{\Delta t^{n,m}}{Re} (\eta \Delta \widetilde{\bm{u}}_1^{n,m+1}-\nabla p_1^{n,m}-\frac{1}{Ca} \phi_1^{n,m} \nabla \tilde \mu_{\phi,1}^{n,m+1}-\frac{\beta}{Ca \epsilon}\sum_{\pm}\zeta_{\pm,1}^{n,m} C_{\pm,1}^{n,m} \nabla \mu_{c_{\pm,1}}^{n,m+1} );
	\end{align}
	\noindent {\bf Step 3:}
	\begin{align}
		-\Delta (p_1^{n,m+1}-p_1^{n,m})&=-\frac{Re}{\Delta t^{n,m}} \nabla \cdot \widetilde{\bm{u}}^{n,m+1}_1,\\
		\bm{u}_1^{n,m+1}&=\widetilde{\bm{u}}^{n,m+1}_1-\frac{\Delta t^{n,m}}{Re}\nabla (p_1^{n,m+1}-p_1^{n,m}).
	\end{align}
	
	{\bf Compute successive corrections:}
	
	For $l=1,...,L$, 
	
	Let $\phi_{l+1}^{n,0}=\phi^n$, $\bm{u}_{l+1}^{n,0}=\bm{u}^n$ and $C_{\pm,l+1}^{n,0}=C_{\pm}^n$.
	
	For $m=0,\ldots,P-1$
	
	\noindent {\bf Step 1:}
	\begin{equation}
		\left\{
		\begin{aligned}
			\phi^{n,m+1}_{l+1}=& \phi^{n,m}_{l+1}+\Delta t^{n,m} (\mathcal{M} \Delta \tilde \mu_{\phi,l+1}^{n,m+1}-\nabla \cdot (\phi_{l+1}^{n,m}\bm{u}^{n,m}_{*,l+1})-\frac{K}{Ca}\tilde \mu_{\phi,l+1}^{n,m+1} |\nabla \phi_{l+1}^{n,m}| \\
			&-\mathcal{M} \Delta \tilde \mu_{\phi,l}^{n,m+1}+\nabla \cdot (\phi_{l}^{n,m}\bm{u}^{n,m}_{*,l})+\frac{K}{Ca}\tilde \mu_{\phi,l}^{n,m+1} |\nabla \phi_{l}^{n,m}|)+I_m^{m+1} (F_{\phi}), \\
			\tilde \mu_{\phi,l+1}^{n,m+1}=&-\epsilon \Delta \phi_{l+1}^{n,m+1}+\frac{S}{\epsilon}(\phi_{l+1}^{n,m+1}-\phi_{l+1}^{n,m})+\frac{1}{\epsilon} f(\phi_{l+1}^{n,m})-\frac{\beta}{\epsilon}\sum_{\pm} \frac{\zeta_{\pm,l+1}^{n,m+1}-\zeta_{\pm,l+1}^{n,m}}{\phi_{l+1}^{n,m+1}-\phi_{l+1}^{n,m}} C_{\pm,l+1}^{n,m+1},\\
			\zeta_{\pm,l+1}^{n,m+1} C_{\pm,l+1}^{n,m+1}=&\zeta_{\pm,l+1}^{n,m} C_{\pm,l+1}^{n,m}+\Delta t^{n,m} (\frac{1}{Pe}\nabla \cdot (\zeta_{\pm,l+1}^{n,m}D_{\pm,l+1}^{n,m}C_{\pm,l+1}^{n,m+1} \nabla \mu_{c_{\pm,l+1}}^{n,m+1})-\nabla \cdot (\bm{u}_{*,l+1}^{n,m}\zeta_{\pm,l+1}^{n,m}C_{\pm,l+1}^{n,m}) \\
			&-\frac{1}{Pe}\nabla \cdot (\zeta_{\pm,l}^{n,m}D_{\pm,l}^{n,m}C_{\pm,l}^{n,m+1} \nabla \mu_{c_{\pm,l}}^{n,m+1})+\nabla \cdot (\bm{u}_{*,l}^{n,m}\zeta_{\pm,l}^{n,m}C_{\pm,l}^{n,m}))+I_m^{m+1}(F_{c_{\pm}}),\\
			\mu_{{c}_{\pm,l+1}}^{n,m+1}=&\ln C_{\pm,l+1}^{n,m+1},
		\end{aligned}
		\right.
	\end{equation}
	with
	\begin{align*}
		& \bm{u}_{*,l+1}^{n,m}=\bm{u}_{l+1}^{n,m}-\frac{\Delta t^{n,m}}{Ca Re} \phi_{l+1}^{n,m} \nabla \tilde \mu_{\phi,l+1}^{n,m+1} -\frac{\beta \Delta t^{n,m}}{Ca \epsilon Re}\sum_{\pm}\zeta_{\pm,l+1}^{n,m} C_{\pm,l+1}^{n,m} \nabla \mu_{c_{\pm,l+1}}^{n,m+1},\\
		&  F_{\phi}=\mathcal{M} \Delta \tilde \mu_{\phi,l}-\nabla \cdot(\phi_l \bm{u}_l)-\frac{K}{Ca} \tilde \mu_{\phi,l} |\nabla \phi_l|,\\
		&F_{c_{\pm}}=\frac{1}{Pe} \nabla \cdot(\zeta_{\pm,l}D_{\pm,l}C_{\pm,l}\nabla \mu_{c_{\pm,l}})-\nabla \cdot (\bm{u}_l \zeta_{\pm,l}C_{\pm,l});
	\end{align*}
	\noindent {\bf Step 2:}
	\begin{align}
		\widetilde{\bm{u}}_{l+1}^{n,m+1}=&\bm{u}_{l+1}^{n,m}-\Delta t^{n,m}((\bm{u}_{l+1}^{n,m} \cdot \nabla) \widetilde{\bm{u}}_{l+1}^{n,m+1}-(\bm{u}_{l}^{n,m} \cdot \nabla) \widetilde{\bm{u}}_{l}^{n,m+1})+\frac{\Delta t^{n,m}}{Re}(\eta \Delta \widetilde{\bm{u}}_{l+1}^{n,m+1}-\nabla p_{l+1}^{n,m} \nonumber \\
		&-\frac{1}{Ca} \phi_{l+1}^{n,m} \nabla \tilde \mu_{\phi,l+1}^{n,m+1}-\frac{\beta}{Ca \epsilon}\sum_{\pm}\zeta_{\pm,l+1}^{n,m} C_{\pm,l+1}^{n,m} \nabla \mu_{c_{\pm,l+1}}^{n,m+1}-\eta \Delta \widetilde{\bm{u}}_{l}^{n,m+1}+\nabla p_{l}^{n,m}  \\
		&+\frac{1}{Ca} \phi_{l}^{n,m} \nabla \tilde \mu_{\phi,l}^{n,m+1}+\frac{\beta}{Ca \epsilon}\sum_{\pm}\zeta_{\pm,l}^{n,m} C_{\pm,l}^{n,m} \nabla \mu_{c_{\pm,l}}^{n,m+1})+I_m^{m+1} (F_{\bm{u}}), \nonumber
	\end{align}
	where
	\begin{equation*}
		F_{\bm{u}}=-(\bm{u}_l \cdot \nabla) \widetilde{\bm{u}}_l+\frac{1}{Re}(\eta \Delta \widetilde{\bm{u}}_l-\nabla p_l-\frac{1}{Ca} \phi_l \nabla \tilde \mu_{\phi,l}-\frac{\beta}{Ca \epsilon} \sum_{\pm} \zeta_{\pm,l}C_{\pm,l} \nabla \mu_{c_{\pm,l}});
	\end{equation*}
	\noindent {\bf Step 3:}
	\begin{align}
		&- (\Delta p^{n,m+1}_{l+1}-\Delta p^{n,m}_{l+1}) = -\frac{Re}{\Delta t^{n,m}} (\nabla \cdot \widetilde{\bm{u}}^{n,m+1}_{l+1}-\nabla \cdot \widetilde{\bm{u}}^{n,m+1}_l) +I_m^{m+1} (F),\\
		&\bm{u}_{l+1}^{n,m+1}=\widetilde{\bm{u}}^{n,m+1}_{l+1}+\frac{\Delta t^{n,m}}{Re}\nabla(p_{l+1}^{n,m+1}-p_{l+1}^{n,m})-\frac{\Delta t^{n,m}}{Re}\nabla(p_{l}^{n,m+1}-p_{l}^{n,m}),
	\end{align}
	where
	\begin{equation*}
		F=-\frac{Re}{(\Delta t^{n,m})^2} \nabla \cdot \widetilde{\bm{u}}_l.
	\end{equation*}
	Here, $I_m^{m+1}(F_{\phi})$ is the integral of the $P$-th degree interpolating polynomial on the $P+1$ points $(t^{n,m},F_{\phi})_{m=0}^P$ over the subinterval $[t^{n,m},t^{n,m+1}]$, which is the numerical quadrature approximation of
	\begin{equation*}
		\int_{t^{n,m}}^{t^{n,m+1}} F_{\phi} dt.
	\end{equation*}
	Finally, we have $\phi^{n+1}=\phi^{n,P}_{L+1}$, $\bm{u}^{n+1}=\bm{u}^{n,P}_{L+1}$ and $C_{\pm}^{n+1}=C_{\pm,L+1}^{n,P}$.
	\begin{remark}(Local truncation error.) 
		The local truncation error obtained with the above semi-implicit SDC scheme \cite{guo-xu-xia} is
		$\mathcal{O}(\tau^{\min[L+1,P+1]}),$
		where $\tau=\max \limits_{n,m} \Delta t^{n,m}$.
	\end{remark}
	\section{Simulation results}
	\label{se:numerical}
	In this section, we present numerical results to validate our proposed schemes and investigate the model's behavior. Specifically, the first example tests the order of convergence to verify the accuracy of our numerical method. The second example demonstrates the energy stability of the proposed scheme. Finally, the last two examples explore the influence of interface permeability and shear flow on the equilibrium profiles of droplets, providing insights into the effects of these factors on transmembrane dynamics.

	\begin{example}(Accuracy tests.)\label{ex1} Consider the model \eqref{eqn: maindimensionless} in the domain $\Omega=[0,2\pi] \times [0,2\pi]$. For the tests, we choose a suitable term such that the exact solution is taken as
		\begin{align*}
			&\phi(x,y,t)=0.3+0.2e^{-2t}\sin(x)\sin(y),\\
			&\bm{u}(x,y,t)=(-0.25e^{-2t}\sin^2(x)\sin(2y),0.25e^{-2t}\sin(2x)\sin^2(y)),\\
			&C_{+}(x,y,t)=0.2+0.1 e^{-2t}\cos(x)\cos(y),~
			C_{-}(x,y,t)=0.2+0.1 e^{-2t}\sin(x)\sin(y).
		\end{align*}
		Choose the parameters $\mathcal{M}=K=\epsilon=\beta=Pe=Ca=Re=\eta=1$.
	\end{example}
	
	To test the temporal accuracy of the scheme \eqref{LDG_scheme}, we let $N=64$ and $\mathcal{P}^2$ approximation to ensure that the spatial discretization error is small enough, and then the temporal discretization error is dominant. The time step refinement path is taken to be $\Delta t=\Delta t_0/2^m$, $\Delta t_0=0.02$ and $m=0,1,2,3$. The $L^2$ and $L^{\infty}$ errors and orders of convergence are presented in Table \ref{temporal_test}, which shows first-order accurate in time.

	\begin{table}[!h]
		\caption{\label{temporal_test} Temporal accuracy test. $L^2$ and $L^\infty$ errors and orders of convergence at $T=0.1$. The time step refinement path is taken to be $\Delta t=\Delta t_0/2^m$, $\Delta t_0=0.02$ and $m=0,1,2,3$.}
		\begin{center}
			\begin{tabular}{|l|c|cc|cc|}
				\hline
				& $m$ & \multicolumn{1}{l}{$L^2$ error } & \multicolumn{1}{l|}{$L^2$ order} & \multicolumn{1}{l}{$L^{\infty}$ error } & \multicolumn{1}{l|}{$L^{\infty}$ order} \\ \hline
				\multicolumn{1}{|c|}{\multirow{4}{*}{$\phi$}} & 0  &   9.64E-03  &  -- & 3.10E-03 &  --\\
				\multicolumn{1}{|c|}{}  & 1  & 5.02E-03  & 0.94 & 1.61E-03 & 0.95 \\
				\multicolumn{1}{|c|}{} & 2 & 2.56E-03  & 0.97  & 8.27E-04   & 0.96  \\
				\multicolumn{1}{|c|}{}  & 3  & 1.29E-03  &  0.99 & 4.18E-04  & 0.98\\ \hline
				\multirow{4}{*}{$\bm{u}$}  & 0  & 8.31E-03  & --  & 3.17E-03  & --  \\
				& 1 & 4.09E-03  & 1.02  & 1.61E-03  & 0.98 \\
				& 2 & 2.07E-03 & 0.98  & 8.18E-04 & 0.98  \\
				& 3 & 1.04E-03  & 0.99  & 4.18E-04 & 0.97 \\ \hline
				\multirow{4}{*}{$C_{+}$} & 0 & 4.55E-04  & --  & 1.44E-04  & -- \\
				& 1 & 2.29E-04 & 0.99  & 7.34E-05  & 0.97  \\
				& 2 & 1.15E-04 & 0.99  & 3.79E-05  & 0.95  \\
				& 3 & 5.82E-04 & 0.98 & 1.93E-05 & 0.97 \\ \hline
			\end{tabular}
		\end{center}
	\end{table}

The \( L^2 \) and \( L^{\infty} \) errors, along with the observed convergence orders, are summarized in Table~\ref{tab:merged_sdc_multicol} for both second-order SDC with \( \mathcal{P}^1 \) elements and third-order SDC with \( \mathcal{P}^2 \) elements. As expected, the second-order SDC scheme exhibits consistent second-order convergence in both norms for all variables when using \( \mathcal{P}^1 \) elements. Similarly, the third-order SDC scheme coupled with \( \mathcal{P}^2 \) elements achieves third-order accuracy in both spatial and temporal directions. These results confirm the effectiveness of the proposed LDG--SDC framework in achieving high-order accuracy while maintaining numerical stability, even under relatively large time step sizes.

\begin{table}[!h]
	\caption{\label{tab:merged_sdc_multicol} Spatial and temporal accuracy for second-order SDC with $\mathcal{P}^1$ and third-order SDC with $\mathcal{P}^2$ elements. $L^2$ and $L^\infty$ errors and convergence orders at $T=0.2$ with $\Delta t = 0.1 \Delta x$.}
	\centering 
	\begin{tabular}{|l|c|cc|cc|cc|cc|}
		\hline
		\multirow{2}{*}{Variable} & \multirow{2}{*}{$N$} & \multicolumn{4}{c|}{$\mathcal{P}^1$ } & \multicolumn{4}{c|}{$\mathcal{P}^2$  } \\
		\cline{3-10}
		& & $L^2$ Err & Order & $L^\infty$ Err & Order & $L^2$ Err & Order & $L^\infty$ Err & Order \\
		\hline
		
		\multirow{4}{*}{$\phi$}
		& 8  & 3.26E-02 & --   & 2.42E-02 & --   & 3.96E-03 & --   & 3.72E-03 & --   \\
		& 16 & 8.74E-03 & 1.90 & 6.80E-03 & 1.83 & 4.94E-04 & 3.01 & 4.63E-04 & 3.01 \\
		& 32 & 2.28E-03 & 1.94 & 1.78E-03 & 1.93 & 6.19E-05 & 3.00 & 5.66E-05 & 3.03 \\
		& 64 & 5.88E-04 & 1.96 & 4.57E-04 & 1.96 & 7.90E-06 & 2.97 & 7.18E-06 & 2.98 \\
		\hline
		
		\multirow{4}{*}{$\bm{u}$}
		& 8  & 9.60E-02 & --   & 7.67E-02 & --   & 2.10E-02 & --   & 1.95E-02 & --   \\
		& 16 & 2.35E-02 & 2.02 & 2.08E-02 & 1.88 & 2.65E-03 & 2.98 & 2.75E-03 & 2.83 \\
		& 32 & 5.98E-03 & 1.98 & 5.45E-03 & 1.94 & 3.33E-04 & 3.00 & 3.39E-04 & 3.02 \\
		& 64 & 1.50E-03 & 1.99 & 1.38E-03 & 1.98 & 4.23E-05 & 2.98 & 4.20E-05 & 3.01 \\
		\hline
		
		\multirow{4}{*}{$C_{+}$}
		& 8  & 1.58E-02 & --   & 1.16E-02 & --   & 2.01E-03 & --   & 1.97E-03 & --   \\
		& 16 & 3.95E-03 & 2.00 & 3.00E-03 & 1.96 & 2.49E-04 & 3.01 & 2.32E-04 & 3.08 \\
		& 32 & 9.77E-04 & 2.02 & 7.30E-04 & 2.04 & 3.17E-05 & 2.97 & 2.90E-05 & 3.00 \\
		& 64 & 2.43E-04 & 2.01 & 1.78E-04 & 2.03 & 4.03E-06 & 2.98 & 3.66E-06 & 2.99 \\
		\hline
	\end{tabular}
\end{table}

\begin{example}(Energy stability test.)
	In this example, we perform stability tests to verify the energy stability of the proposed scheme, using the same initial conditions and parameters as in Example~\ref{ex1}.
\end{example}



\begin{figure} \centering
	\subfigure[First order scheme] {
		\includegraphics[width=2.0in]{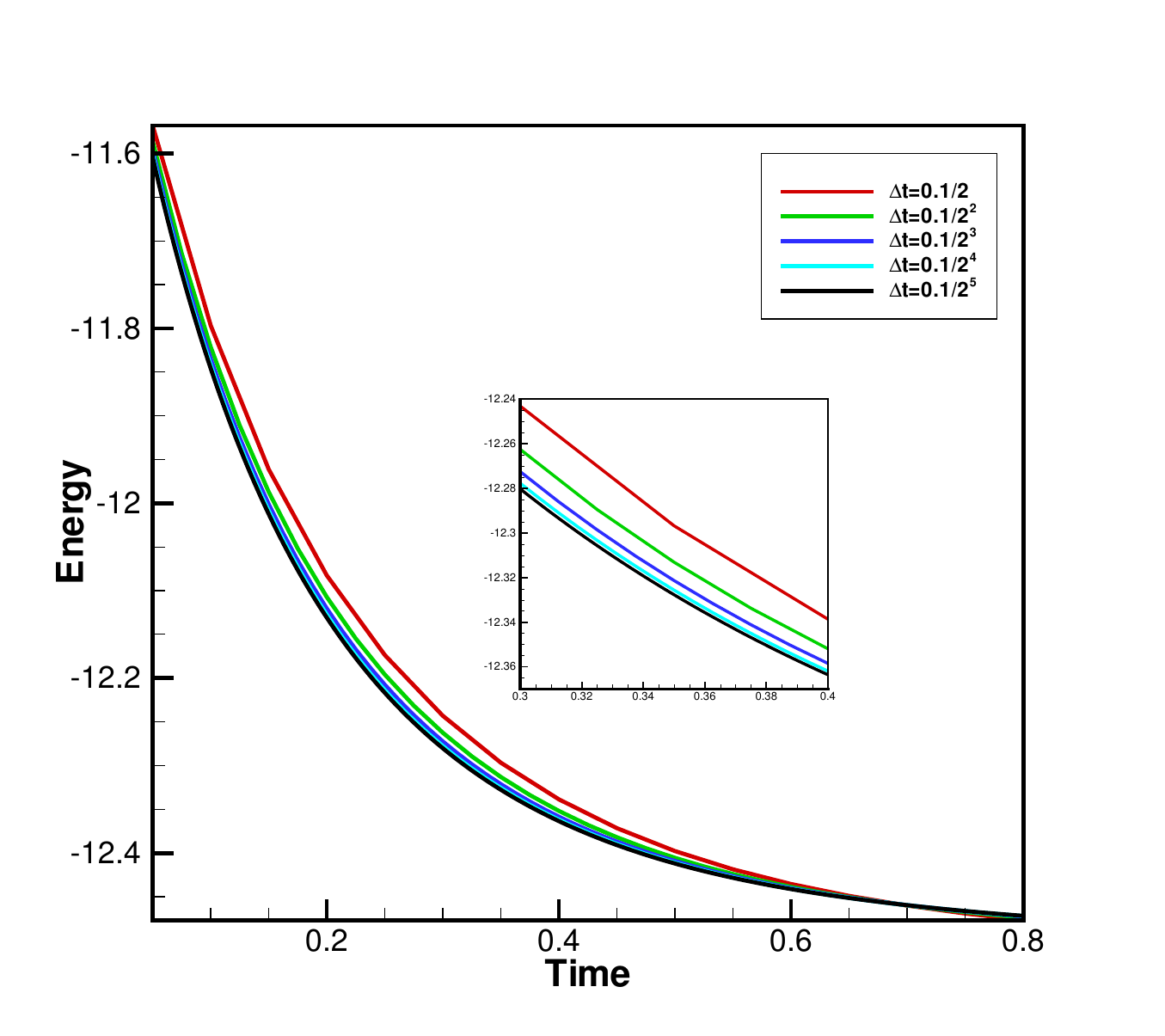}
	}
	\subfigure[Second order SDC scheme] {
		\includegraphics[width=2.0in]{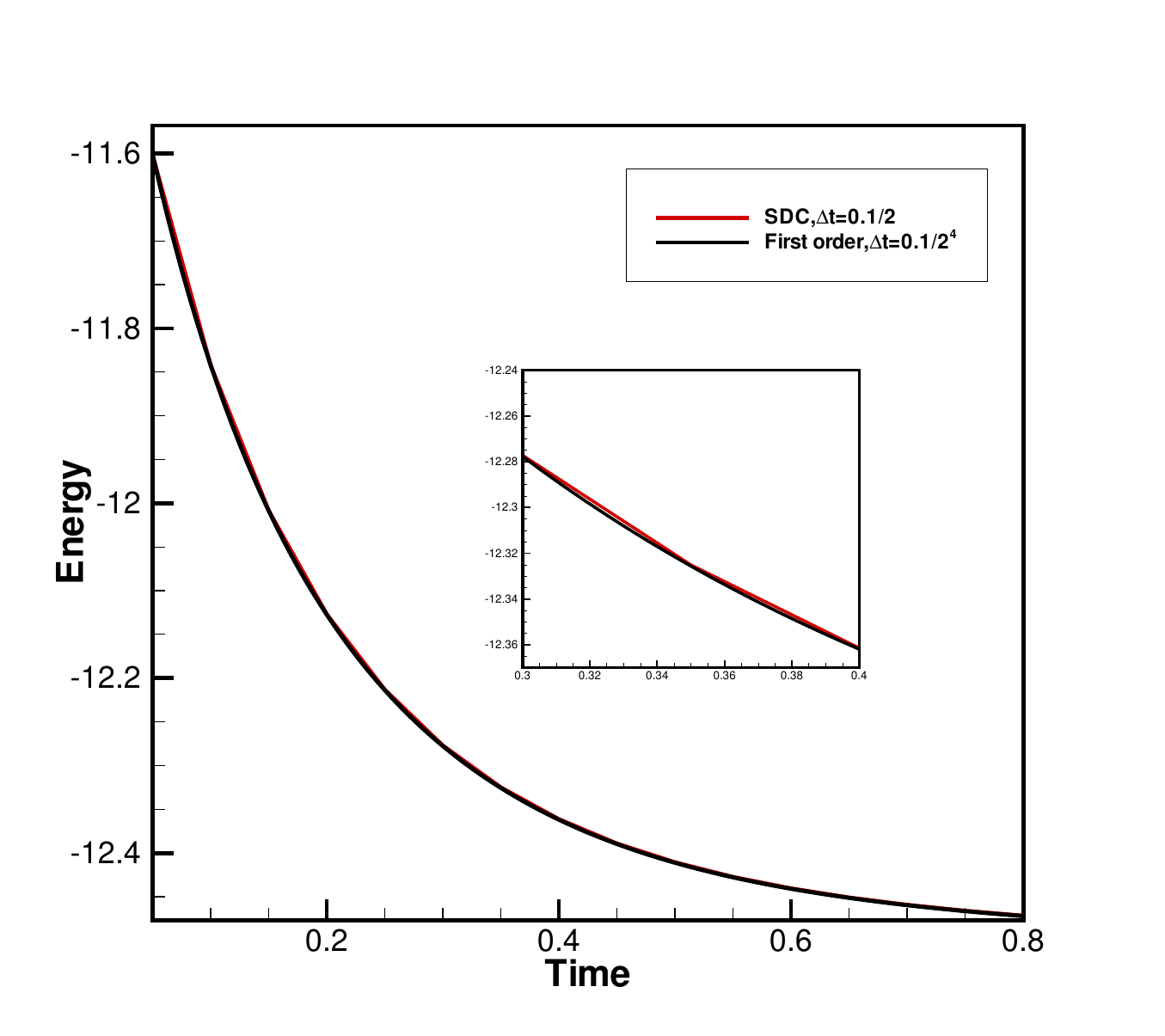}
	}
	\caption{Energy evolution of the first order scheme and the second order SDC scheme.}
	\label{energy-test}
\end{figure}

Figure~\ref{energy-test}(a) presents the energy evolution of the scheme \eqref{LDG_scheme} for various time steps, ranging from \(\Delta t = 0.1/2\) to \(\Delta t = 0.1/2^5\), using a \(\mathcal{P}^1\) approximation with a fixed mesh size \(\Delta x = \Delta y = 2\pi/64\). The energy curves consistently exhibit monotonic decay across all time step sizes, in agreement with the theoretical guarantee of unconditional energy stability.

For larger time steps (\(\Delta t = 0.1/2, 0.1/2^2, 0.1/2^3\)), noticeable deviations from the reference energy curve (computed with \(\Delta t = 0.1/2^5\)) are observed, reflecting the reduced accuracy of the first-order temporal scheme. This highlights the necessity of smaller time steps to achieve high-fidelity solutions. 

By contrast, our semi-implicit SDC scheme significantly improves temporal accuracy. As illustrated in Figure~\ref{energy-test}(b), the energy curve of the second-order SDC scheme with \(\Delta t = 0.1/2\) closely matches the reference energy curve computed with \(\Delta t = 0.1/2^4\), demonstrating the effectiveness of the SDC method in achieving both stability and high-order accuracy.

\begin{figure} \centering
	\subfigure[Initial condition of $\phi$] {
		\includegraphics[width=2.0in]{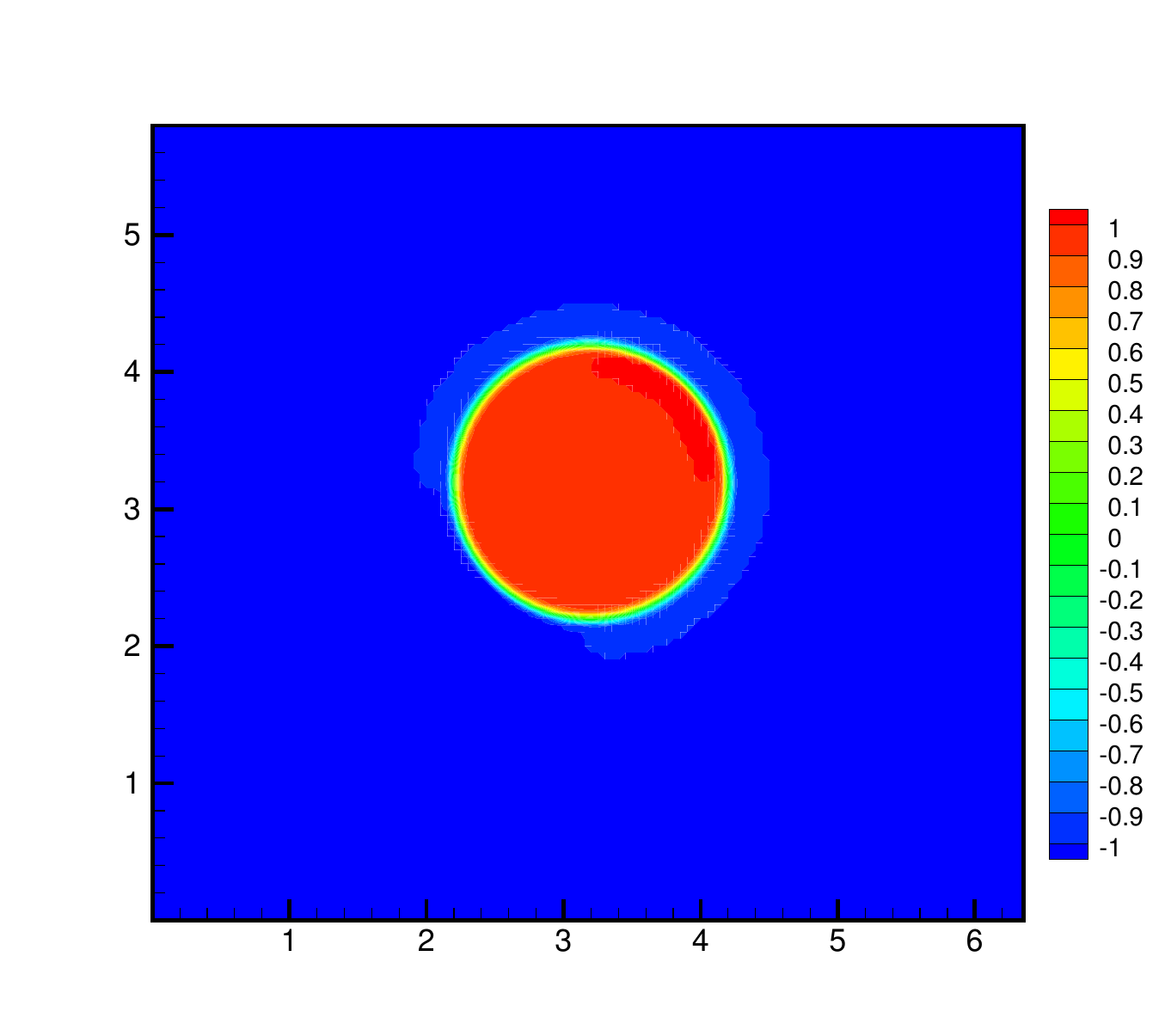}
	}
	\subfigure[Initial condition of $C_{+}$] {
		\includegraphics[width=2.0in]{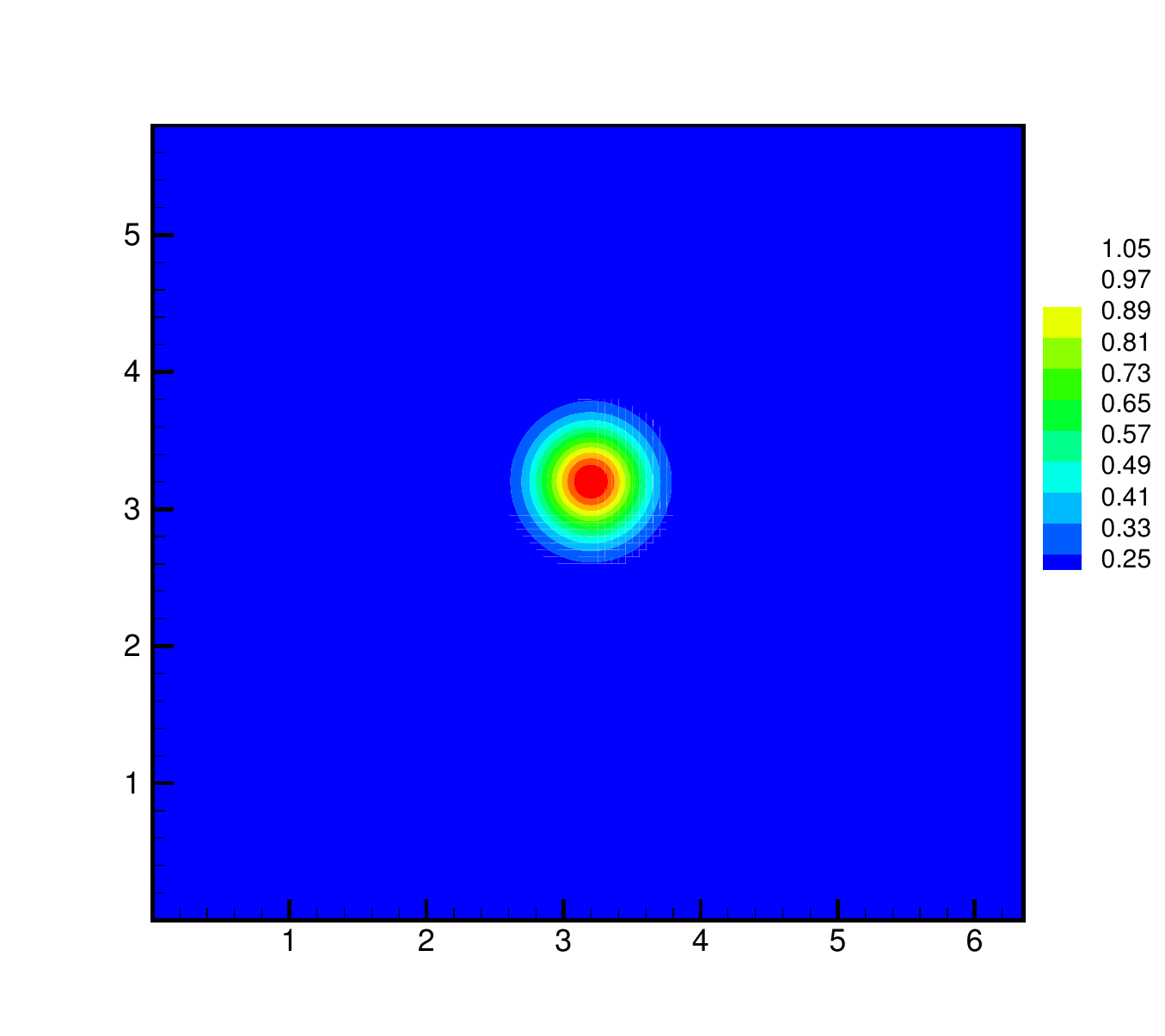}
	}
	\caption{Initial condition for Example \ref{example-ex3} with $c_0 =0.15$.}
	\label{initial}
\end{figure}

\begin{example}(Effects of permeability $K$.)\label{example-ex3}
	In this example, we consider how the permeability affects the equilibrium shapes of droplets.  
	Consider the model \eqref{eqn: maindimensionless} in the domain $\Omega=[0,6.4] \times [0,6.4]$. The initial profiles of concentration and interface are given as follows (see Figure \ref{initial})
	\begin{align*}
		\phi(x,y,0)&=\tanh\left(\frac{1.0-\sqrt{(x-3.2)^2+(y-3.2)^2}}{\sqrt{2}\epsilon}\right),\\
		\bm{u}(x,y,0)&= \bm{0},\\
		C_{+}(x,y,0)&=c_0+\exp\left(-\frac{(x-3.2)^2+(y-3.2)^2}{0.15}\right), \\
		C_{-}(x,y,0)&=0.02,
	\end{align*}
	with parameters 
	\[\mathcal{M}=0.03, \epsilon =0.03,  Re=1, Pe =1, \eta=1, Ca = 1, \beta =2\epsilon,  \]
	where the substance is concentrated inside the droplet center  (see Figure \ref{initial} (b)).  
	Two different initial concentration cases are considered: \( c_0 = 0.15 \) and \( c_0 = 0.1 \), each with three different permeabilities: \( K = 0 \), \( 0.1 \), and \( 0.3 \). Additionally, a reference case with no solute---where transmembrane flow is driven purely by hydrostatic pressure---is included for comparison to highlight the effects of osmotic pressure.
\end{example}

We employ the second-order SDC method with a piecewise \( \mathcal{P}^1 \) polynomial basis. The computational domain consists of a uniform mesh with \( 128 \times 128 \) elements, and the time step is set to \( \Delta t = 0.01 \).

\begin{figure} \centering
	\subfigure[$K=0$] {
		\includegraphics[width=2.0in]{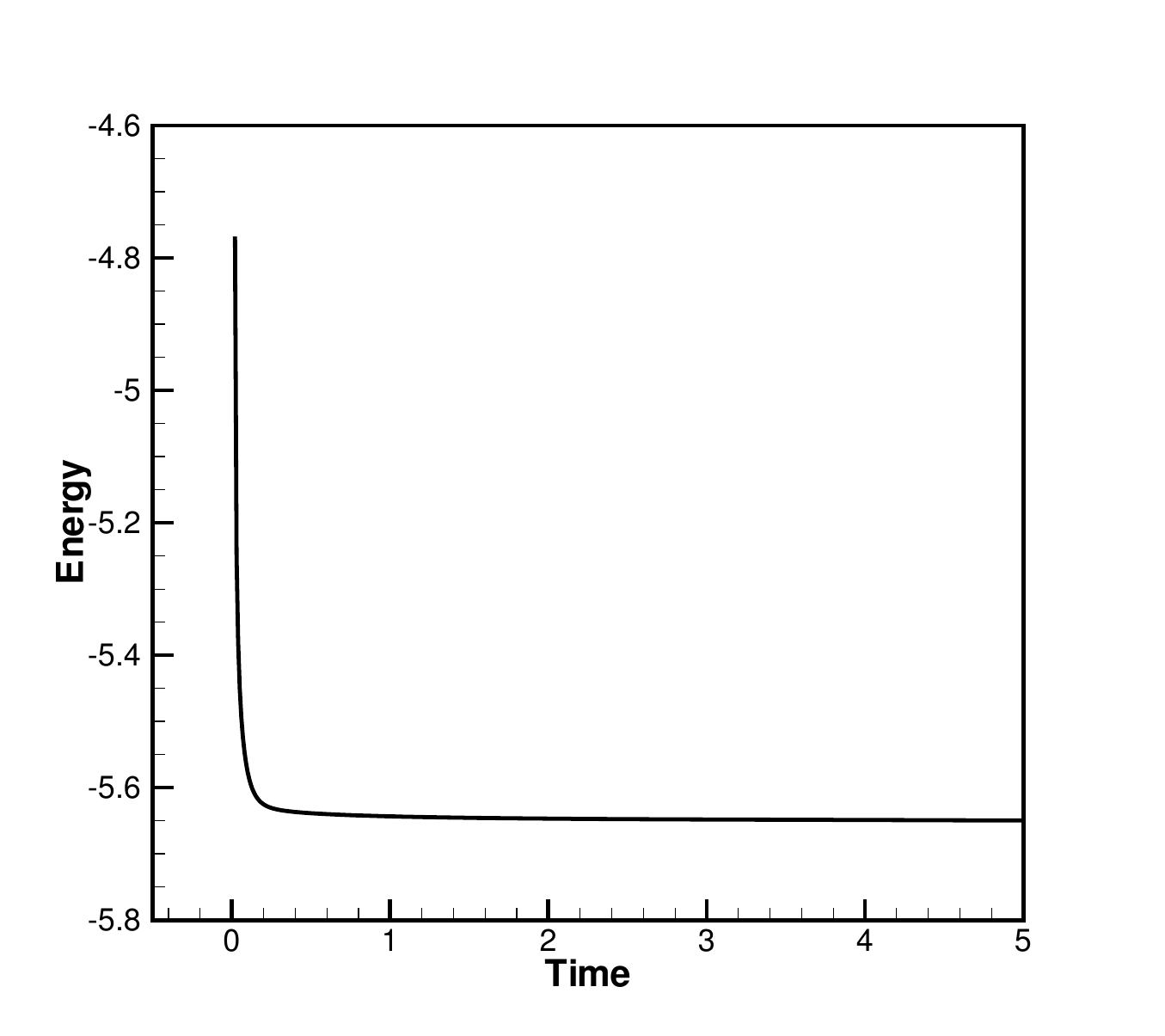}
	}
	\subfigure[$K=0.1$] {
		\includegraphics[width=2.0in]{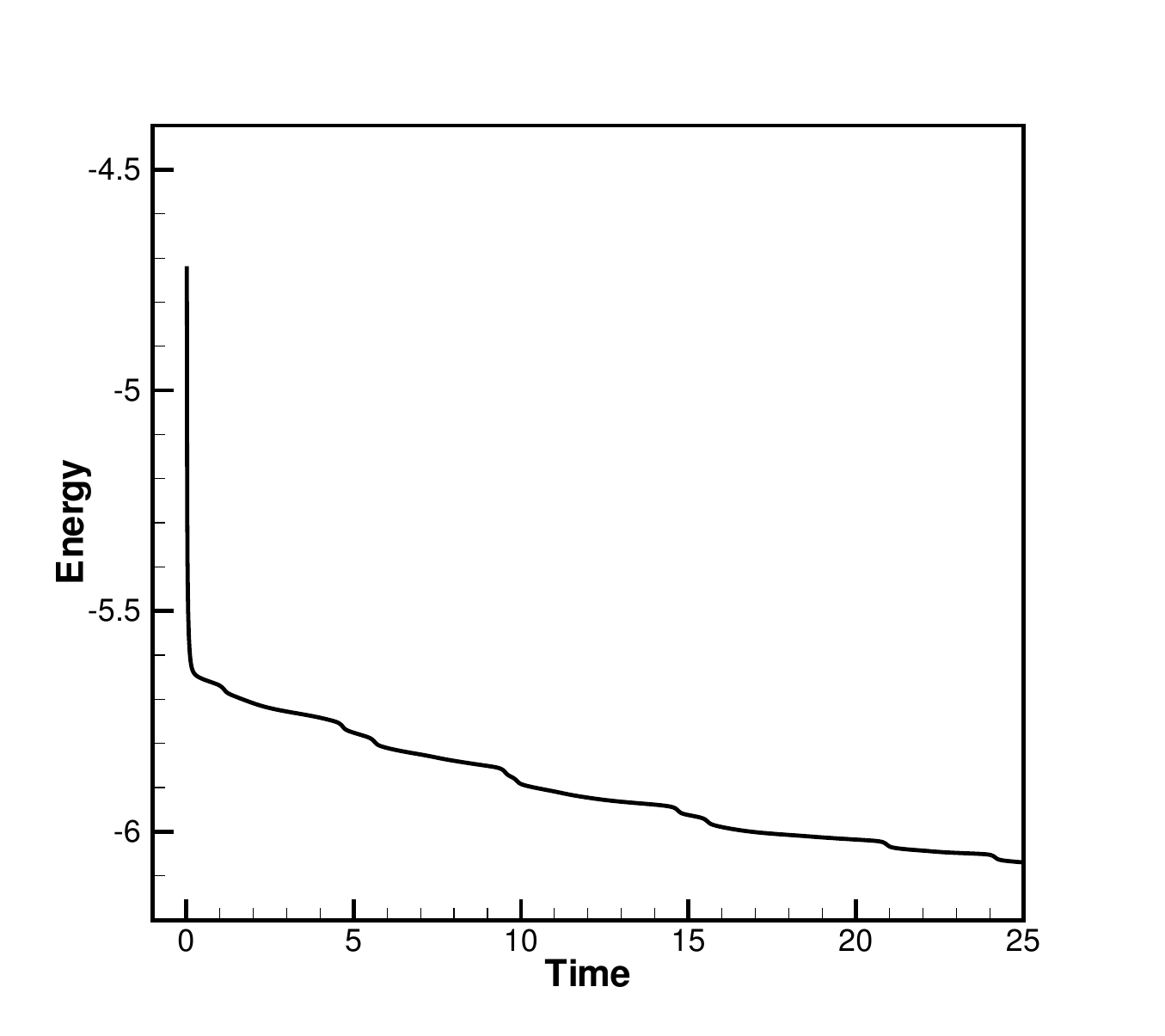}
	}
	\subfigure[$K=0.3$] {
		\includegraphics[width=2.0in]{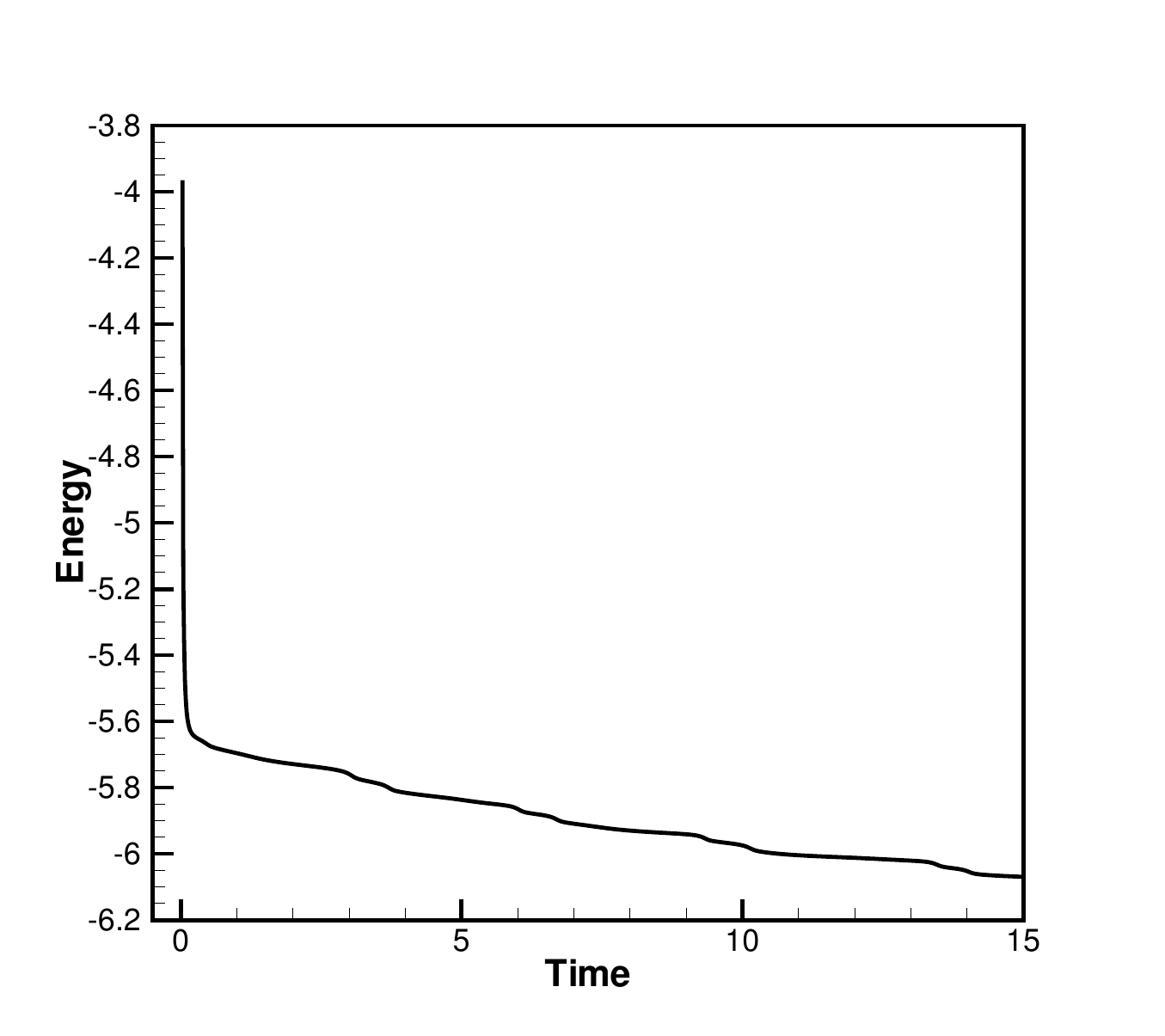}
	}
	\caption{Time evolution of the free energy functional with different permeability: $K=0,0.1$ and $0.3$ with $c_0 =0.15$.}
	\label{energy-ex3}
\end{figure}

\begin{figure} \centering
	\subfigure[$K=0$] {
		\includegraphics[width=2.0in]{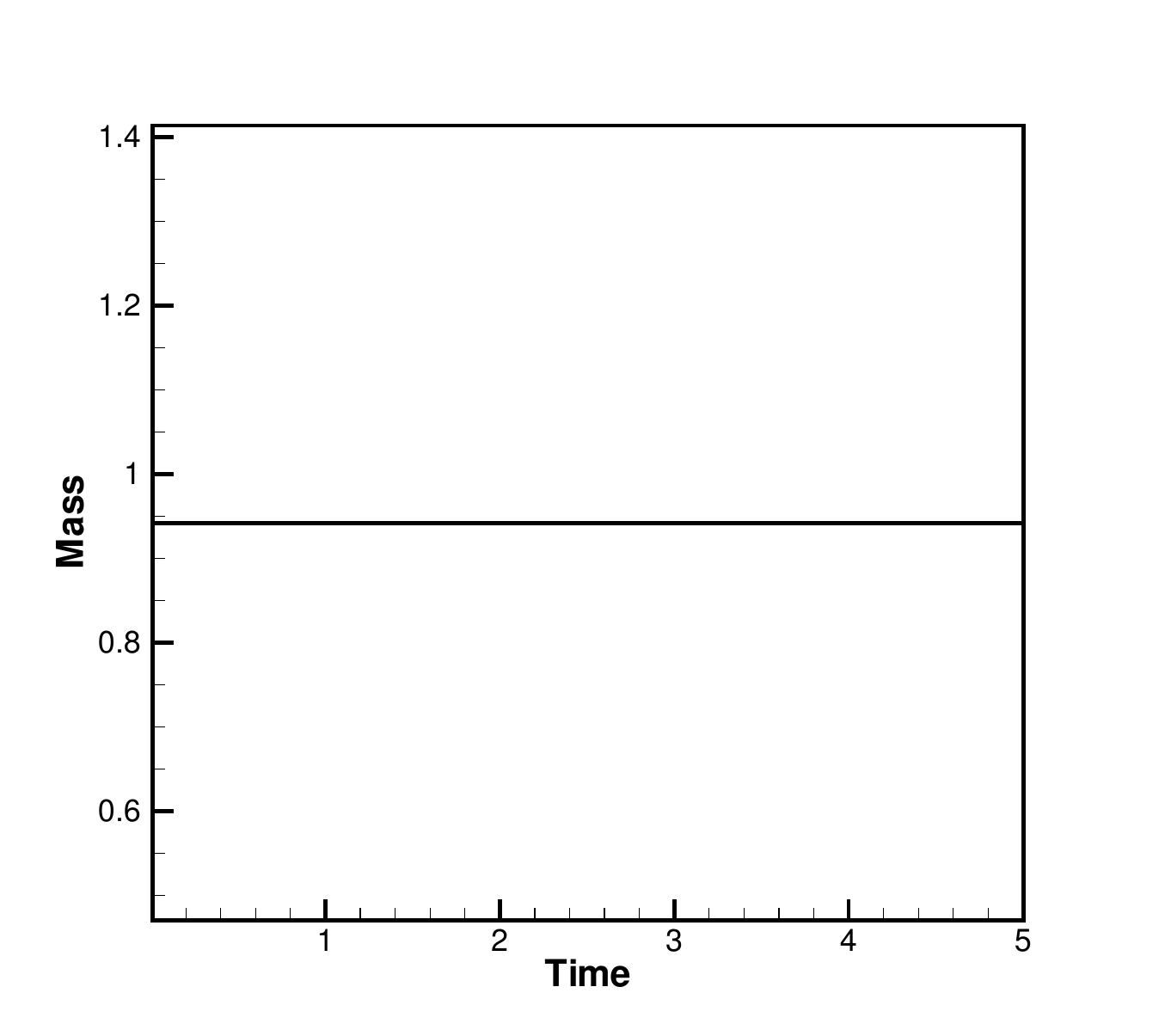}
	}
	\subfigure[$K=0.1$] {
		\includegraphics[width=2.0in]{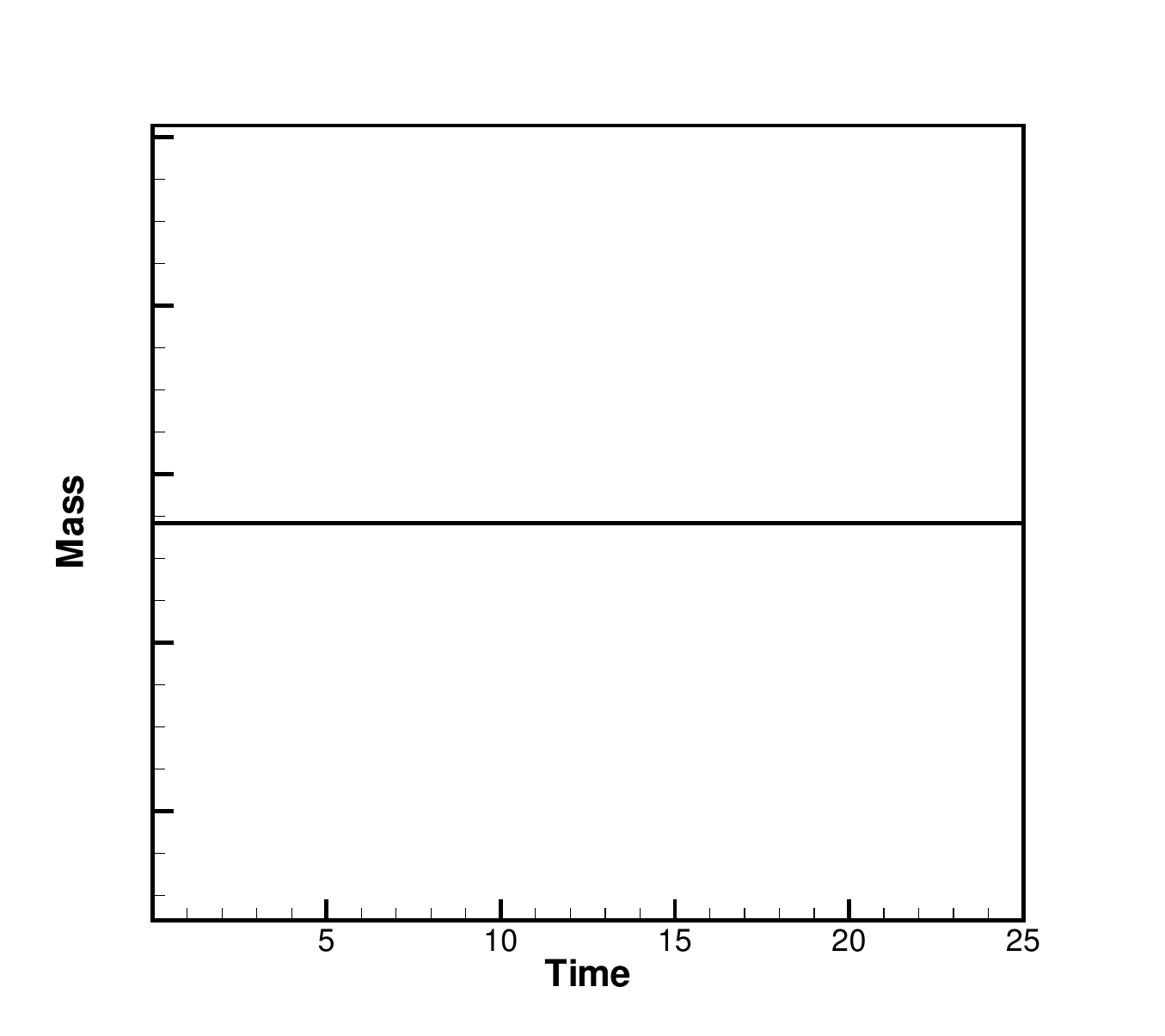}
	}
	\subfigure[$K=0.3$] {
		\includegraphics[width=2.0in]{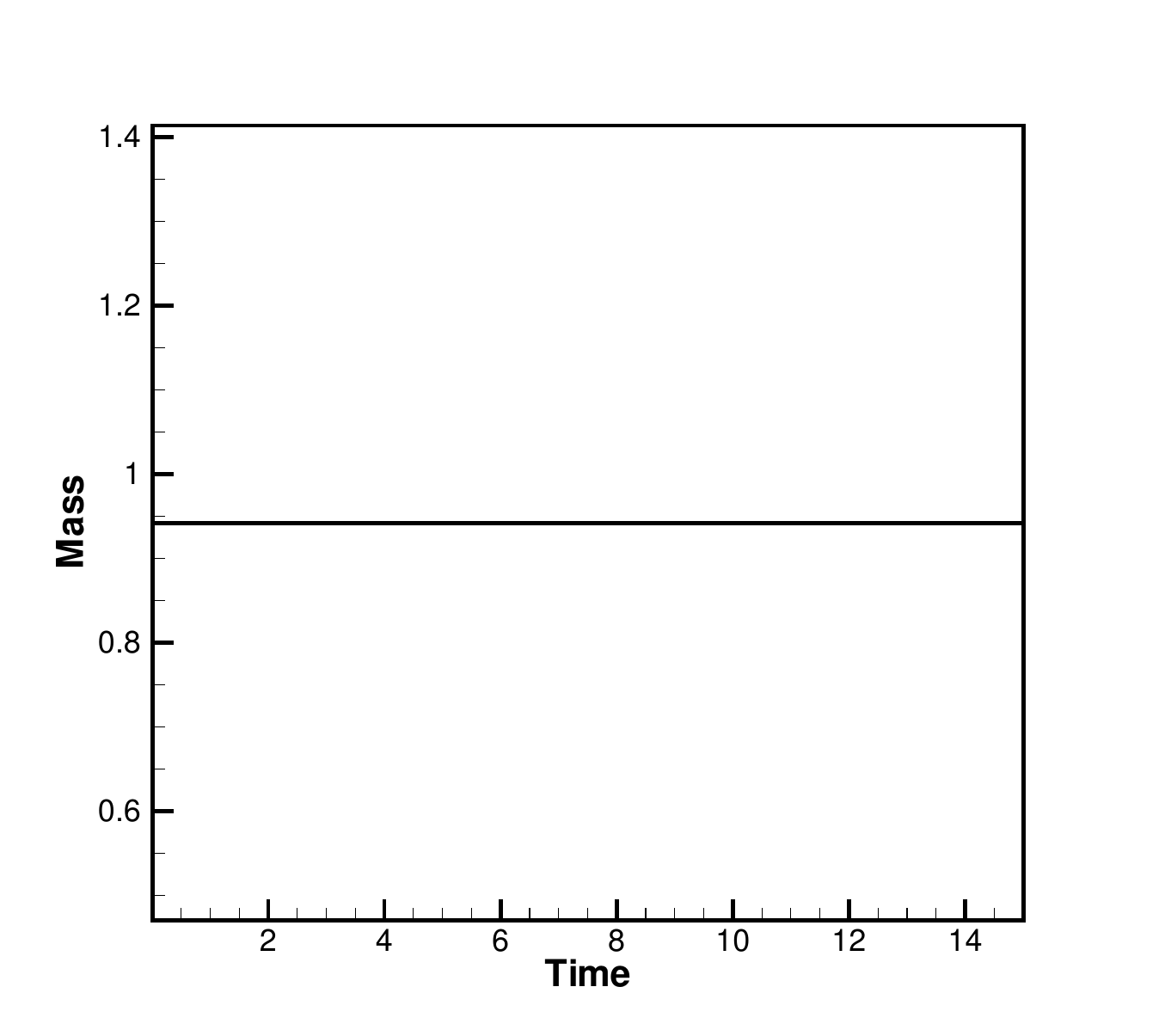}
	}
	\caption{Time evolution of the mass ($\int_{\Omega} \zeta_{+}C_{+} d\x$) with different permeability: $K=0,0.1$ and $0.3$ with  $c_0 =0.15$.}
	\label{mass-ex3}
\end{figure}

For the case $c_0 = 0.15$, the evolutions of energy and mass are presented in Figures \ref{energy-ex3}-\ref{mass-ex3}.  As expected, the energy decreases over time, confirming the stability of the scheme, while the total mass remains conserved throughout the simulation.

\begin{figure}\centering
	\subfigure[$K=0$] {
		\includegraphics[width=2.0in]{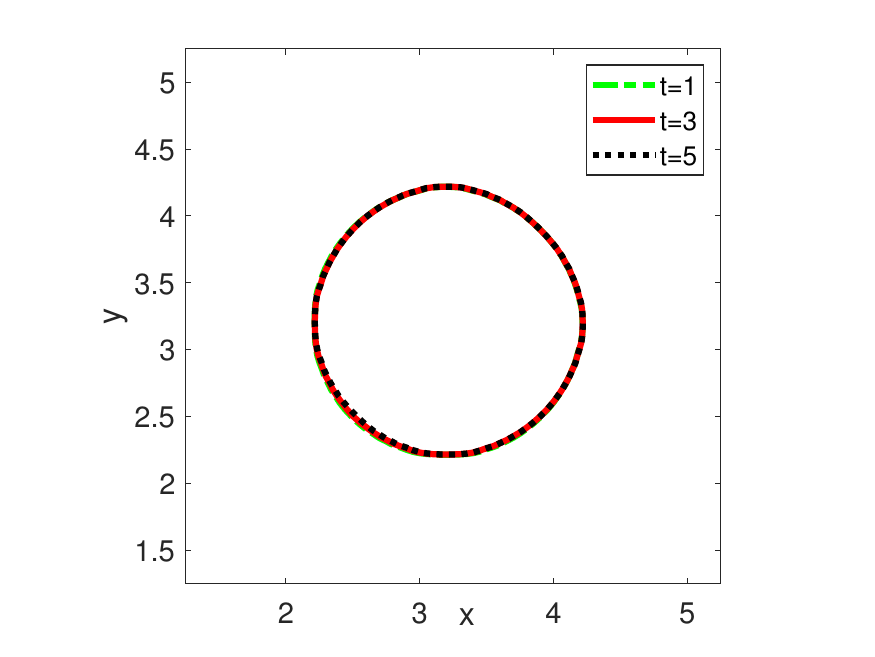}
	}
	\subfigure[$K=0.1$] {
		\includegraphics[width=2.0in]{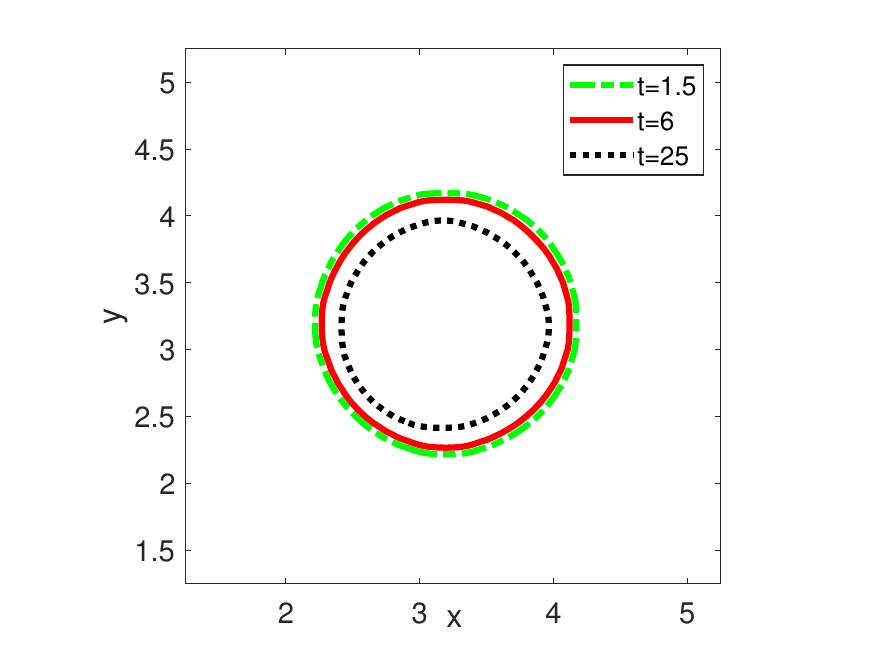}
	}
	\subfigure[$K=0.3$] {
		\includegraphics[width=2.0in]{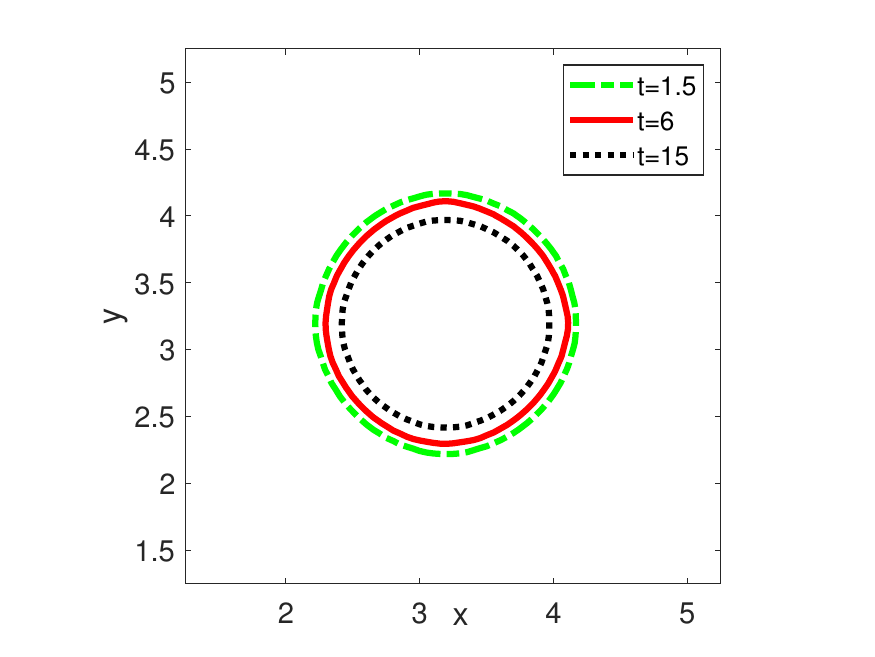}
	}
	\caption{Snapshots of interface $\phi=0$ at different time slots with different permeability: $K=0,0.1$ and $0.3$ with  $c_0 =0.15$.}
	\label{phi-equili_diffk}
\end{figure}

\begin{figure} \centering
	\subfigure[$K=0$] {
		\includegraphics[width=2.0in]{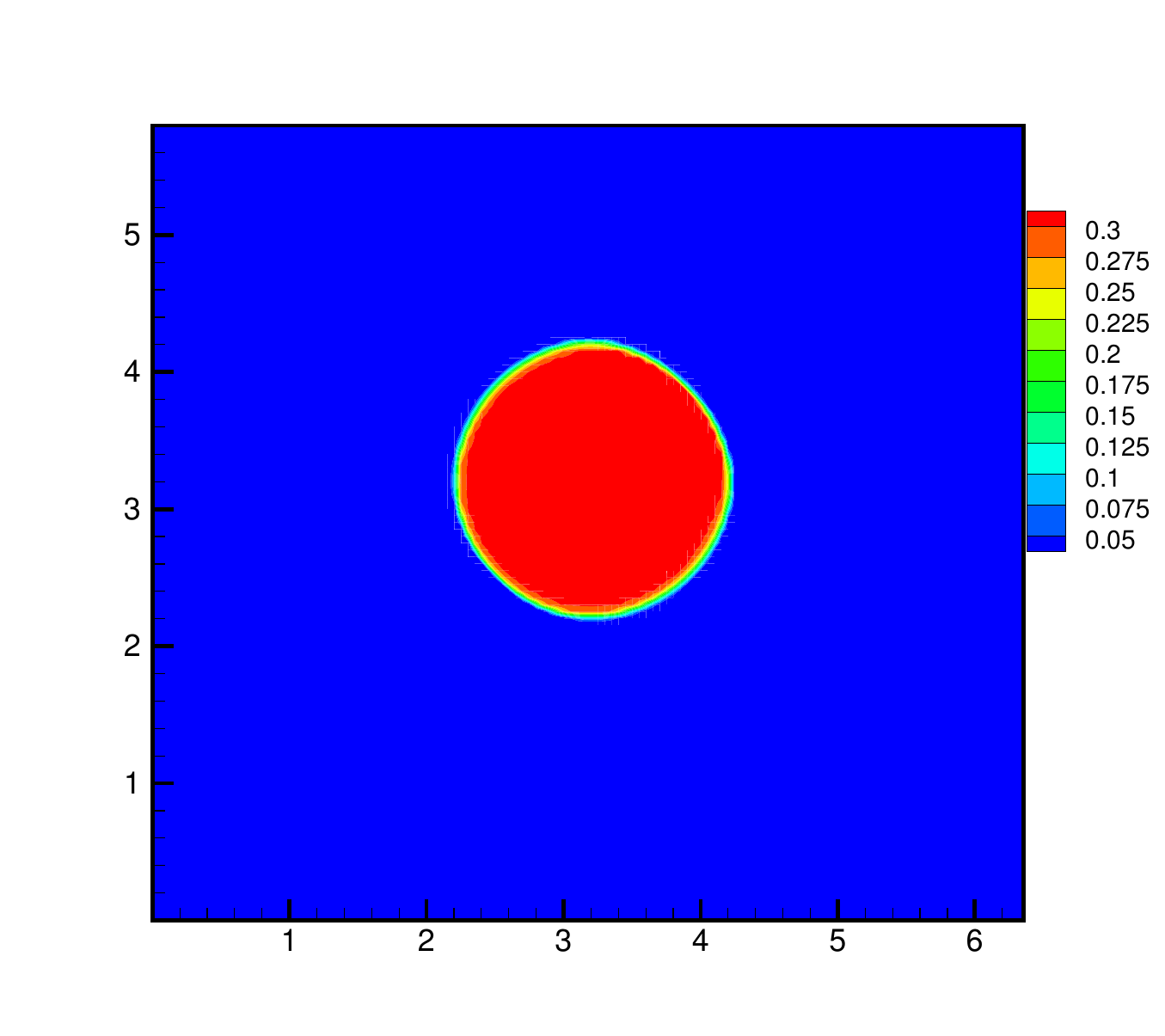}
	}
	\subfigure[$K=0.1$] {
		\includegraphics[width=2.0in]{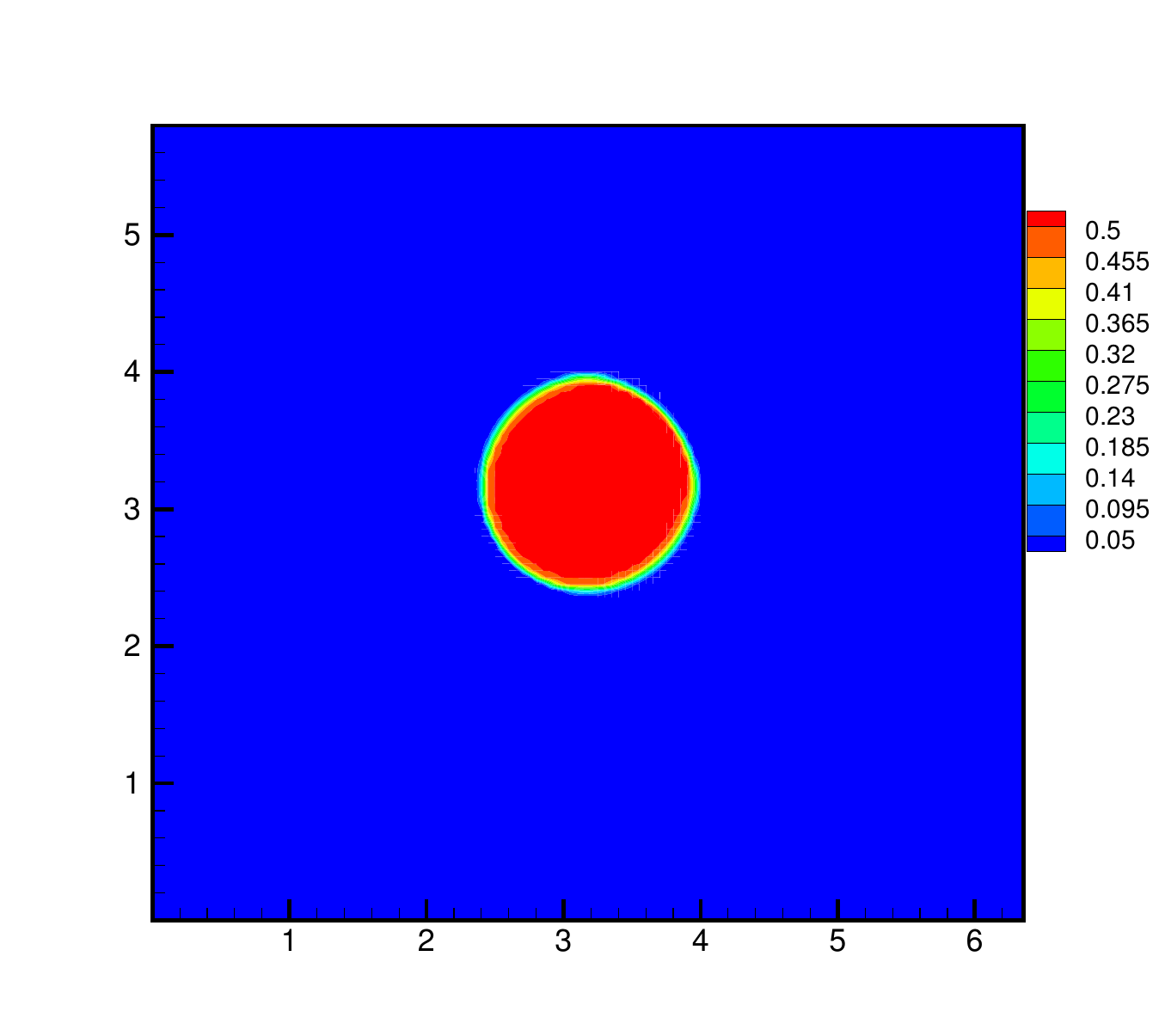}
	}
	\subfigure[$K=0.3$] {
		\includegraphics[width=2.0in]{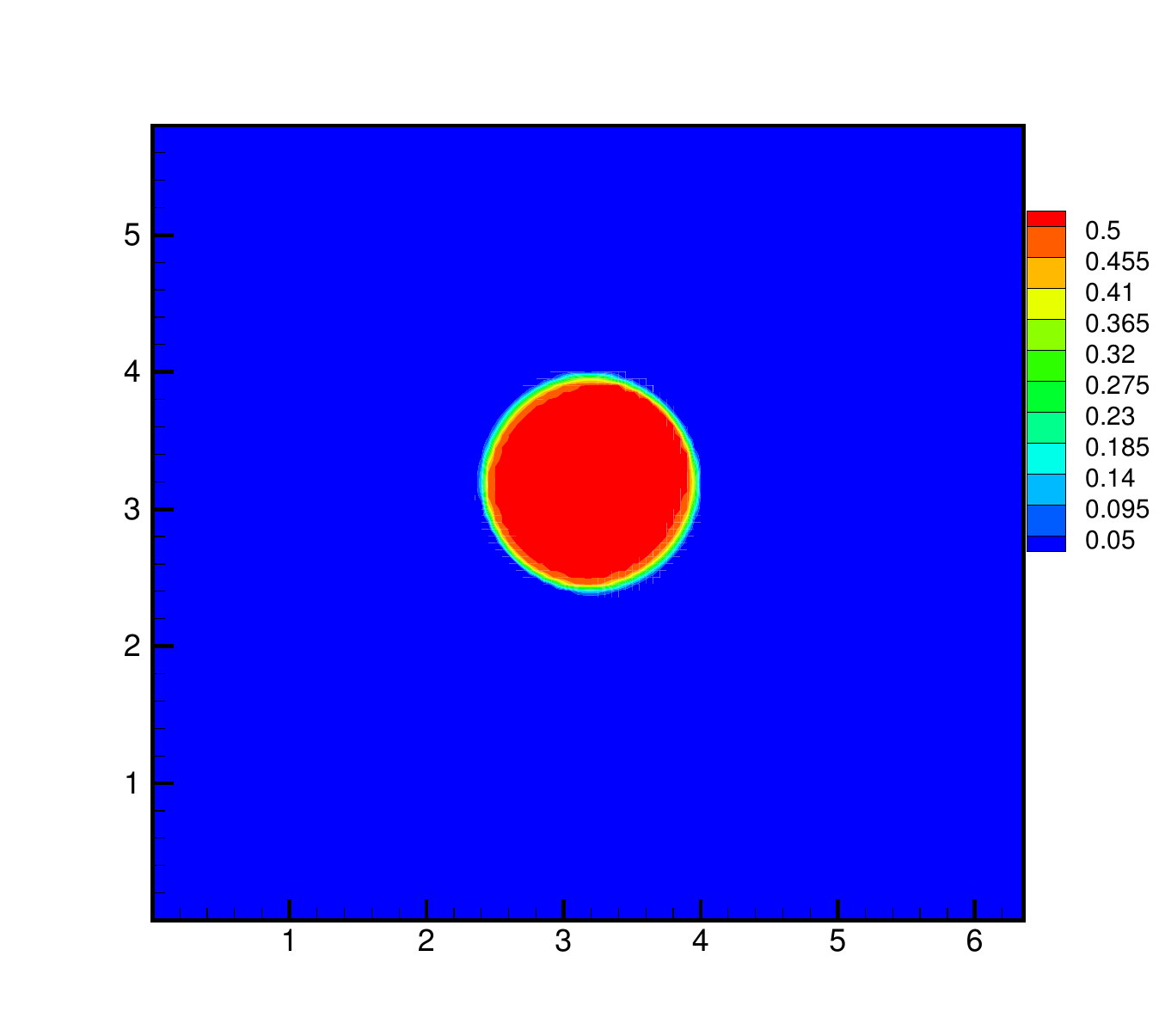}
	}
	\caption{Snapshots of $\zeta_{+}C_{+}$ at equilibrium with different permeability: $K=0,0.1$ and $0.3$ with   $c_0 =0.15$.}
	\label{cp-ex3}
\end{figure}

Figure~\ref{phi-equili_diffk} displays the evolution of droplet interfaces under different membrane permeabilities. The corresponding concentration profiles at equilibrium are shown in Figure~\ref{cp-ex3}. When the permeability \( K = 0 \), water cannot cross the membrane, so the droplet retains its initial volume. In contrast, when \( K \neq 0 \), the initial hydrostatic pressure inside the droplet exceeds the osmotic pressure outside, driving water outward and causing the droplet to shrink. Since the solute remains confined within the droplet, the volume loss increases the internal solute concentration, thereby raising the osmotic pressure. This elevated osmotic pressure gradually draws water back into the droplet until equilibrium is reached, where the hydrostatic and osmotic pressures are balanced.

According to the energy dissipation law stated in Theorem~\ref{energy law theorem}, for a closed system, the equilibrium state is characterized by the conditions
\[
\bm{u} = 0, \quad \tilde{\mu}_{\phi} = 0, \quad \mu_c^{\pm} = 0,
\]
which implies that the droplet shape at equilibrium is independent of the membrane permeability \( K \), and is determined solely by surface tension and the solute concentration difference across the interface. At equilibrium, the concentrations \( C^{\pm} \) inside and outside the droplet become spatially uniform, with their values determined by the total initial solute mass and the geometry of the final droplet. These behaviors are clearly observed in Figures~\ref{phi-equili_diffk}–\ref{cp-ex3}. The role of permeability primarily affects the rate at which equilibrium is reached. Higher permeability leads to faster transmembrane water exchange, accelerating droplet shrinkage and energy dissipation. Consequently, systems with larger \( K \) values reach equilibrium more rapidly, as demonstrated in Figures~\ref{energy-ex3} and \ref{phi-equili_diffk}.

To further investigate the role of osmotic pressure, we consider two cases: one with an initial concentration \( c_0 = 0.1 \), and another where the solute is entirely neglected. In the latter case, the system reduces to the degenerate NSCHAC system, where the Allen–Cahn term 
\(
-\frac{K}{Ca} \tilde{\mu}_{\phi} |\nabla \phi|
\)
governs the transmembrane flow driven purely by hydrostatic pressure. 

\begin{figure} \centering
	\subfigure[$t=1.5$] {
		\includegraphics[width=2.0in]{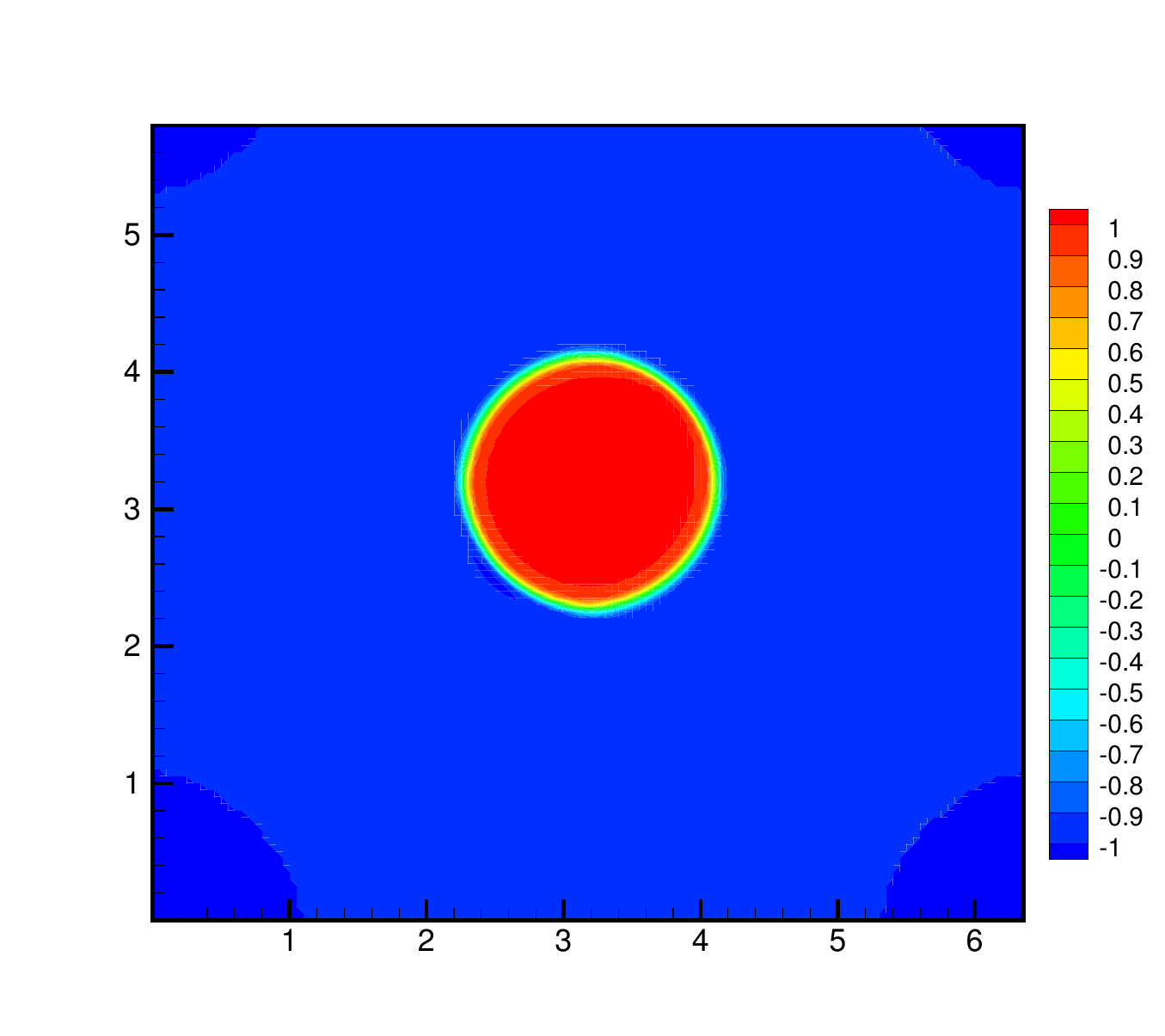}
	}
	\subfigure[$t=6$] {
		\includegraphics[width=2.0in]{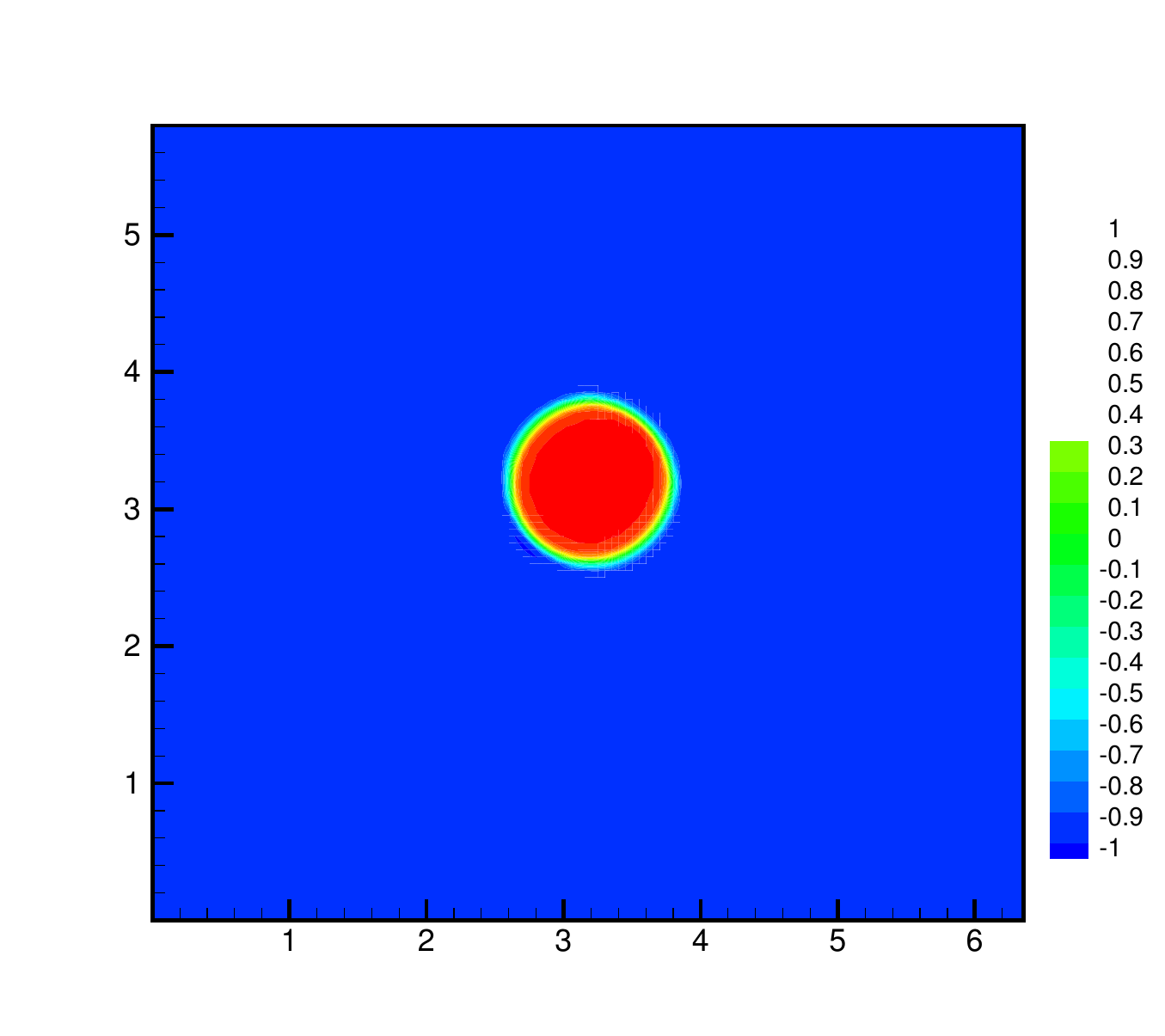}
	}
	\subfigure[$t=12$] {
		\includegraphics[width=2.0in]{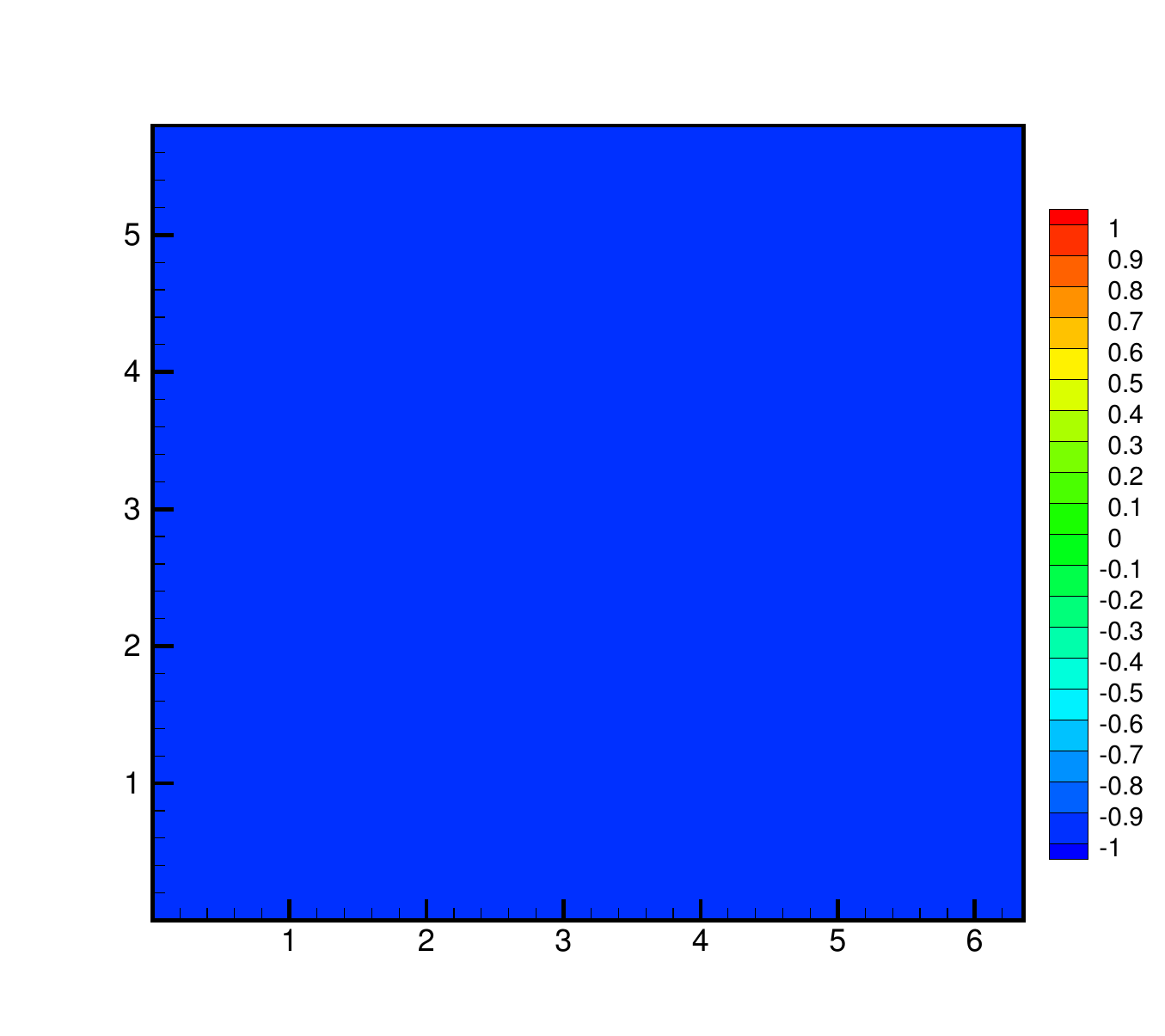}
	}
	\caption{The profiles of the interface when the osmotic pressure is neglected at time slots $t=1.5, 6, 12$ with permeability $K=0.3$.}
	\label{fig:interface_nocon_difftime}
\end{figure}

\begin{figure}
	\centering
	\includegraphics[width=0.5\linewidth]{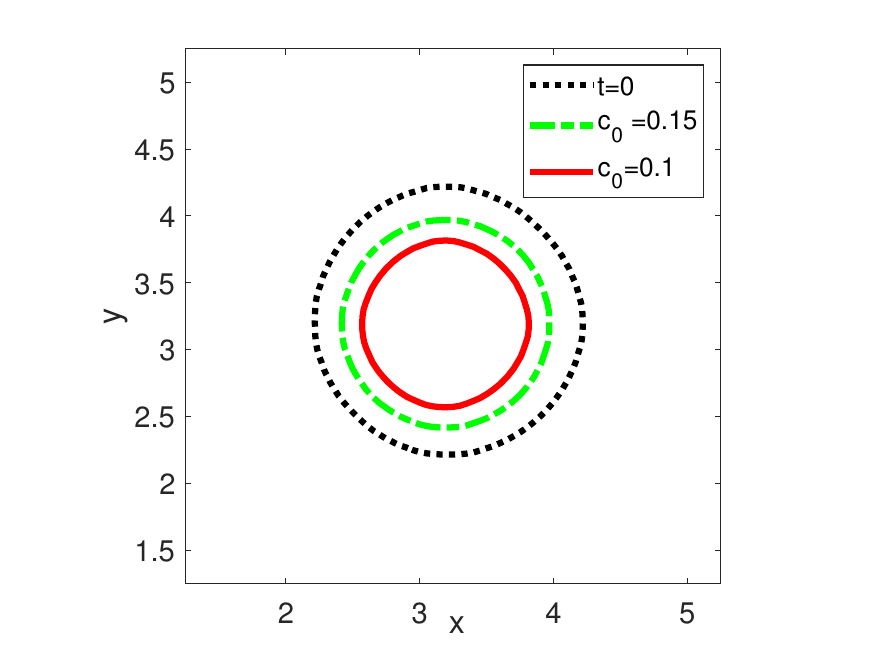}
	\caption{The interface profiles of the droplet at equilibrium for different initial concentrations \( c_0 = 0.1 \) and \( c_0 = 0.15 \) with permeability \( K = 0.3 \). The black dotted line represents the initial interface. The green dashed line shows the equilibrium interface profile for \( c_0 = 0.15 \) at \( t = 15 \), while the red solid line depicts the equilibrium interface profile for \( c_0 = 0.1 \) at approximately \( t = 20 \).
	}
	\label{fig:interfacediffc0}
\end{figure}

As shown in Figure~\ref{fig:interface_nocon_difftime}, when osmotic effects are neglected, the droplet eventually vanishes—a behavior consistent with previous studies. In contrast, when osmotic pressure is taken into account, the droplet evolves toward an equilibrium configuration, where the final interface profile is determined by the initial solute concentration \( c_0 \), as illustrated in Figure~\ref{fig:interfacediffc0}. A higher initial concentration inside the droplet leads to a stronger osmotic pressure, which counteracts the shrinkage earlier in the evolution and results in a larger equilibrium droplet size.


\begin{example}(Shear flow.)
	In this example, we study the dynamics of two droplets suspended in a shear flow within a square domain \( \Omega = [0, 6.4] \times [0, 6.4] \). The initial conditions for the phase-field variable \( \phi \), the solute concentrations \( C_{\pm} \)  and velocity $\bm{u}$ are given by
	{\scriptsize \begin{align*} 
			& \phi(x,y,0) = \tanh\left(\frac{1.0 - \sqrt{(x - 2.35)^2 + (y - 4.05)^2}}{\sqrt{2} \epsilon}\right)
			+ \tanh\left(\frac{1.0 - \sqrt{(x - 4.05)^2 + (y - 2.35)^2}}{\sqrt{2} \epsilon}\right) + 1.0, \\
			&C_{+}(x,y,0) = 0.1 + \exp\left(-\frac{(x - 2.35)^2 + (y - 4.05)^2}{0.15}\right)
			+ \exp\left(-\frac{(x - 4.05)^2 + (y - 2.35)^2}{0.15}\right), \\
			&C_{-}(x,y,0)=0.02, ~~   \bm{u} = \bm{0}. 
	\end{align*}}
	
	We consider two cases: impermeable membranes (\( K = 0 \)) and semi-permeable membranes (\( K = 0.3 \)), while keeping all other parameters identical to those in Example~\ref{example-ex3}. The boundary conditions are specified as
	\begin{align*}
		& \nabla \phi \cdot \bm{n}|_{y=0} = \nabla \phi \cdot \bm{n}|_{y=6.4} = 0, \quad \phi(0,y,t) = \phi(6.4,y,t), \\
		& \nabla \tilde{\mu}_{\phi} \cdot \bm{n}|_{y=0} = \nabla \tilde{\mu}_{\phi} \cdot \bm{n}|_{y=6.4} = 0, \quad \tilde{\mu}_{\phi}(0,y,t) = \tilde{\mu}_{\phi}(6.4,y,t), \\
		& \nabla C_{\pm} \cdot \bm{n}|_{y=0} = \nabla C_{\pm} \cdot \bm{n}|_{y=6.4} = 0, \quad C_{\pm}(0,y,t) = C_{\pm}(6.4,y,t), \\
		& \bm{u}|_{y=0} = (-1, 0)^{T}, \quad \bm{u}|_{y=6.4} = (1, 0)^{T},
	\end{align*}
	with periodic boundary conditions applied on the left and right boundaries.
\end{example}

\begin{figure} \centering
	\subfigure{
		\includegraphics[width=1.25in]{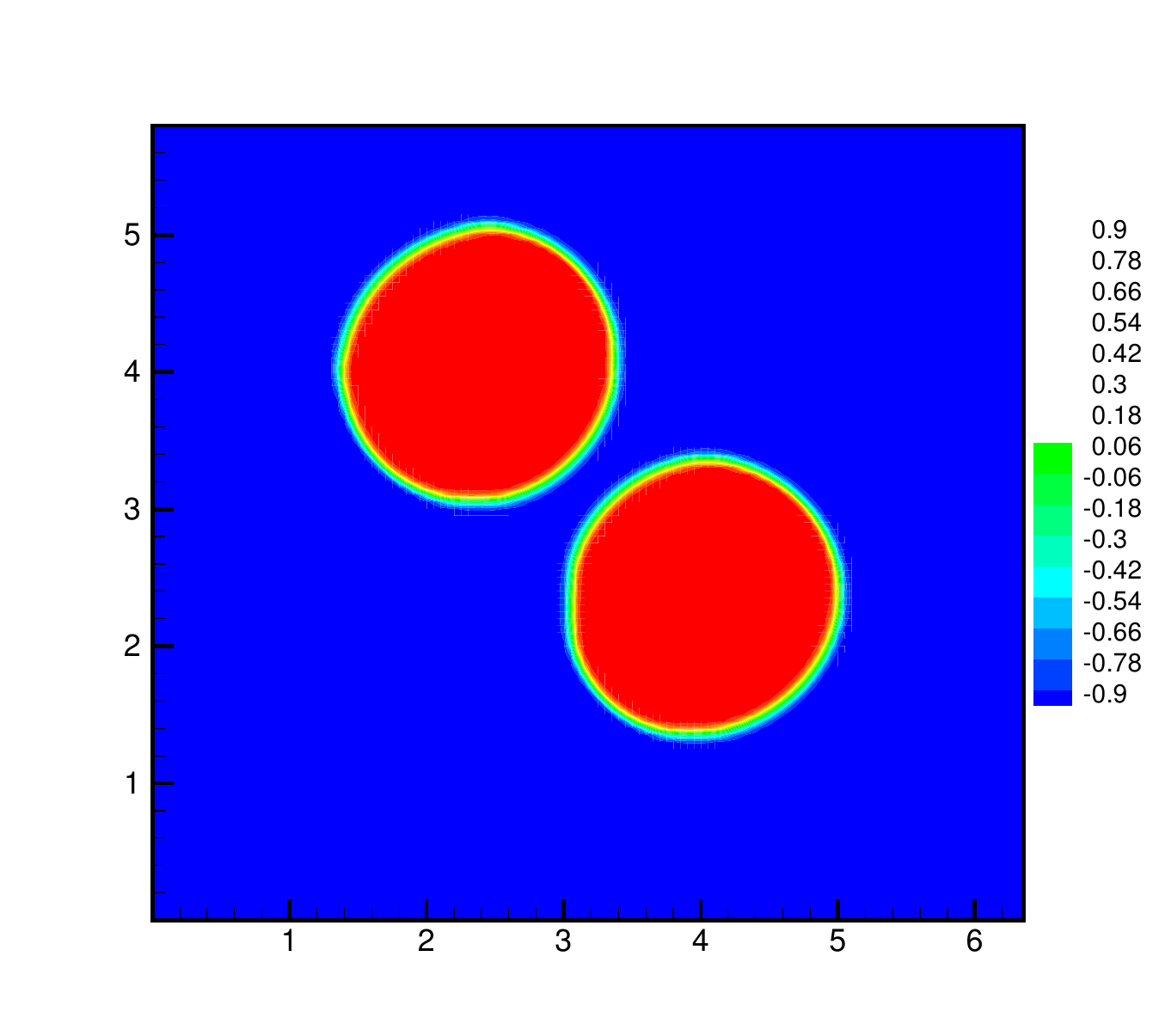}
	}
	\subfigure{
		\includegraphics[width=1.25in]{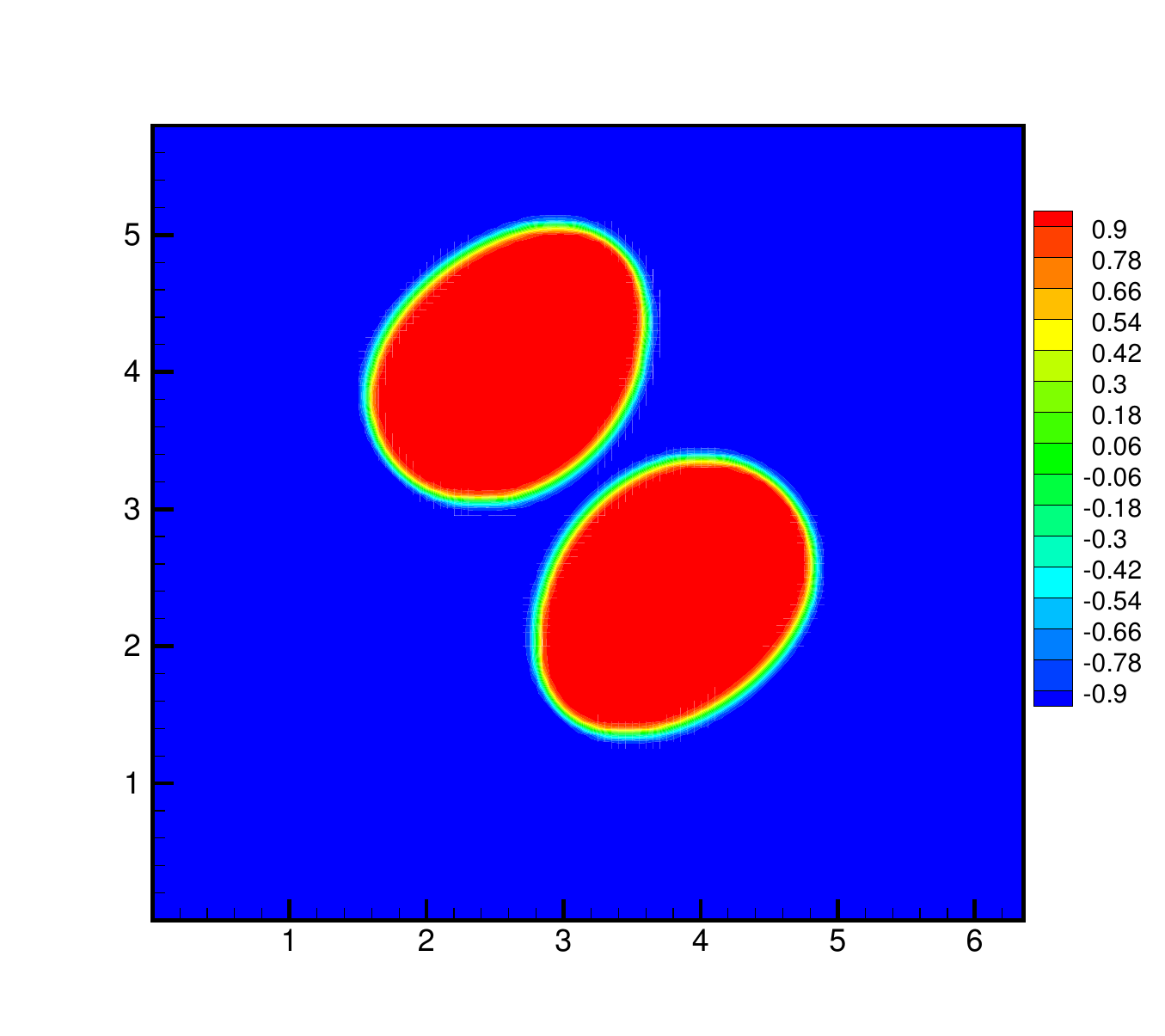}
	}
	\subfigure {
		\includegraphics[width=1.25in]{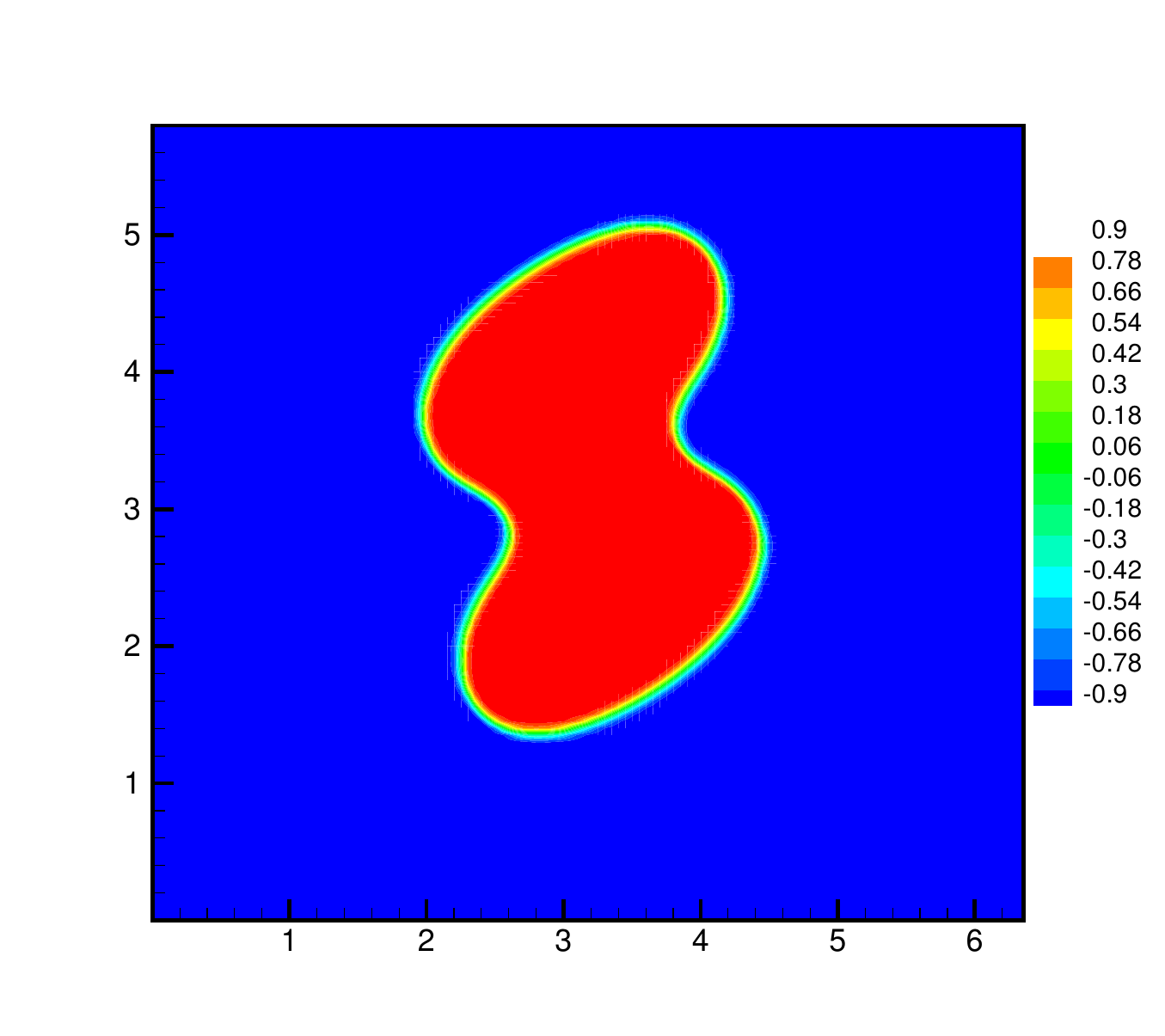}
	}
	\subfigure{
		\includegraphics[width=1.25in]{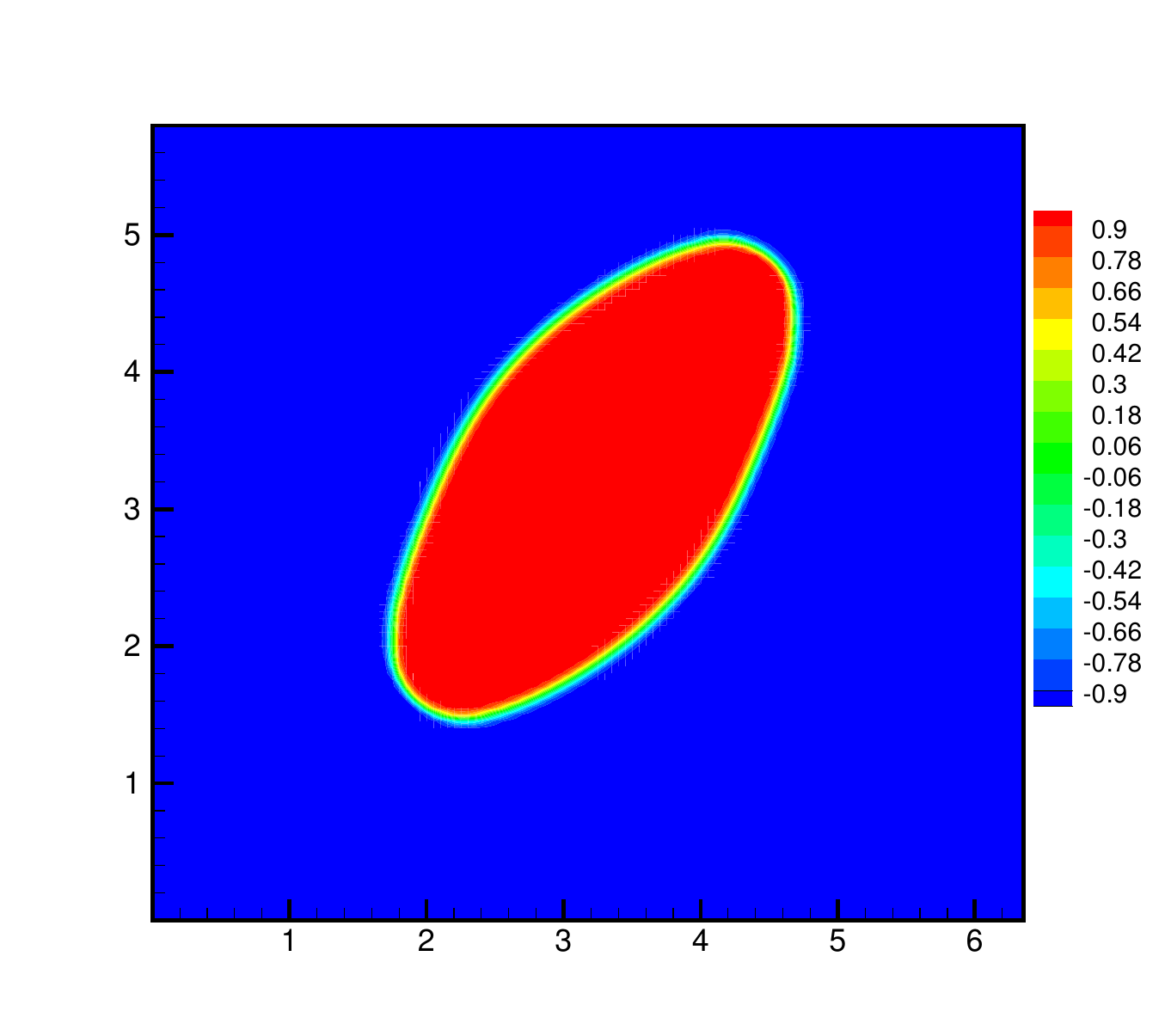}
	}
	\subfigure {
		\includegraphics[width=1.25in]{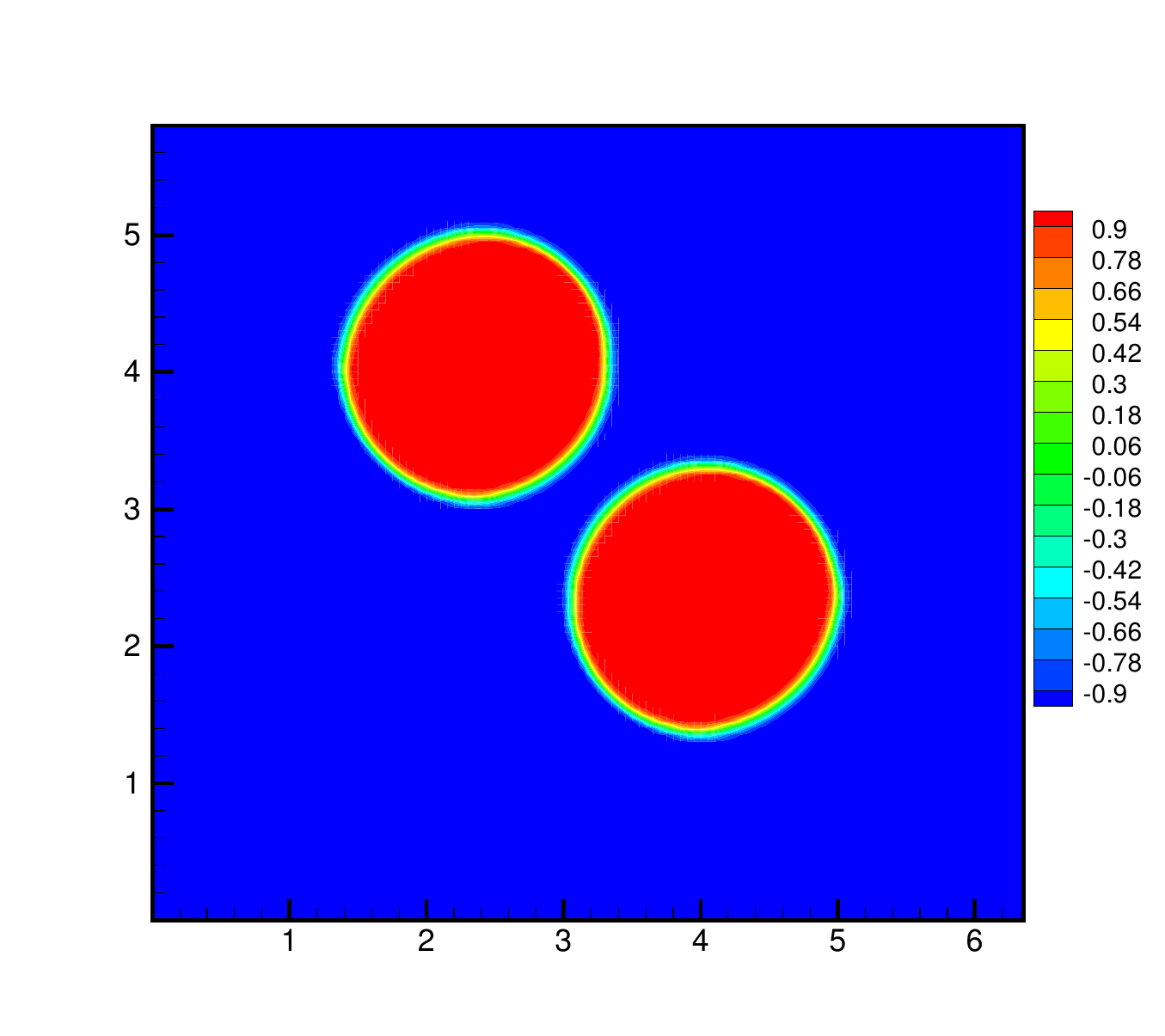}
	}
	\subfigure {
		\includegraphics[width=1.25in]{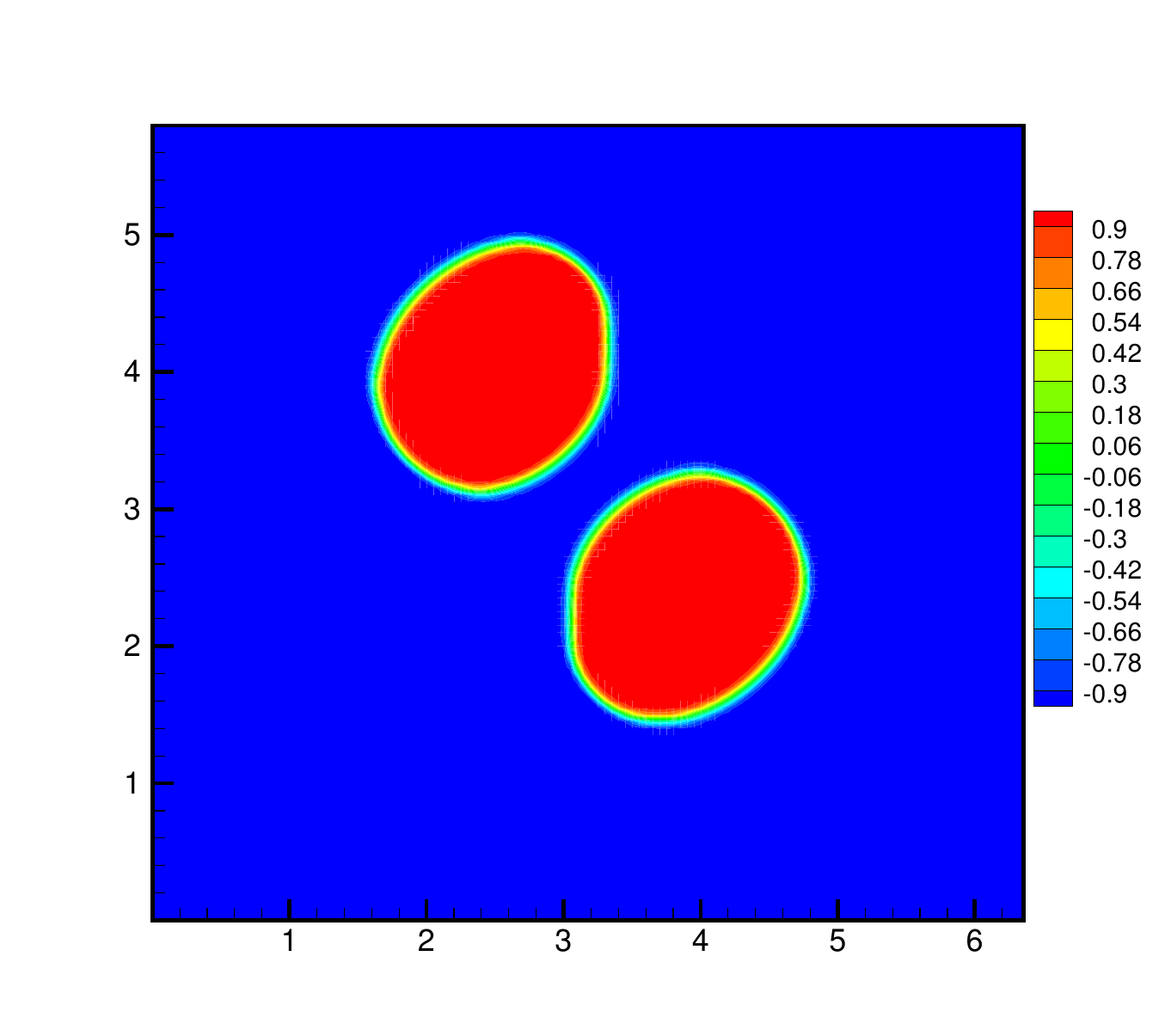}
	}
	\subfigure {
		\includegraphics[width=1.25in]{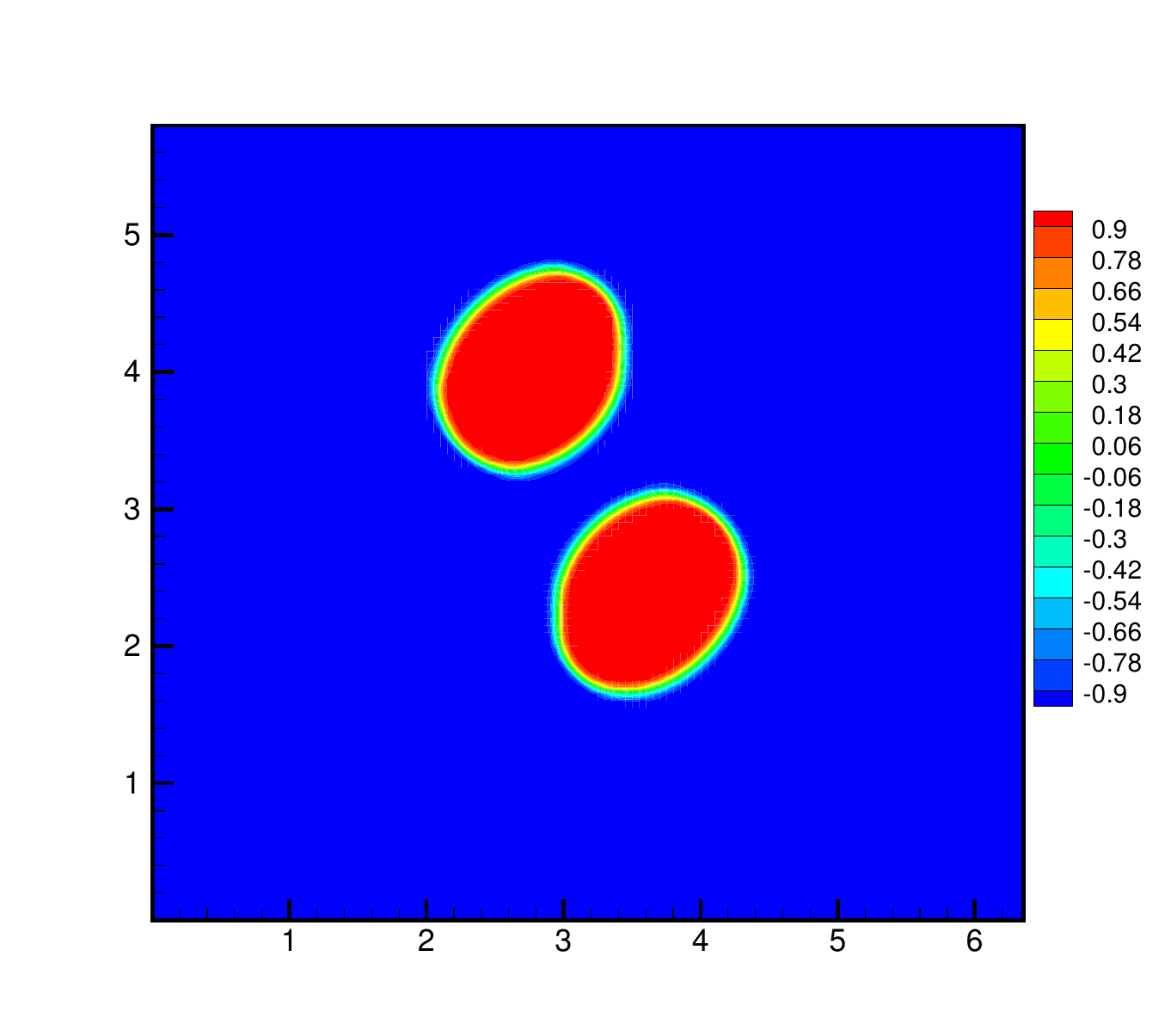}
	}
	\subfigure {
		\includegraphics[width=1.25in]{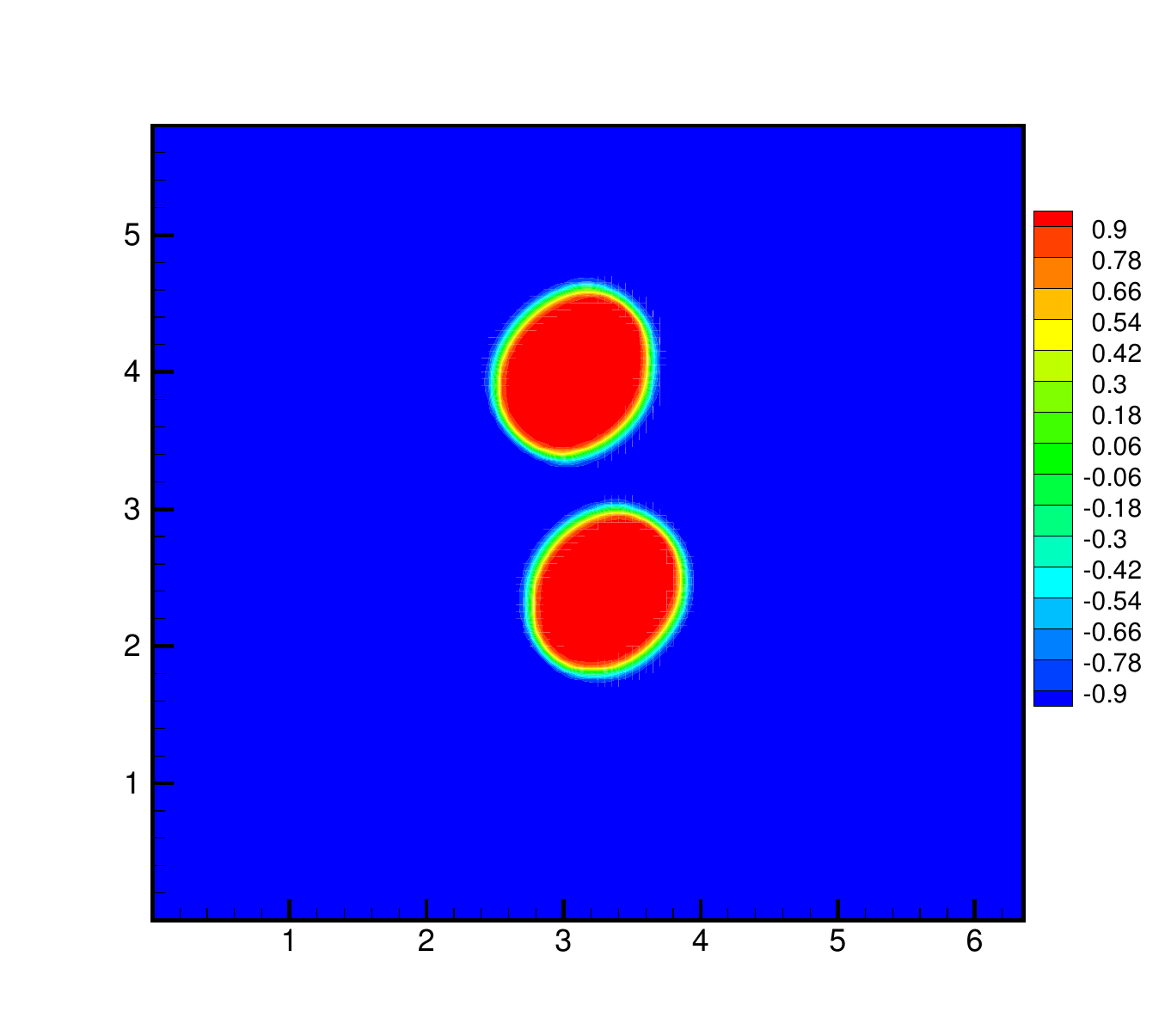}
	}
	\caption{  Snapshots of $\phi$ with $K=0$ (first row) and $K=0.3$ (second row) at different time $t=1.5$ (first column), $t=6$ (second column), $t=15$ (third column) and $t=25$ (fourth column).}
	\label{shear-two-phi-K0}
\end{figure}

Figure~\ref{shear-two-phi-K0} shows the evolution of the droplet interfaces  at various time instances for the two different permeability settings. When \( K = 0 \), the droplets are impermeable, and their enclosed volume remains constant. Under the shear flow, the two initially separated droplets move closer, eventually coalesce into a single elongated droplet, and rotate with the background flow.
In contrast, when \( K = 0.3 \), the membranes become semi-permeable to water. The imbalance between osmotic and hydrostatic pressures leads to transmembrane water flux, causing the droplets to gradually shrink. As the droplet volume decreases, the spatial extent of each droplet reduces, preventing the two droplets from coming into contact and merging. This behavior highlights how permeability not only affects the rate of deformation but also alters the qualitative topological outcome—suppressing coalescence under shear flow due to volume reduction induced by osmotic regulation.

\section{Conclusions}

In this work, we developed a thermodynamically consistent phase-field model for simulating water transport across semi-permeable membranes driven by osmotic pressure and hydrostatic pressure differences. Our formulation extends the classical Navier–Stokes–Cahn–Hilliard (NSCH) system by incorporating a transmembrane water flux governed by solute gradients and membrane permeability. This extension leads to a highly nonlinear Navier–Stokes–Cahn–Hilliard–Allen–Cahn (NSCHAC) system, where volume changes induced by osmotic flow dynamically feed back into the solute distribution, resulting in a tightly coupled and highly nonlinear interaction between fluid motion, interfacial dynamics, and solute transport.

To address the computational challenges posed by this strongly coupled system, we proposed a suite of energy-stable and high-order numerical schemes. Spatial discretization is performed using the local discontinuous Galerkin (LDG) method, which provides high-order accuracy, nonlinear stability, and flexibility for adaptive mesh refinement. For time integration, we first constructed a first-order decoupled temporal scheme and rigorously proved its unconditional energy stability. To further improve temporal accuracy, we incorporated a semi-implicit spectral deferred correction (SDC) method, achieving high-order convergence in both time and space while preserving the model's thermodynamic consistency.

Our framework provides a flexible and robust tool for simulating osmotic-driven fluid transport through deformable interfaces, offering significant advantages for modeling complex biological and industrial systems. Numerical experiments validate the theoretical properties of the proposed schemes and demonstrate the effects of membrane permeability, solute concentration, and shear flow on droplet deformation and equilibrium morphology.
Future work will focus on extending the model to account for additional biophysical processes, such as ionic transport, electrochemical coupling, and active membrane mechanics.

\label{se:concluding}


\section*{Acknowledgment} 
The research results of this article are partially supported by the National Natural Science Foundation of China No. 12271492 (R. Guo and S. Xu), and the Natural Science Foundation of Henan Province, China grant No. 252300420322 (R. Guo).

\section*{Conflicts of interest}
The authors declare that they have no known competing financial interests or personal relationships that could have appeared to influence the work reported in this paper.

\section*{Declaration of generative AI and AI-assisted technologies in the writing process}
During the preparation of this work, the authors used ChatGPT to improve the language and readability. After using this tool/service, the author(s) reviewed and edited the content as needed and take(s) full responsibility for the content of the published article.

\bibliographystyle{myplain}
\bibliography{sample}

\newpage
\appendix
\section{Model Derivation Details}
We calculate the time derivative of the total energy given in \eqref{energyfunction}.
For the first term, by using the first two equations of equation \eqref{kimematic assume oss}, we have 
\begin{align}\label{eqn: kinetic dissipation}
I_1 
=&\frac{d}{dt}(\frac{1}{2}\int_{\Omega}\rho|\bm{u}|^{2}d\bm{x})
\nonumber\\
=&\frac{1}{2}\int_{\Omega}\frac{\partial\rho}{\partial t}|\bm{u}|^{2}d\bm{x}
+\int_{\Omega}\rho\bm{u}\cdot\frac{\partial \bm{u}}{\partial t}d\bm{x}
\nonumber\\
=&\frac{1}{2}\int_{\Omega}\frac{\partial\rho}{\partial t}|\bm{u}|^{2}d\bm{x}
+\int_{\Omega}\rho\bm{u}\cdot\frac{\textbf{D}\bm{u}}{\textbf{D}t}d\bm{x}
+\int_{\Omega}\nabla\cdot(\rho\bm{u})\frac{|\bm{u}|^{2}}{2}d\bm{x}
\nonumber\\
=&\int_{\Omega}\bm{u}\cdot(\nabla\cdot\bm\sigma_{\eta}+\nabla\cdot\bm\sigma_{\phi})d\bm{x}
-\int_{\Omega}p\nabla\cdot\bm{u}d\bm{x}
\nonumber\\
=&-\int_{\Omega}\nabla\bm{u}:(\bm\sigma_{\eta}+\bm\sigma_{\phi})d\bm{x}
-\int_{\Omega}p\nabla\cdot\bm{u}d\bm{x}. 
\end{align}
where pressure is induced as a Lagrange multiplier for incompressibility. 

For the second term,  the equation of the concentrations and label functions  yields 
\begin{eqnarray}
I_2+I_3 &=& \frac{d}{dt}\left\{  \int_{\Omega}RT \left( \zeta_+   C_+(\ln{\frac{ C_+}{c_\infty}}-1) +  \zeta_- C_-(\ln{\frac{C_-}{c_\infty}}-1)\right) d\bm{x}+ \int_{\Omega} \lambda \left(\frac{l^2|\nabla\phi|^2 }{2}+F(\phi)\right) d\bm{x}\right\}\nonumber\\
&=& \int_{\Omega}RT \left(\zeta_+\ln \frac{ C_+}{c_\infty}\frac{\partial  C_+}{\partial t} +\zeta_-\ln \frac{ C_+}{c_\infty}\frac{\partial C_-}{\partial t} \right)d\bm{x} \nonumber\\
&&+ \int_{\Omega}\left(RT\left( \frac{\partial \zeta_+}{\partial \phi}   C_+(\ln{\frac{ C_+}{c_\infty}}-1) +  \frac{\partial\zeta_-}{\partial\phi} C_-(\ln{\frac{C_-}{c_\infty}}-1)\right)+\mu_{\phi}\right)\frac{\partial\phi}{\partial t}d\bm{x} \nonumber\\
&=& \int_{\Omega} RT \left(\ln \frac{ C_+}{c_\infty}\frac{\partial (\zeta_+ C_+)}{\partial t} +\ln \frac{ C_+}{c_\infty}\frac{(\partial \zeta_-C_-)}{\partial t} \right) d\bm{x}  \nonumber\\
&& +  \int_{\Omega}\left(RT\left( \frac{\partial \zeta_+}{\partial \phi}   C_+(\ln{\frac{ C_+}{c_\infty}}-1) +  \frac{\partial\zeta_-}{\partial\phi} C_-(\ln{\frac{C_-}{c_\infty}}-1)\right)+\mu_{\phi}- C_+\ln{\frac{ C_+}{c_\infty}}\frac{\partial \zeta_+}{\partial \phi} -C_-\ln{\frac{C_-}{c_\infty}}\frac{\partial \zeta_-}{\partial \phi}\right)\frac{\partial\phi}{\partial t}d\bm{x} \nonumber\\
&=&  \int_{\Omega} RT \left(\ln \frac{ C_+}{c_\infty}\frac{\partial (\zeta_+ C_+)}{\partial t} +\ln \frac{ C_+}{c_\infty}\frac{\partial (\zeta_-C_-)}{\partial t} \right) d\bm{x}   
+   \int_{\Omega}\left(RT\left(- \frac{\partial \zeta_+}{\partial \phi}   C_+ - \frac{\partial\zeta_-}{\partial\phi} C_-  \right)+\mu_{\phi} \right)\frac{\partial\phi}{\partial t}d\bm{x} \nonumber\\
&=& \int_{\Omega} RT \left(-\ln \frac{ C_+}{c_\infty}  \nabla\cdot(\zeta_+ C_+\bm{u})  -\ln \frac{ C_+}{c_\infty}\nabla\cdot (\zeta_-C_-\bm{u}) \right) d\bm{x}  \nonumber\\
&&+ \int_{\Omega} RT \left(-\ln \frac{ C_+}{c_\infty}  \nabla\cdot(\zeta_+\bm{J}_+)  -\ln \frac{C_-}{c_\infty}\nabla\cdot (\zeta_-\bm{J}_-) \right) d\bm{x}   \nonumber\\
&&+\int_{\Omega}\tilde{\mu}_{\phi}(-\nabla\cdot(\bm{u}\phi)-\nabla\cdot\bm{J}_{\phi}+S_{\phi}|\nabla\phi|)d\bm{x} \nonumber\\
&=& \int_{\Omega}(RT\zeta_+\nabla  C_+ \cdot\bm{u} +RT\zeta_-\nabla C_- \cdot\bm{u} )d\bm{x}  +\int_{\Omega} (\zeta_+\bm{J}_+\cdot\nabla\mu_c^++\zeta_-\bm{J}_-\cdot\nabla\mu_c^-) d\bm{x} \nonumber\\
&&+\int_{\Omega} (\nabla\tilde{\mu}_{\phi}\phi \cdot\bm{u}+ \bm{J}_{\phi}\cdot\nabla\tilde{\mu}_{\phi}) d\bm{x}  +\int_{\Omega} \tilde{\mu}_{\phi} S_{\phi}|\nabla\phi|d\bm{x} \nonumber\\
&=& \int_{\Omega}(RT\zeta_+\nabla  C_+  +RT\zeta_-\nabla C_-  +\nabla\tilde{\mu}_{\phi}\phi)\cdot\bm{u}d\bm{x} \nonumber\\
&&+\int_{\Omega} (\zeta_+\bm{J}_+\cdot\nabla\mu_c^++\zeta_-\bm{J}_-\cdot\nabla\mu_c^-)d\bm{x}  +\int_{\Omega}   \bm{J}_{\phi}\cdot\nabla\tilde{\mu}_{\phi}  d\bm{x} +\int_{\Omega} \tilde{\mu}_{\phi} S_{\phi}|\nabla\phi|d\bm{x}  \nonumber\\
&=& \int_{\Omega}(-(RT  \frac{\partial \zeta_+}{\partial \phi}   C_+  + RT\frac{\partial \zeta_-}{\partial \phi} C_-)-\tilde{\mu}_{\phi} )\nabla\phi\cdot\bm{u}d\bm{x}  +\int_{\Omega} (\zeta_+\bm{J}_+\cdot\nabla\mu_c^++\zeta_-\bm{J}_-\cdot\nabla\mu_c^-)d\bm{x} \nonumber \\
&&+\int_{\Omega}   \bm{J}_{\phi}\cdot\nabla\tilde{\mu}_{\phi}  dx +\int_{\Omega} \tilde{\mu}_{\phi} S_{\phi}|\nabla\phi|d\bm{x}  \nonumber\\
&=& -\int_{\Omega}\mu_{\phi}\nabla\phi\cdot\bm{u}d\bm{x}  +\int_{\Omega} (\zeta_+\bm{J}_+\cdot\nabla\mu_c^++\zeta_-\bm{J}_-\cdot\nabla\mu_c^-) d\bm{x} \\
&&+\int_{\Omega}   \bm{J}_{\phi}\cdot\nabla\tilde{\mu}_{\phi}  d\bm{x} +\int_{\Omega} \tilde{\mu}_{\phi} S_{\phi}|\nabla\phi|d\bm{x}   \nonumber\\ 
&=&  \int_{\Omega} -\lambda  l^2 (\nabla\phi\otimes\nabla\phi):\nabla\bm{u}d\bm{x}  +\int_{\Omega} (\zeta_+\bm{J}_+\cdot\nabla\mu_c^++\zeta_-\bm{J}_-\cdot\nabla\mu_c^-) d\bm{x} +\int_{\Omega}   \bm{J}_{\phi}\cdot\nabla\tilde{\mu}_{\phi}  d\bm{x}   +\int_{\Omega} \tilde{\mu}_{\phi} S_{\phi}|\nabla\phi|d\bm{x} \nonumber 
\end{eqnarray}

In summary, we have 
\begin{align*}
&\frac{dE}{dt} \\
=&   -\int_{\Omega}\nabla\bm{u}:(\bm\sigma_{\eta}+\bm\sigma_{\phi})d\x
-\int_{\Omega}p\nabla\cdot\bm{u}d\x  - \int_{\Omega} \lambda l^2 (\nabla\phi\otimes\nabla\phi):\nabla\bm{u}d\x +\int_{\Omega} (\zeta_+\bm{J}_+\cdot\nabla\mu_c^++\zeta_-\bm{J}_-\cdot\nabla\mu_c^-)d\x \\
& +\int_{\Omega}   \bm{J}_{\phi}\cdot\nabla\tilde{\mu}_{\phi}  d\x+\int_{\Omega} \tilde{\mu}_{\phi} S_{\phi}|\nabla\phi|d\x  \\
=&-\Delta.  
\end{align*}

\end{document}